\documentclass[aps,prd, twocolumn,amsmath, superscriptaddress, amssymb,showpacs, floatfix,nofootinbib, preprintnumbers]{revtex4-1}

\usepackage{graphicx}
\usepackage{dcolumn}
\usepackage{xcolor}
\usepackage{bm}
\usepackage{natbib}
\usepackage{calligra}
\usepackage[T1]{fontenc}
\usepackage{egothic}
\usepackage[T1]{fontenc}
\newfont{\rsfsten}{rsfs10 scaled 1200}
\newfont{\rsfsseven}{rsfs10 scaled 1200}
\newfont{\rsfsfive}{rsfs10 scaled 1200}
\usepackage{epsfig}
\usepackage{units}
\usepackage[utf8]{inputenc}

\newcommand{\beq}{\begin{equation}}
\newcommand{\eeq}{\end{equation}}
\newcommand{\bea}{\begin{eqnarray}}
\newcommand{\eea}{\end{eqnarray}}

\newcommand{\alg}[1]{\begin{align} \begin{split} #1 \end{split}  \end{align}}

\def\lsim{\mathrel{\raise.3ex\hbox{$<$\kern-.75em\lower1ex\hbox{$\sim$}}}}
\def\gsim{\mathrel{\raise.3ex\hbox{$>$\kern-.75em\lower1ex\hbox{$\sim$}}}}

\renewcommand{\deg}{^\circ}

\begin{document}

\preprint{FERMILAB-PUB-18-057-A-T}

\title{Analyzing the Gamma-ray Sky with Wavelets}

\author{Bhaskaran Balaji}
\email[Corresponding author: ]{bbalaji1@jhu.edu}
\affiliation{Department of Physics and Astronomy, The Johns Hopkins University, Baltimore, Maryland, 21218, USA}
\author{Ilias Cholis}
\affiliation{Department of Physics and Astronomy, The Johns Hopkins University, Baltimore, Maryland, 21218, USA}
\author{Patrick J.~Fox}
\affiliation{Theoretical Physics Department, Fermi National Accelerator Laboratory, Batavia, Illinois, 60510, USA}
\author{Samuel D.~McDermott}
\affiliation{Center for Particle Astrophysics, Fermi National Accelerator Laboratory, Batavia, Illinois, 60510, USA}

\date{\today}

\begin{abstract}

We analyze the gamma-ray sky at energies of 0.5 to 50 GeV using the undecimated wavelet transform on the sphere. Focusing on the inner $60^{\circ} \times 60^{\circ}$  
of the sky, we identify and characterize four separate residuals beyond the expected Milky Way diffuse emission. We detect the \textit{Fermi} 
Bubbles, finding compelling evidence that they are diffuse in nature and contain very little small-scale structure. We detect the ``cocoon'' inside 
the Southern Bubble, and we also identify its northern counterpart above 2 GeV. The Northern Cocoon lies along the same axis but is $\sim 30 \%$ 
dimmer than the southern one. We characterize the Galactic center excess, which we find extends up to $20^{\circ}$ in $|b|$. At latitudes $|b| \leq 5^{\circ}$ 
we find evidence for power in small angular scales that could be the result of point-source contributions, but for $|b| \geq 5^{\circ}$ the Galactic center 
excess is dominantly diffuse in its nature. Our findings show that either the Galactic center excess and {\it Fermi} Bubbles connect smoothly or that the 
Bubbles brighten significantly below $15^\circ$ in latitude. We find that the Galactic center excess appears off-center by a few degrees towards negative $\ell$.
 Additionally, we find and characterize two emissions along the Galactic disk centered at $\ell \simeq +25^{\circ}$ and $-20^{\circ}$. These emissions are 
 significantly more elongated along the Galactic disk than the Galactic center excess. 
\end{abstract}

\maketitle
	
\section{Introduction}

Electromagnetic radiation has allowed us a gateway to the mysteries of the Universe since time 
immemorial. Over the ages, we have become sensitive to radiation of increasingly higher energy.
The highest energy photons are classified as gamma rays. Gamma-ray astronomy started in 1961 
with 22 events observed by \textit{Explorer 11} \cite{1965ApJ...141..845K}. This was followed by 
\textit{OSO-3}, which observed 621 photons and provided the first proof of emission from our own  
Milky Way \cite{1972ApJ...177..341K}. Observations ensued with the \textit{SAS-2} \cite{1975ApJ...200L..79T} 
and \textit{COS-B} \cite{1981ApJ...243L..69S, 1989ARA&A..27..469B} instruments, and, upon the 
launching of the Compton Gamma Ray Observatory (CGRO), the \textit{BATSE} 
\cite{Band:1993eg, Kaneko:2006ru}, \textit{OSSE} \cite{Gierlinski:1996az}, \textit{COMPTEL} 
\cite{Schoenfelder:2000bu, Kuiper:2001ev}, and \textit{EGRET} 
\cite{Sreekumar:1997un, Hunger:1997we, Hartman:1999fc} instruments continued this exploration.
At the highest energies, CGROs' \textit{EGRET} was followed by the Fermi Large Area Telescope 
(\textit{Fermi-LAT}), aboard the Fermi Gamma Ray Space Telescope. The number of recorded events  
from the end of the X-ray spectrum up to tens of GeV has grown exponentially with time. The LAT 
instrument alone has collected 160 million {\tt CLEAN} class events above 50 MeV since 2008.
This growth in recorded events has allowed us to develop statistical techniques for analyzing both 
the spectral and the morphological information in the gamma-ray data.

Gamma ray photons are produced in some of the most energetic phenomena in the Universe.
This emission can be categorized by whether or not the source is large compared to the angular 
resolution of the observing instrument: a source is either a ``point source,'' localized well within the 
point spread function (PSF), or ``diffuse.'' Diffuse gamma-ray emission is expected to originate from 
cosmic rays (CRs) propagating in the Galaxy and interacting with the interstellar medium (ISM). 
The mechanism of diffuse emission is conventionally broken down into three classes, depending 
on the type of CR and the type of target it impinges upon. The dominant contribution to diffuse 
emission is from inelastic collisions of CR {\it nuclei} with ISM gas; these collisions produce neutral 
particles, predominantly $\pi^{0}$ and $\eta$ mesons, whose decay products include photons. This 
emission is conventionally referred to as $\pi^{0}$-emission \cite{1977ApJ...212...60S, Mori:1996aq}. 
CR {\it electrons} can also interact with the ISM gas \cite{Blumenthal:1970gc}. The resulting photons 
are collectively referred to as bremsstrahlung radiation. Finally, CR electrons may up-scatter low-energy 
background photons, from the Cosmic Microwave Background (CMB) as well as infrared and starlight, 
a process known as inverse Compton scattering (ICS) \cite{Blumenthal:1970gc}. Other sources of diffuse 
gamma-rays include extragalactic sources, such as the Andromeda Galaxy, the Large and Small 
Magellanic Clouds, and the isotropic gamma-ray background. The isotropic gamma-ray background is 
known to include starforming galaxies, distant and misaligned active galactic nuclei, and misidentified 
CRs \cite{Gehrels:1999ri, FermiSite}. In addition to these diffuse gamma rays, the gamma-ray sky also 
contains a multitude of point sources \cite{Acero:2015hja}, which may be blazars, millisecond pulsars, 
or, more exotically, annihilating dark matter inside galactic substructures 
\cite{Abazajian:2010zy, Cholis:2013ena, DiMauro:2013zfa, Ackermann:2014usa, DiMauro:2015tfa}. 
	
The standard method for analyzing gamma-ray data involves the use of ``templates,'' whereby one fits the 
observed gamma-ray emission as a linear combination of model ``maps'' characterizing the three Milky 
Way diffuse emission components. The model maps in turn rely on observations at larger wavelengths 
and models of CR injection and propagation at high energy. The product of these observations and models 
are turned into gamma-ray maps by codes such as 
\cite{Strong:1998fr, Strong:2015zva, GALPROPSite, Evoli:2008dv, DRAGONweb}. Such codes allow the 
observer to calculate the spatial and spectral characteristics of the emission from each physical process. 
Building these templates requires a detailed knowledge of both the ISM and the CR injection pattern and 
propagation behavior. Thus, this type of analysis depends on many complementary assumptions that 
characterize these aspects of the Milky Way. 

Despite having caveats on the expected background and foreground emission the template analysis technique has 
had some significant successes. One of the spectacular features of the Milky Way detected in the \textit{Fermi} data 
with the template-based approach is known as the \textit{Fermi} Bubbles: two lobes of gamma-ray radiation with energies 
0.5 GeV $\lsim E_\gamma \lsim 300$ GeV extending orthogonally to the disk from the Galactic center, about $50^\circ$ 
north and south of the Galactic disk \cite{Dobler:2009xz, Su:2010qj}. Likewise, gamma-ray emission from Loop I has been 
identified using template techniques \cite{2009arXiv0912.3478C, Su:2010qj}. Great attention has also been placed on 
an excess of GeV gamma-rays seen towards the Galactic Center and the Inner Galaxy 
\cite{Goodenough:2009gk,  Hooper:2010mq, Hooper:2011ti, Abazajian:2012pn, 
Hooper:2013rwa, Gordon:2013vta, Daylan:2014rsa, Calore:2014xka, TheFermi-LAT:2015kwa}, which were 
discovered using templates. The template technique has been implemented in disentangling the extragalactic 
gamma-ray background emission \cite{Abdo:2010nz, Ackermann:2014usa} as well as the study of background/foreground 
emission around point sources \cite{Casandjian:2015ura, 2010ApJS..188..405A, 2012ApJS..199...31N, Acero:2015hja,  
Ackermann:2015uya, TheFermi-LAT:2015kwa, Fermi-LAT:2017yoi} from the gamma-ray data.

In this work we revisit the gamma-ray emission from the \textit{Fermi} Bubbles and the Galactic center excess (GCE). 
We also characterize two additional components of gamma ray emission along the Galactic plane, identified 
previously by the {\it Fermi} collaboration \cite{DiskExtendedEmissions}. Different template-based approaches 
have been used to characterize the \textit{Fermi} Bubbles \cite{Dobler:2009xz, Su:2010qj, Fermi-LAT:2014sfa} 
and the GCE \cite{Abazajian:2012pn, Hooper:2013rwa, Gordon:2013vta, Huang:2013pda, Daylan:2014rsa, 
Zhou:2014lva, Calore:2014xka, TheFermi-LAT:2015kwa, Carlson:2016iis, Linden:2016rcf}.
The goal of this work is to extract information about these excesses with fewer assumptions than required with 
templates. Instead, we develop a method of analyzing gamma-ray maps directly based on the morphology of their 
underlying constituents. For this purpose, we use a {\it wavelet transform}. Wavelet transforms are widely used in other 
applications in image processing including image denoising \cite{1998A&AS..128..397S, 2000ApJ...534..490K, 
2002PASP..114.1051S, Starck:2005fb, 2006aida.book.....S, 2012MNRAS.422.1674M, PATIL2015849}, but they have 
not been applied to the analysis of gamma-ray data in place of, or alongside template-based methods to characterize 
large-scale features on the sky.  
The approach we take is to decompose a map using the isotropic undecimated wavelet transform on the sphere 
(IUWTS) \cite{Starck:2005fb}. We extend our previous work in \cite{McDermott:2015ydv}, where this method was tested 
on mock data. Each of our decomposed maps has the same number of pixels as the initial map. These output maps 
display features of increasing minimum angular size centered at each pixel. 
	
In section~\ref{sec:method} we discuss the data that we use, our initial assumptions, 
and the implementation of the wavelet transform. In section~\ref{sec:results} we present our results, while 
in~\ref{sec:PrevWork} we compare these results with results from template analyses and 
initiate discussion of possible physical interpretations of these emissions. Finally in 
section~\ref{sec:Conclusions} we conclude and discuss future directions for gamma-ray analysis that we can
pursue with such techniques.
	
\begin{table*}[t]
\begin{center}
\begin{tabular}{c|cccccccc} 
(all in units of GeV) & $E_1$ & $E_2$ & $E_3$ & $E_4$ & $E_5$ & $E_6$ \\ \hline
$E$ range & 0.465-1.021\; & 1.021-2.041\; & 2.041-4.919\; & 4.919-10.799\; & 10.799-23.707\; &  23.707-52.043\; \\ 
mean $E$ & 0.68 & 1.5 & 3.3 & 7.3 & 16 &  35  \\ 
\end{tabular}
\end{center}
\caption{Energy range and central value for the energy binning described in Sec.~\ref{subsec:data}.}
\label{tab:E_bins}
\end{table*}%

\section{Methodology}
\label{sec:method}

In this section we discuss the \textit{Fermi-LAT} data selection that we use, all the basic aspects of 
our wavelet analysis, and any additional assumptions we make before implementing the wavelet transform. 
The complete description of our technique and the motivations behind certain choices are given in 
\cite{McDermott:2015ydv}, where tests were performed using mock gamma-ray maps. The reader interested 
in our findings regarding the \textit{Fermi} Bubbles and the GCE can go directly to section~\ref{sec:results}. 

\subsection{Data Selection}
\label{subsec:data}

We use Pass 8 gamma-ray data taken from\footnote{The \textit{Fermi-LAT} data are publicly available at https://fermi.gsfc.nasa.gov/ssc/data/access/} 
August 4th 2008 through November 2nd 2017. We use all {\tt CLEAN} class events, with additional filters 
{\tt DATA$\_$QUAL==1}, {\tt LAT$\_$CONFIG==1}, and {\tt ABS(ROCK$\_$ANGLE) < 52}. We avoid data 
collected when the LAT passes through the South Atlantic Anomaly. We use Fermi {\tt ScienceTools P8v10r0p5} 
for selection event-cuts and to calculate the relevant exposure cube-files and exposure maps
\footnote{https://fermi.gsfc.nasa.gov/ssc/data/analysis/}. Our event maps are in {\tt HEALPix}\footnote{http://healpix.sf.net} 
projection \cite{Gorski:2004by} with {\tt NSide=128} (resolution index of 7), so that each map has 196,608 
equal area pixels covering the full sky.  We have cross-checked our findings with Pass 7 reprocessed data taken 
between August 4th 2008 and August 14 2014, see Appendix~\ref{app:Pass7}. For the Pass 7 data we use 
{\tt ScienceTools P7v9r33p0} to obtain {\tt FRONT}-converted {\tt CLEAN} events, using the same cuts otherwise 
as above.

Because the wavelet decomposition splits the original map into maps of emission at different angular scales 
(see discussion in subsection~\ref{subsec:WaveletTr} and \cite{McDermott:2015ydv}), the starting maps must 
include a large number of photons. This necessitates wide energy bins. We use six energy bins shown in 
Table~\ref{tab:E_bins}. For the reminder of  the paper we refer to these energy bins by their geometric mean 
values or their label as given in Table~\ref{tab:E_bins}.

\subsection{Templates and Point Sources}
\label{subsec:StartingAss}
\begin{figure*}[htbp]
\centering
\includegraphics[width=.45\textwidth]{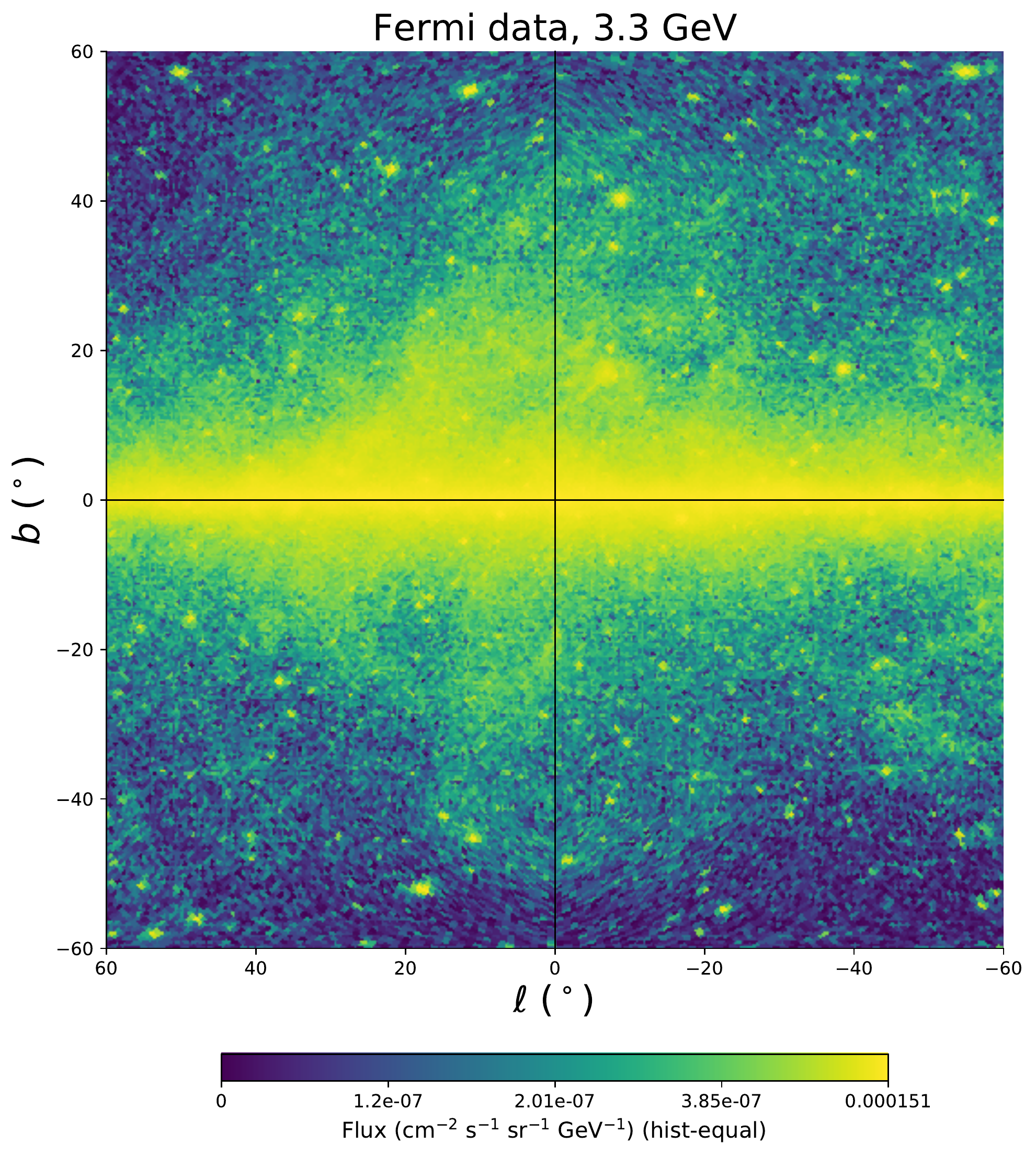} \qquad~~
\includegraphics[width=.45\textwidth]{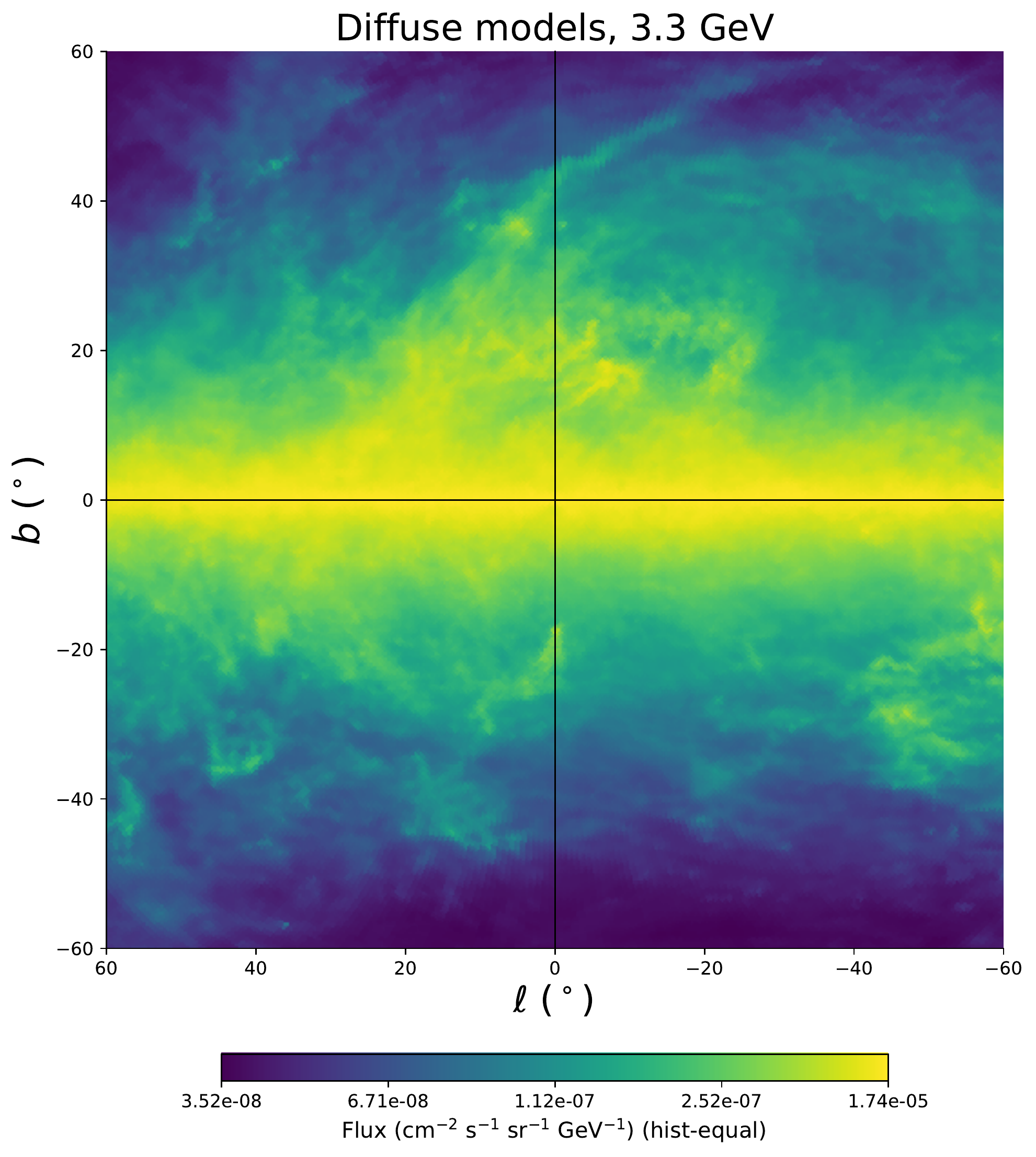}
\caption{\textit{Left}, the flux map $D(E_3)$ for \textit{Fermi} data from $2.041-4.919$ GeV. \textit{Right}, 
the average of our 19 diffuse models at the same energy. We show a region of $120^{\circ} \times 120^{\circ}$
around the Galactic center.}
\label{D0}
\end{figure*}

In this work, we analyze the gamma-ray emission in each of the six energy bins independently. We do not do 
a combined fit across all energy bins. As a result, our findings at any given energy bin rely only on the observations 
and instrument characterization for that particular energy range. For each energy bin, we start with a map of 
\textit{Fermi} data, $D(E)$.  As an example, we show the data for the energy bin centered at 3.3 GeV in 
Figure~\ref{D0} (left). Next, we subtract the expected diffuse emission given in Figure~\ref{D0} (right). To evaluate 
the expected Galactic diffuse emission, we average 19 diffuse models developed in \cite{Ackermann:2012pya, Calore:2014xka}, 
each of which accounts for $\pi^0$ decay, ICS, and bremsstrahlung emission. We take models A-D, F-R, W 
and GXI in the listing of \cite{Calore:2014xka} (Appendix A), which is the same set that was used in \cite{McDermott:2015ydv}. 
We do not select the 19 diffuse emission models to best fit the $D$ map. The 19 model maps were selected to 
envelop the diffuse emission uncertainties in the CR sources, the CR propagation, and the ISM gas and 
radiation field distributions. For further details see \cite{Ackermann:2012pya, Calore:2014xka, McDermott:2015ydv}. 
The systematic uncertainty that arises as a result of this choice of diffuse -and also point source- backgrounds 
is discussed in more detail in section \ref{sec:results}. In brief, we repeat the first step in our analysis by
replacing the average of the 19 models by alternative sets of diffuse emission models.
	
\begin{figure}[htbp]
\centering
\begin{minipage}{.45\textwidth}
\includegraphics[width=\linewidth]{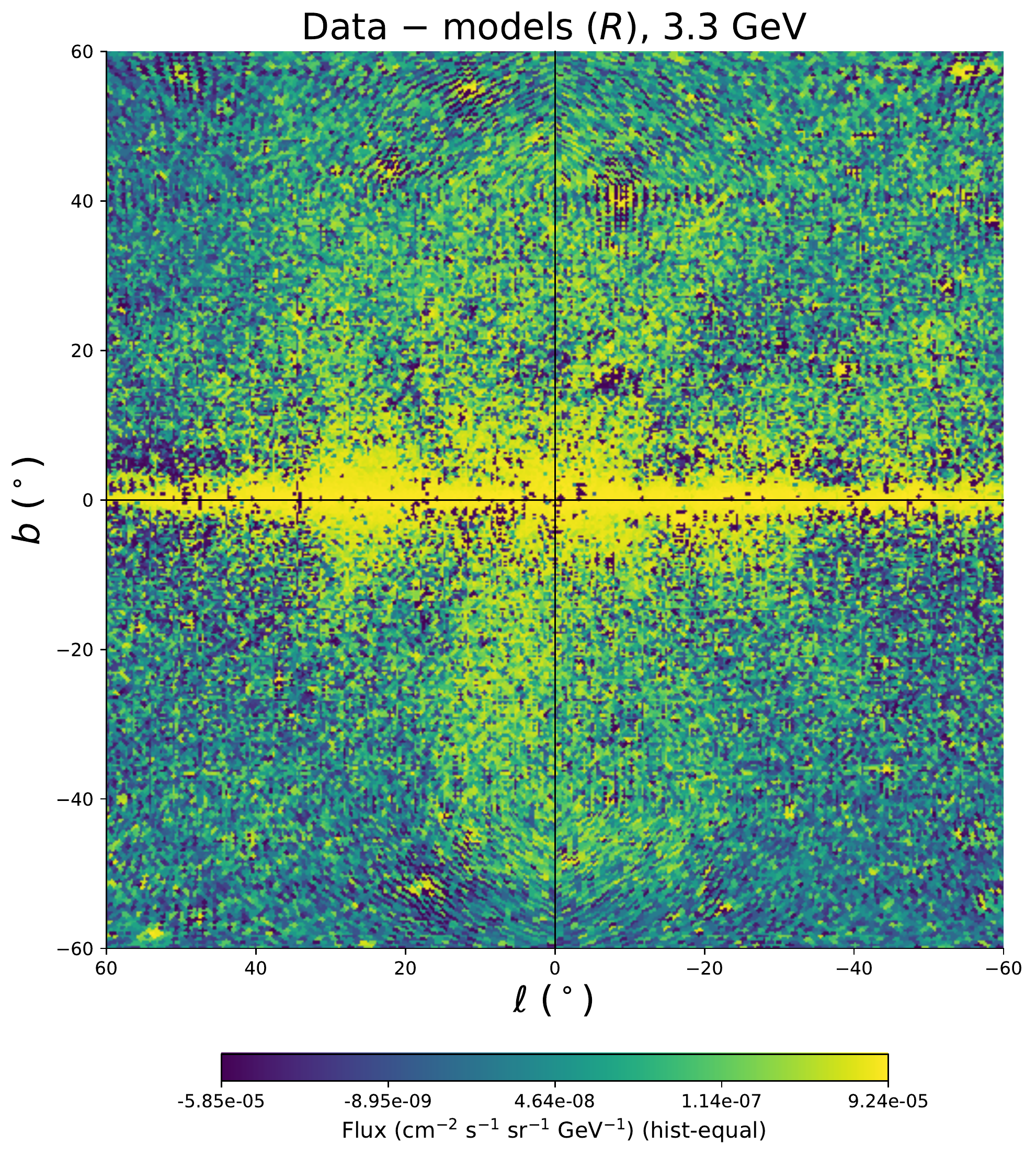}
\end{minipage}
\caption{The residual flux map $R(E_3)$: the difference of the fluxes in Figure~\ref{D0}.
}
\label{R0}
\end{figure}

We also subtract the emission of established gamma-ray point sources. The point sources we use for our main 
analysis are given by the {\tt 1FIG} \cite{TheFermi-LAT:2015kwa} catalog inside a $15^\circ \times 15^\circ$ 
square centered at the Galactic Center and the {\tt 3FGL} \cite{Acero:2015hja} catalog outside of this region 
(with the exception of {\tt 3FGL J1709.7-4429}, a very bright source, which we discuss in more detail in 
Appendix~\ref{sec:bright-PS}). We also test our results using only the {\tt 3FGL} point-source catalog. As 
shown in Appendix~\ref{app:15by15}, our results are robust against these uncertainties, as expected 
from \cite{Calore:2014xka}. 

We repeat this procedure for each energy bin to produce residual maps that we refer to as $R(E)$. In this way, 
the first step in our analysis is a template procedure. By removing the main diffuse emission of the data map 
$D$, we can focus on unexplained features in the residual map. The residual map $R(E_3)$ is shown in 
Figure~\ref{R0}.

\subsection{Using the Wavelet Transform}
\label{subsec:WaveletTr}

Our analysis relies in an essential way on the wavelet transform. The wavelet transform decomposes the data 
in such a way that the decomposed data simultaneously retain information about the position and the angular 
scale of the initial data. This wavelet decomposition is performed after the template-based step described in the 
preceding subsection. The output of the wavelet decomposition can either be in the form of a partition of the initial 
data, referred to as the ``discrete wavelet transform,'' or in a redundant form referred to as the ``stationary'' or 
``undecimated'' wavelet transform, which is translationally invariant. An undecimated wavelet transform therefore 
allows direct comparison of different wavelet ``levels'' against one another. Because the wavelet transform 
decomposes the map into different angular scales, differences between data and predictions are revealed as a 
function of characteristic angular scale. This is the key feature of the wavelet transform we wish to exploit.

Specifically, we use the Isotropic Undecimated Wavelet on the Sphere (IUWTS) \cite{Starck:2005fb}. The IUWTS 
is a spherical harmonic decomposition convolved with a special window function $\psi$, defined below. Because 
the IUWTS operates on the spherical harmonics of the initial image, it is inherently nonlocal: each level provides 
information about structures of different angular size with support at a given point. For more details on the history, 
uses, and varieties of wavelet transforms, we refer the interested reader to \cite{2006aida.book.....S, 1992tlw..conf.....D} 
and references therein.

The output of the IUWTS applied to a residual map $R(E)$ is $j_{\rm max}$ different angular ``wavelet levels'' plus 
a monopole term; here and in what follows we will always take $j_{\rm max}=9$. It is desirable that the window function 
that defines the wavelet decomposition is isotropic, so that azimuthal angular information is provided only by the initial data.
The IUWTS relies on a ``window function'' $\psi_{\ell_c}$ whose spherical harmonics are defined as the difference of 
``smoothing functions.'' Letting hats indicate the spherical harmonic transform, the smoothing functions are cubic splines,
\alg{
\hat \phi (x)=\frac18 \big( |x+&2|^3 - 4|x+1|^3 + 6|x|^3  \\ &-4|x-1|^3+|x-2|^3 \big) \Theta(4-x^2),
}
where $\Theta(x)$ is the Heaviside step function. The window functions are $\hat \psi_{\ell_{\rm max},j} = \hat\phi\left(2^j\ell/\ell_{\rm max}\right) - \hat\phi\left(2^{j+1}\ell/\ell_{\rm max}\right)$
and the wavelet levels are $\hat w_j(E; R) \propto \hat \psi_{\ell_{\rm max},j} \hat R(E)$ \cite{McDermott:2015ydv, Starck:2005fb}.

\begin{figure*}[ht]
\begin{centering}
\includegraphics[width=2.3in,angle=0]{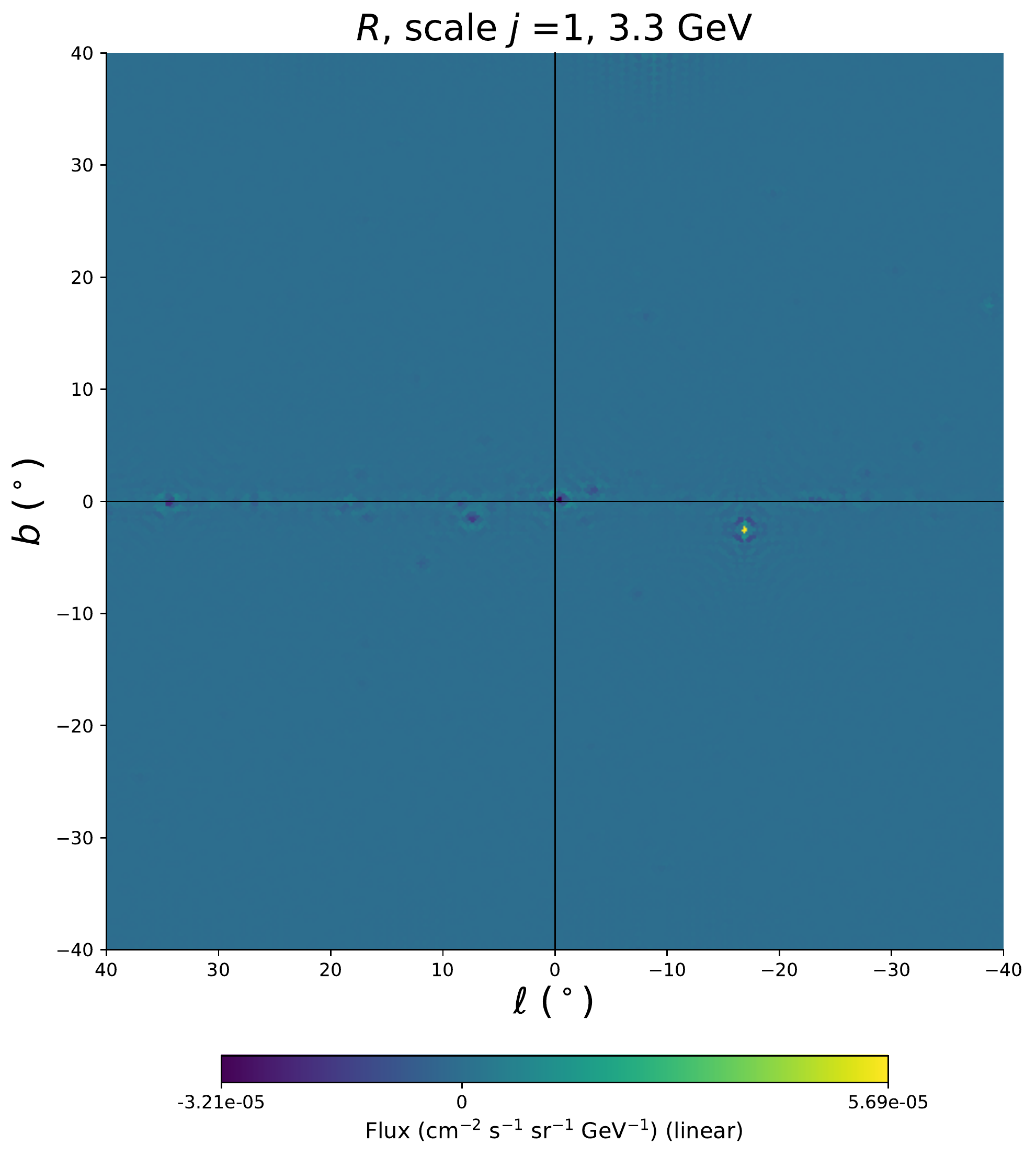}
\includegraphics[width=2.3in,angle=0]{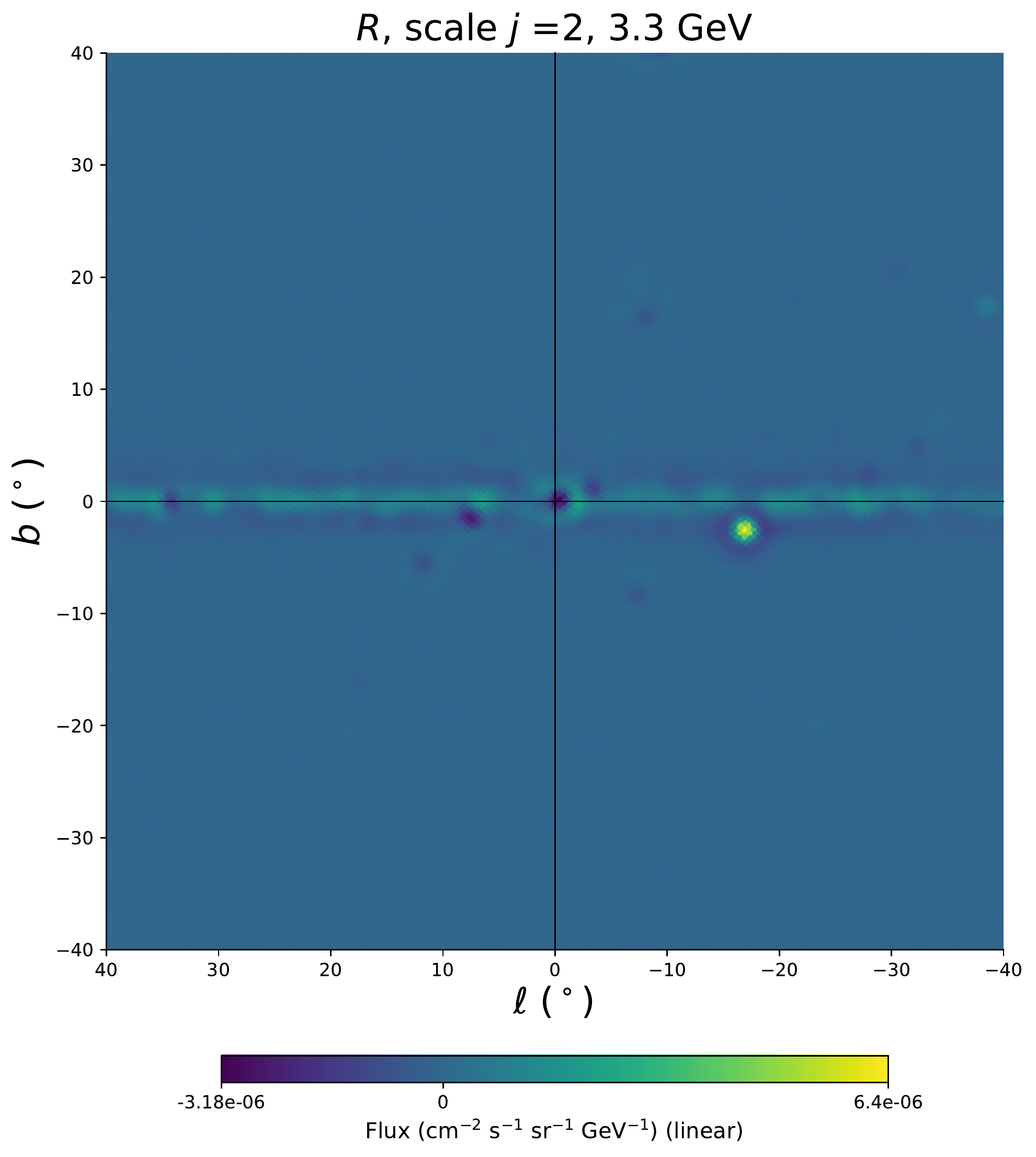}
\includegraphics[width=2.3in,angle=0]{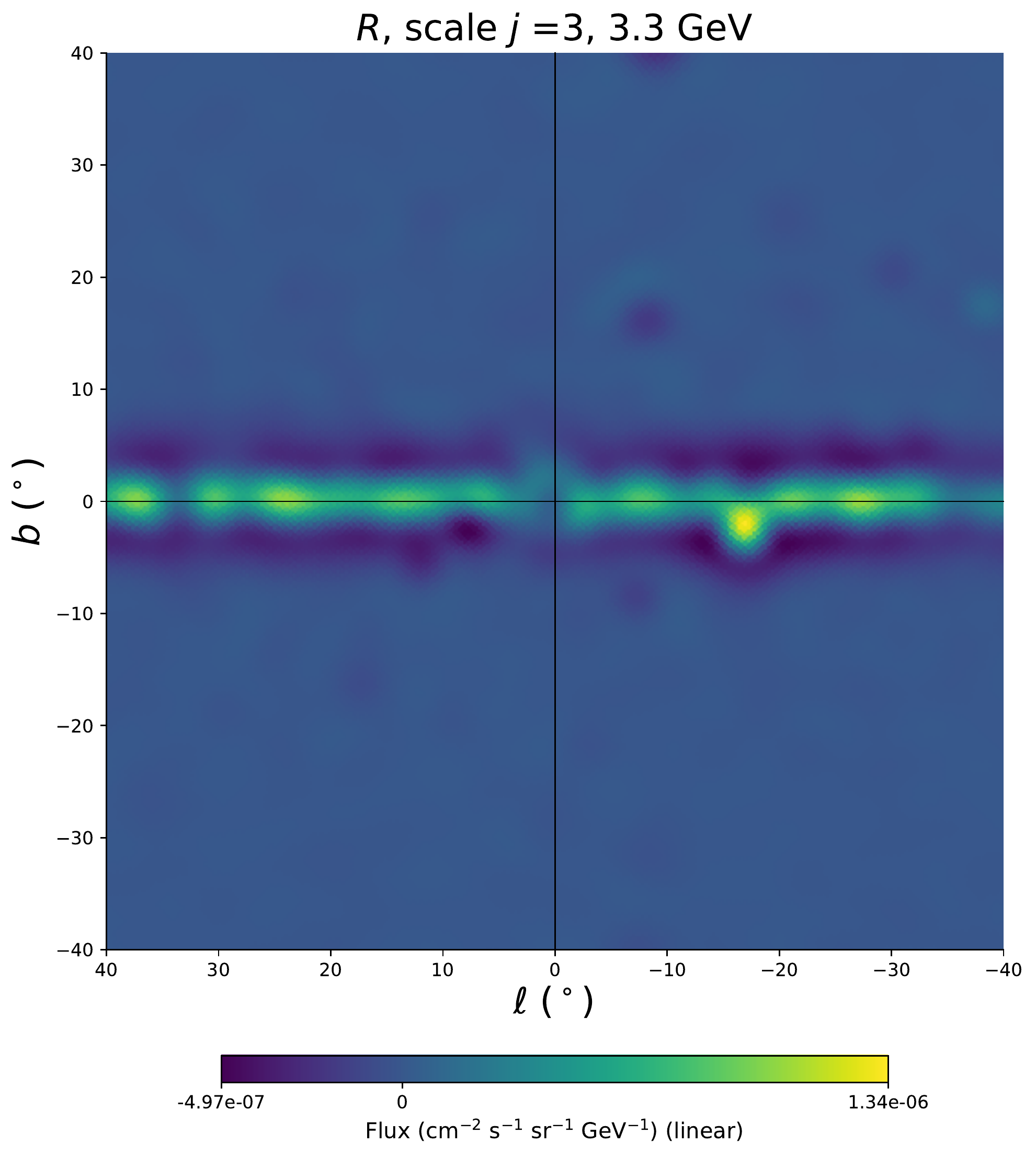} \\
\includegraphics[width=2.3in,angle=0]{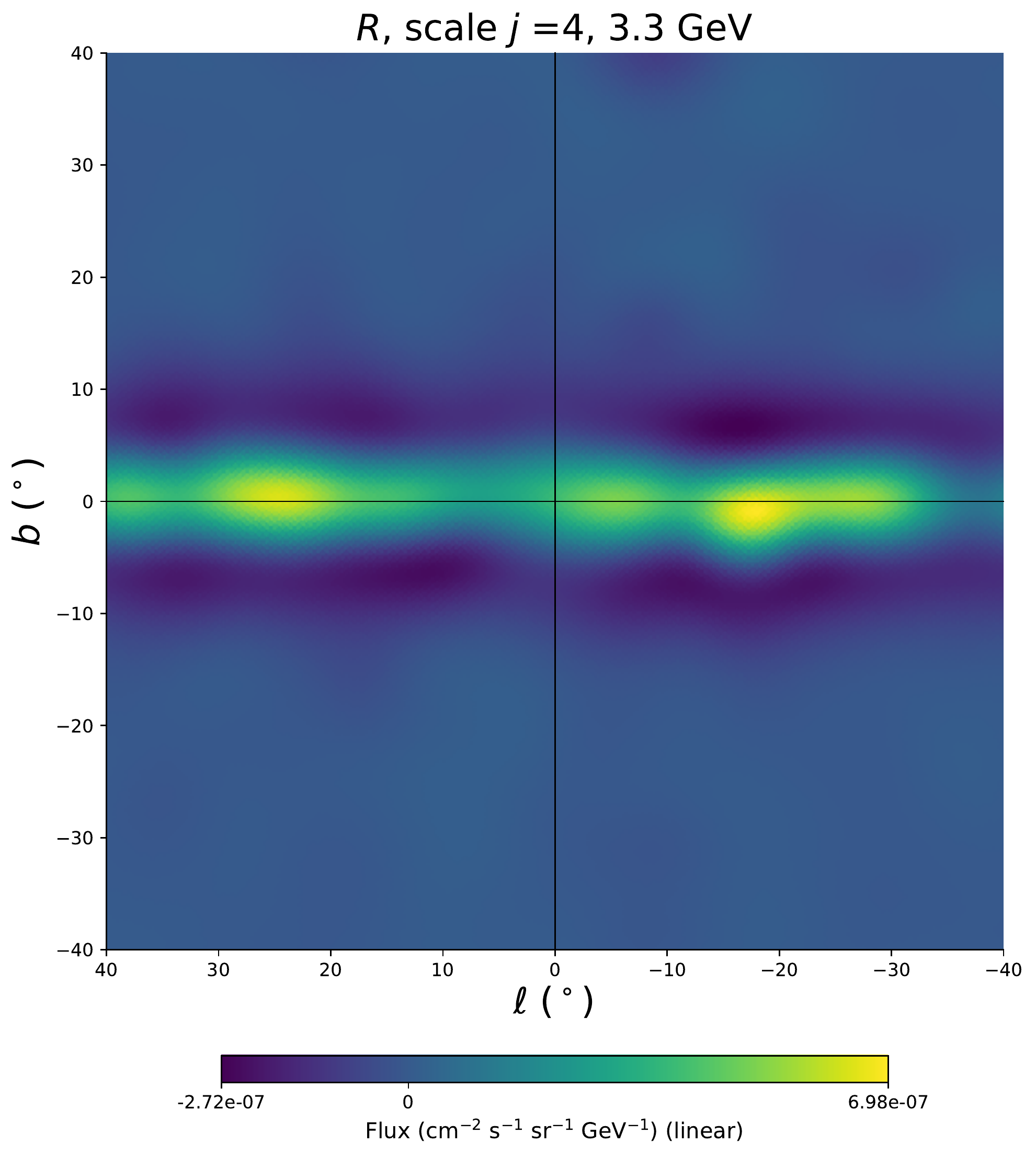}
\includegraphics[width=2.3in,angle=0]{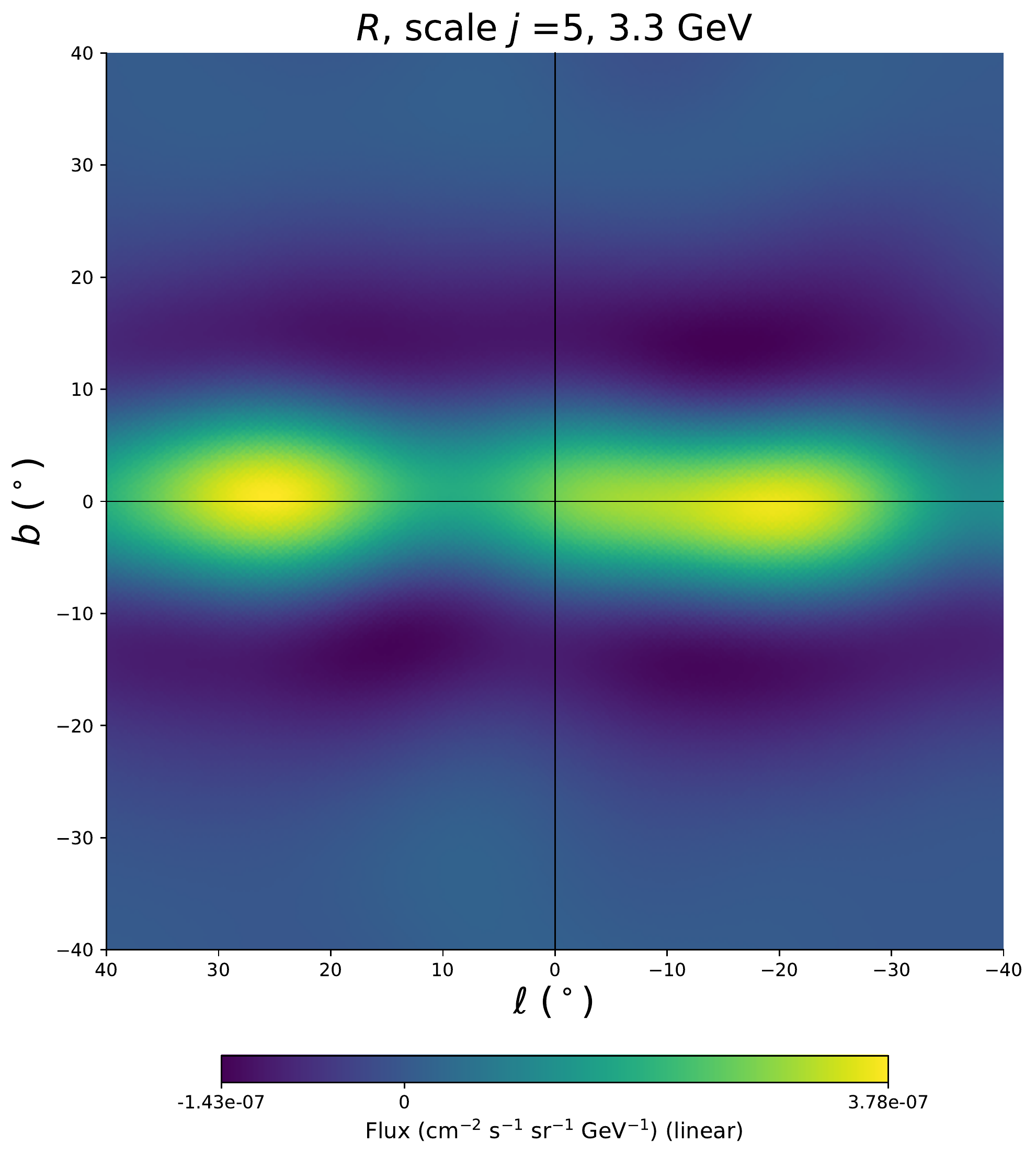}
\includegraphics[width=2.3in,angle=0]{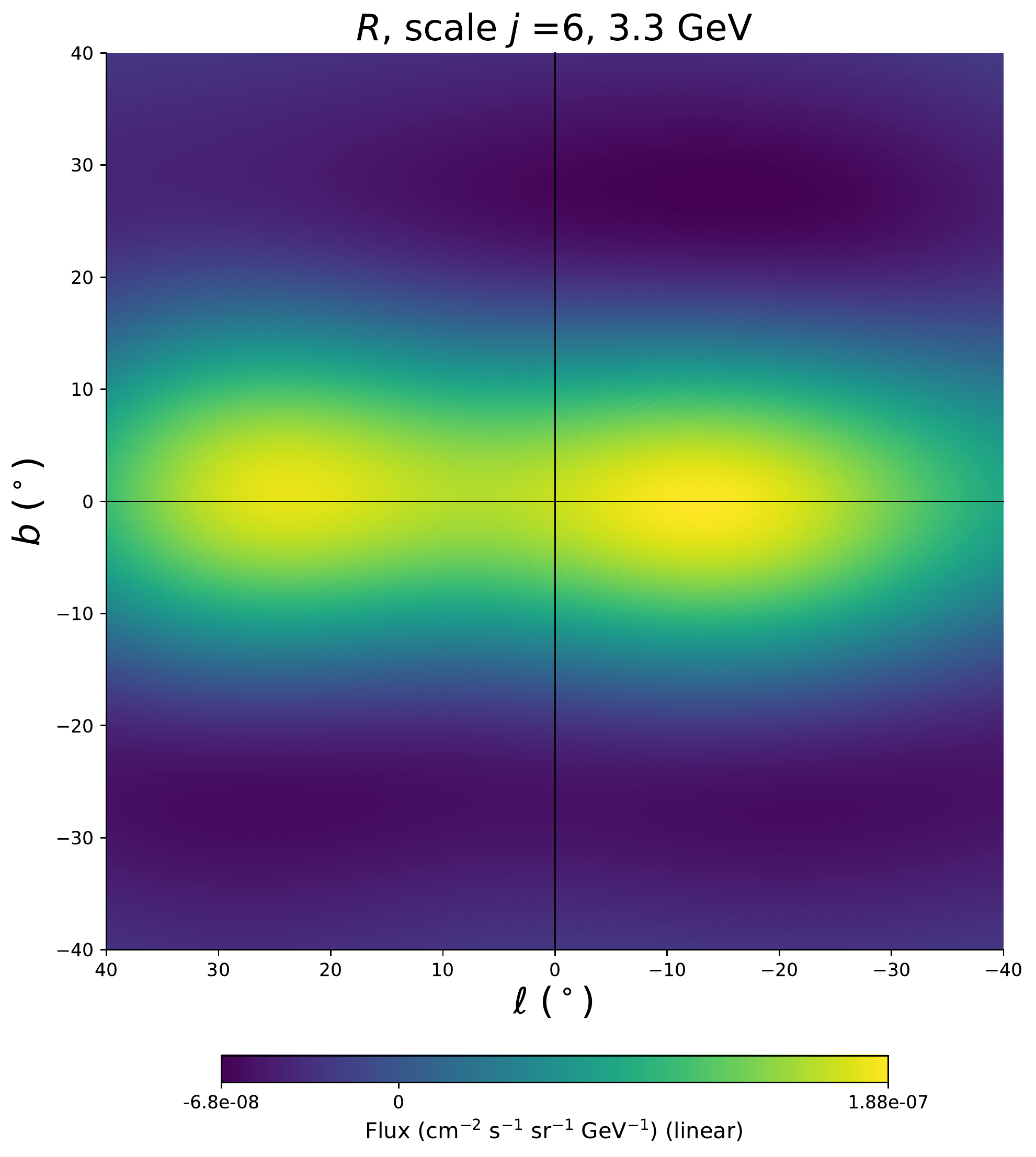}
\end{centering}
\vspace{-0.2cm}
\caption{The decomposition of the residual map $R(E_3)$ into the first six wavelet 
levels. Areas near $b=0$ and with $\ell$ of about $-20^\circ$ and $+25^\circ$  show 
power at high wavelet levels, with relatively low normalization compared to lower levels.}
\vspace{-0.3cm}
\label{fig:R0_decomp}
\end{figure*}

An initial map $R(E)$ is reconstituted from its wavelet levels by,
\beq
R(E) = \sum_{j=1}^{j_{\rm max}} w_j(R; E) + c_{j_{\rm max}}(R; E),
\eeq
where the levels $w_j(R; E)$ and the average emission $c_{j_{\rm max}}(R; E)$ each have the same dimensionality 
as the original map. In what we follows we will usually drop the auxiliary labels $R$ and $E$. Summing up every 
pixel from a given wavelet level $w_j$ will give 0; contiguous regions within $w_j$ with positive pixels arise from the 
presence of features of angular size $\theta_j \sim 2^j \times \theta_{\rm pix}$ in the original image, where 
$\theta_{\rm pix} \simeq 0.5^\circ$ for our choice of {\tt HEALPix} parameters. The angular scales that provide most 
of the support for each wavelet scale are recorded in Table~\ref{tab:Wavelet_scales}. As an example, the result of the 
wavelet decomposition on the map $R(E_3)$ is shown in Figure~\ref{fig:R0_decomp}.

There are several motivations for analyzing gamma-ray data with the wavelet transform. Most importantly, uncertainties in 
the expected gamma-ray background that arise from uncertainties in the interstellar medium properties and the galactic 
cosmic-ray distribution are more severe on smaller angular scales, while uncertainties from the ICS are relatively most 
important far from the Galactic disk. Also, our understanding of the Milky-Way gamma-ray point-source distribution is 
limited to the ones that are bright enough to be detected at high significance. Finding evidence for non-standard extended 
sources has inherent value for our understanding of the gamma-ray sky.

\begin{table*}[t]
\begin{center}
\begin{tabular}{c|ccccccccc} 
  & $w_1$ & $w_2$ & $w_3$ & $w_4$ & $w_5$ & $w_6$ & $w_7$ & $w_8$ & $w_9$\\ \hline
$\theta$ & $[0.7^{\circ},1.4^{\circ}$] & $[1.4^\circ,2.8^\circ]$ & $[2.8^\circ,5.6^\circ]$ & $[5.6^\circ,11.3^\circ]$ & $[11.3^\circ,22.5^\circ]$ & $[22.5^\circ,45^\circ]$ & $[45^\circ,90^\circ]$ & $[90^\circ,180^\circ]$ & $[180^\circ,360^\circ]$ 
\end{tabular}
\end{center}
\caption{Angular scales that dominate the wavelet levels $w_j$ for $\ell_{\rm max}=512$.}
\label{tab:Wavelet_scales}
\end{table*}%

As we will shortly discuss in much more detail, our joint template- and wavelet-based analysis indicates four candidate 
regions of interest in the sky. After identifying these regions, we will focus on that part of the sky to further characterize 
the gamma-ray spectrum and morphology. For a region of interest denoted ROI from our residual $R$, we will use the 
shorthand,
\beq \label{roidef}
\textrm{ROI}^{a-b}(E) = \sum_{j=a}^b w_j(E) \times \Theta(\textrm{pixels within ROI}).
\eeq
The wavelet transform is performed on the entire sky once per energy bin. We identify regions of interest qualitatively 
using this decomposed data, though this can be done using a multiscale resolution analysis or a hard-thresholding 
procedure \cite{2006aida.book.....S} if desired.
	
Due to its inherently nonlocal nature, the method for finding statistical error bars from the wavelet decomposition is 
nontrivial: the statistical error on a pixel $\Omega_p$ from wavelet level $w_j$ is not simply $\propto \sqrt{w_j|_{\Omega_p}}$, 
since this is nonzero only if there is a structure of size $2^j\times \theta_{\rm pix}$ with support at $\Omega_p$.
The procedure for extracting a well-defined error bar on any linear combination of wavelet levels is defined explicitly 
in Appendix~A of \cite{McDermott:2015ydv}. Unfortunately, this procedure is computationally expensive.
In the present work, we will use the fact that the statistical error bar on any individual wavelet level is bounded from 
above by the statistical error bar on the entire flux of that pixel. We will show this upper limit on the statistical error bar 
as the statistical error bar on all wavelet levels and combinations thereof. We will see that the systematic error bars typically 
exceed these error bars, and thus that this analysis is systematics limited.

\subsection{Systematic errors}
\label{sec:sys}

The choice of the diffuse templates that we average together to produce our residual maps $R(E)$ acts as a source 
of systematic uncertainty in our results. The 19 models (A-D, F-R, W and GXI) that form the base set that we average 
over are chosen to encompass the outside envelope of systematic errors derived in \cite{Calore:2014xka}. This combination 
of models encompasses uncertainties associated with the galactic diffusion of CRs, with the significance of convective 
winds perpendicular to the galactic disk, and with diffusive reacceleration of CRs at GeV energies. Moreover these models 
account for uncertainties associated to the position- and energy-dependence of CR energy losses. These models use 
different assumptions for the distribution and type of CR sources in the Milky Way, as well as for the spectral characteristics 
of the injected CRs into the ISM. Finally, the models used account for different assumptions regarding the distribution of the 
ISM gas and the magnetic and interstellar radiation fields.

To generate the systematic error bands that we show below, we repeat the first step in our analysis by replacing the average 
of these 19 models by certain well-motivated alternative sets.  In addition to the base set of models we consider the average of two 
subsets of models from \cite{Calore:2014xka}: {\it (i)} A-C, F, I, M-O and R and {\it (ii)} A-D, W, GI, GIX, GXI, GXXI and GXXIX. 
Our three different sets of models are chosen to include varying ranges for the combination of astrophysical uncertainties 
discussed above. Our first set of models includes all the above mentioned astrophysical uncertainties. Our second option of 
models focuses more on modeling the uncertainties associated with the ISM gas distribution, while the third set of models 
envelops the CR propagation uncertainties.

\section{Results}
\label{sec:results}

The pixelization of the initial maps and the maximum angular mode of the spherical harmonic decomposition inherent to the 
IUWTS each provide a minimum angular scale for our analysis. We take {\tt NSide=128} in our {\tt HEALPix} projection, giving 
pixels on a side of order $0.5^\circ$, and we take $j_{\rm max}=9$ for the wavelet transform, retaining spherical 
harmonics up to $2^{j_{\rm max}} = 512$, corresponding to angular scales of $\theta_{\rm min} = 0.7^\circ \approx \theta_{\rm pix}$. 
Point sources can contaminate our residual image, especially at the lower levels of the wavelet decomposition, since 
the individual point sources will be as large as the instrumental PSF. At low energies the PSF is non-negligible, which sets 
the pixel and minimum angular sizes. Thus, we rely on the accuracy of the point-source catalog, before using the wavelet 
transform. We then determine whether a source can truly be taken as a point source by observing and removing any large-scale 
features around point sources of interest. In Appendix~\ref{app:AuxEff} we show an example of such an identification. 

\begin{figure}[t]
\begin{minipage}{\linewidth}
\includegraphics[width=\linewidth]{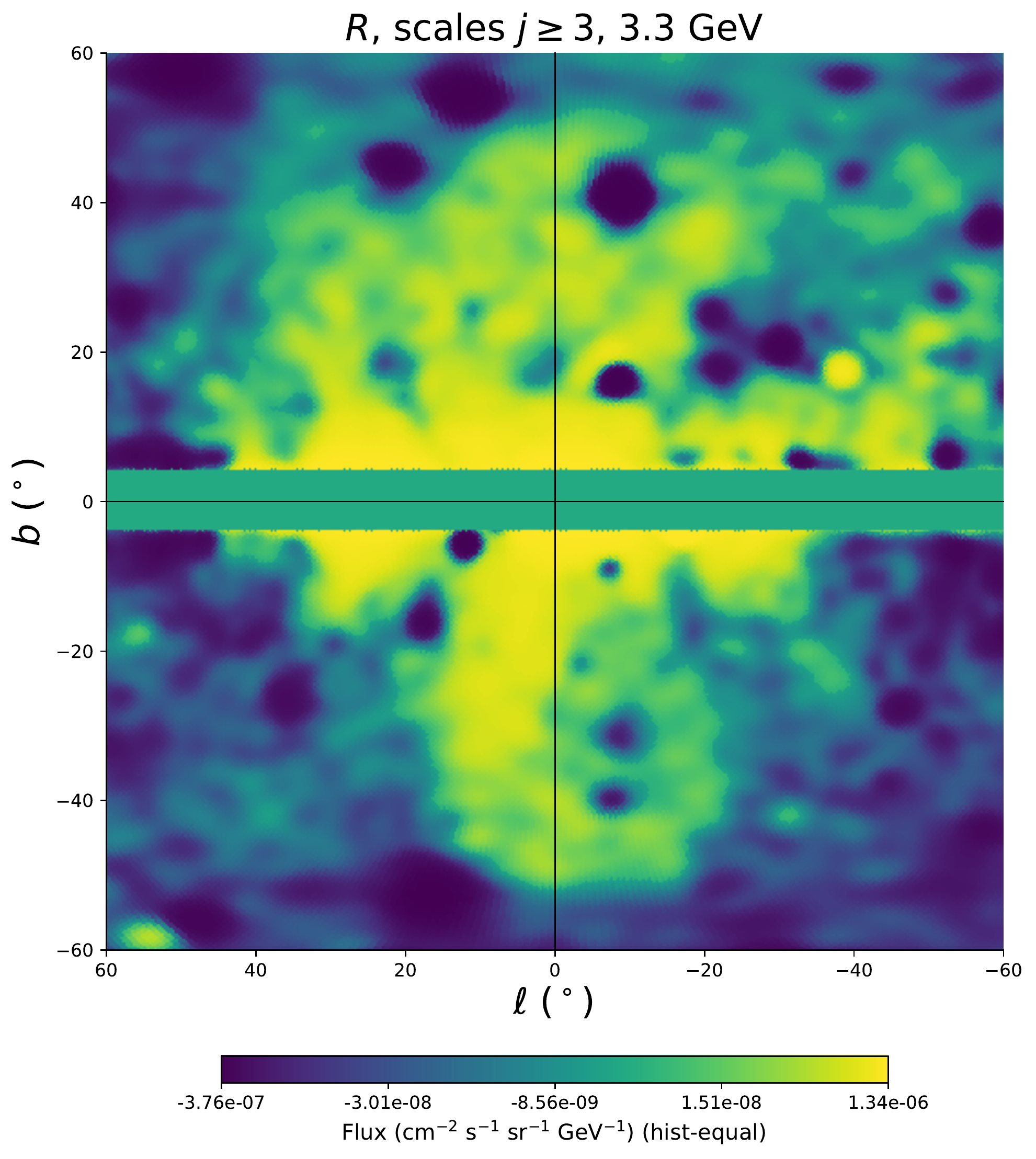}
\end{minipage}
\caption{Map of the inner $120^{\circ} \times 120^{\circ}$ gamma-ray sky at 3.3 GeV including wavelet scales of 
$j \geq 3$, {\it i.e.}~removing structures with support below $2.8^\circ$. The \textit{Fermi} Bubbles are quite distinct. The region 
with $|b|<2^\circ$ has been masked to make the interesting higher latitude emission more apparent.}
\label{fig:finalbubblese3}
\end{figure}

Using the wavelet decomposition of the residual maps as described in section~\ref{subsec:WaveletTr}, we now characterize the 
gamma-ray emission in the inner Galaxy. We identify four distinct emissions: the \textit{Fermi} Bubbles, the GCE, and extended 
emissions on the Galactic disk centered at $\ell \approx 25^{\circ}$ and at $\ell \approx -20^{\circ}$, which we henceforth refer to 
as West Diffuse Emission (WDE) and East Diffuse Emission (EDE), respectively. The presence of the EDE and WDE is clearly 
seen in the bottom row of Figure~\ref{fig:R0_decomp}: they are particularly bright on wavelet levels 4, 5, and 6 ({\it i.e.}~angular 
scales between $\sim$$5^{\circ}$ and $45^{\circ}$). These emissions are worth examining since their physical extent, their 
separation from the Galactic center, and the claimed size of the Galactic center excess are all comparable, so they could in 
principle mutually contaminate each other on the wavelet levels of interest for understanding the emission from the Galactic center. 
All four emissions are also visible in Figure~\ref{fig:finalbubblese3}, where we have masked the inner $|b|\leq 2^{\circ}$ to highlight 
these new diffuse emission components. We highlight here our result, discussed more below, that the three emission components 
with $b=0^\circ$ {\it do not} substantially overlap in longitude and, regardless of similarities in their underlying mechanisms, must 
originate in different parts of the Milky Way.  

\subsection{The \textit{Fermi} Bubbles}
\label{subsec:FB}
We first study the \textit{Fermi} Bubbles. The two Bubbles have been distinctly identified 
above $10^{\circ}$ \cite{Su:2010qj, Fermi-LAT:2014sfa}. We confirm the well-known feature that there is a sharp drop in 
the residual emission at $|b|\gsim 50^{\circ}$ and $|\ell| \gsim 20^{\circ}$, as is clearly visible Figure~\ref{fig:finalbubblese3}.

\begin{figure*}[t]
\centering
\includegraphics[width=.44\textwidth]{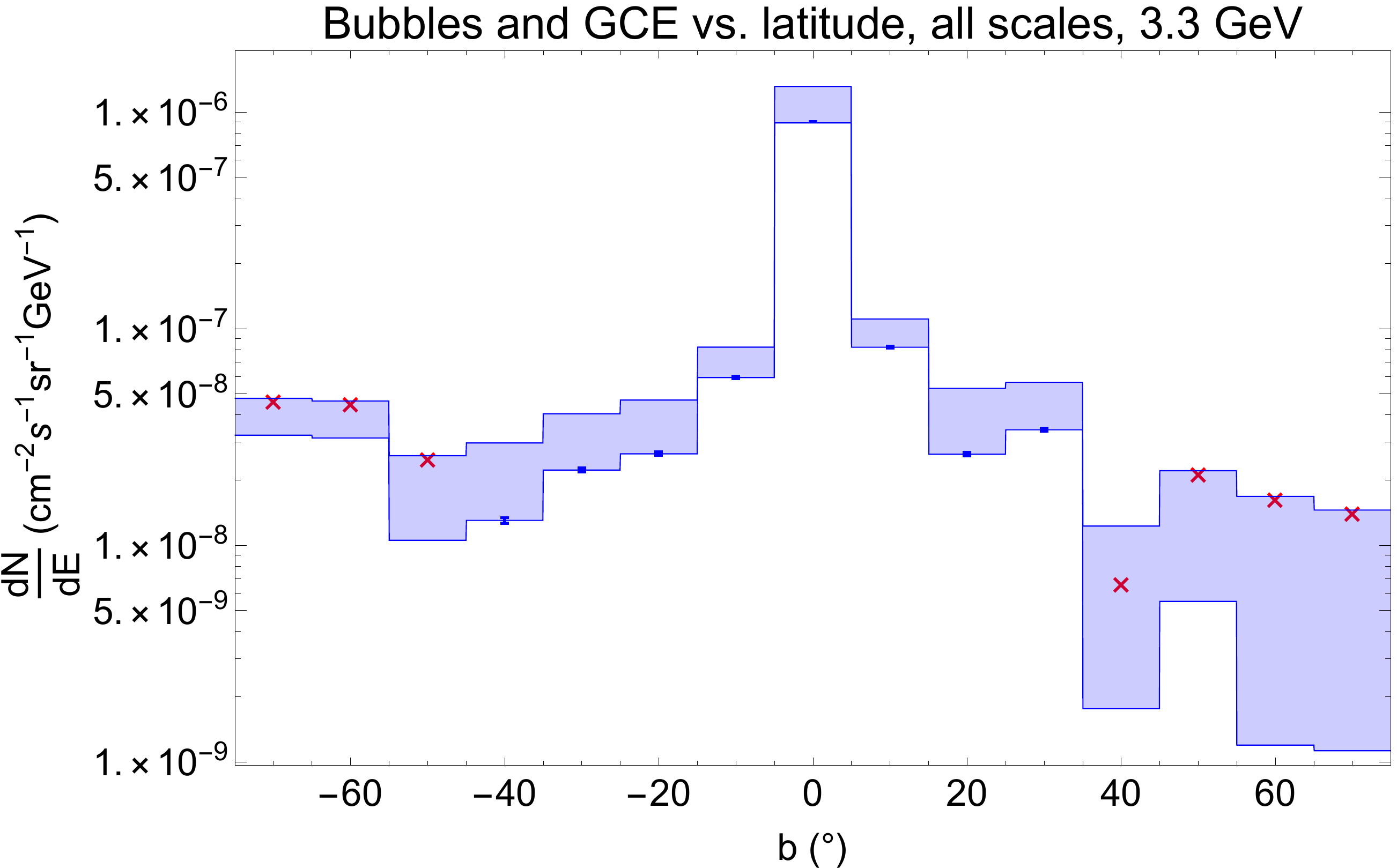}\qquad~~
\includegraphics[width=.49\textwidth]{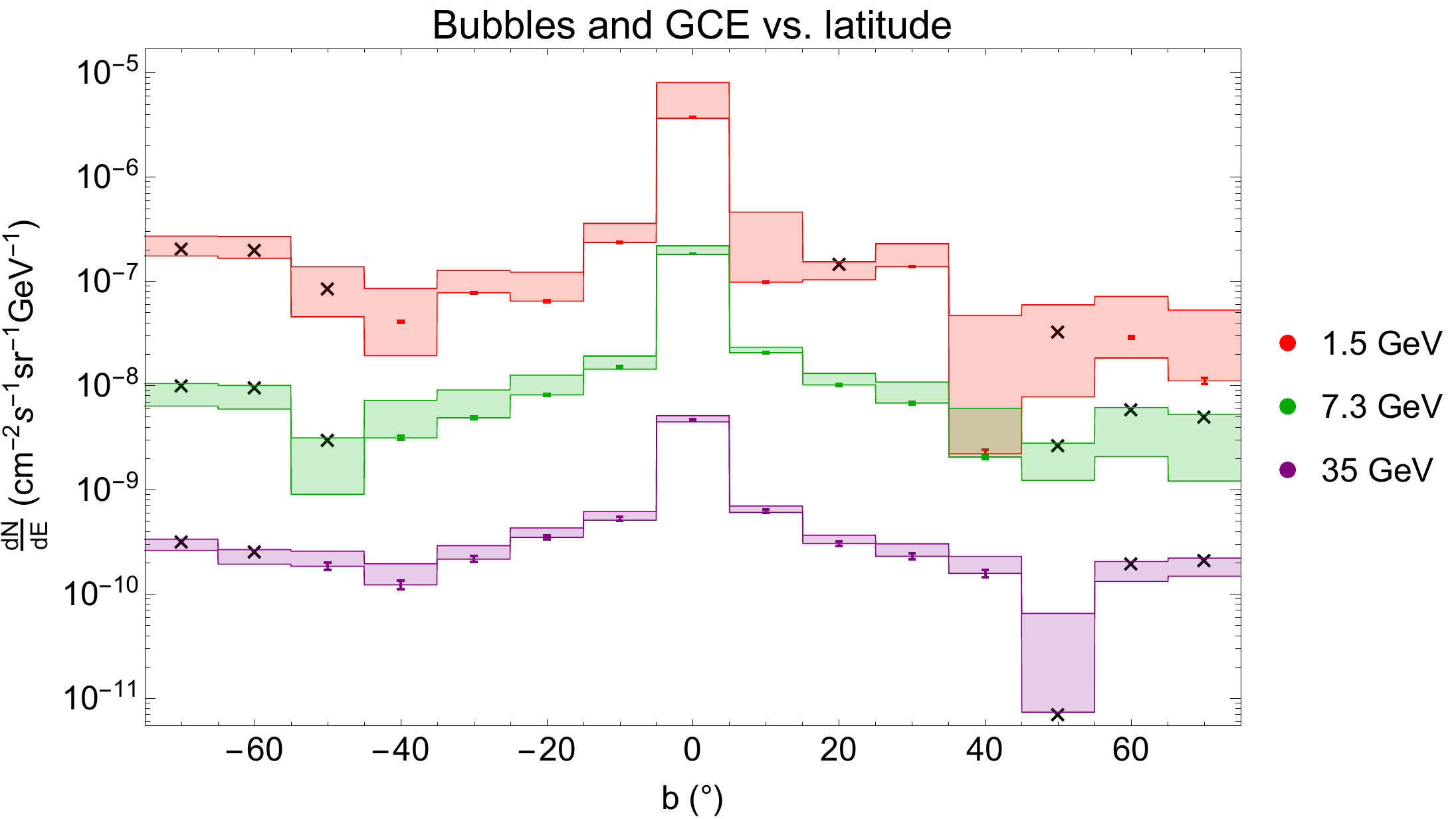}
\caption{Average gamma-ray flux as a function of latitude, in the energy bin centered at 3.3 GeV (\textit{left}), in windows that cover 10$\deg$ 
in latitude and with $|\ell| < 20\deg$. Systematic errors are presented as bands, while statistical errors are displayed with error-bars 
(associated to the inferred number of photons in each region). Red "$\times$" indicate regions where the flux was calculated to be negative 
in which case the magnitude of the flux is plotted. \textit{Right}, same as \textit{left} for the energy bins centered at 1.5, 7.3, and 35 GeV.}
\label{fig:bubblelatfluxebin2}
\end{figure*}

To characterize the Bubbles' emission as a function of latitude, we calculate the average flux in boxes of $\Delta b = 10^{\circ}$ 
for $|\ell| < 20^{\circ}$ including wavelet levels 1 through 9 ({\it i.e.}~having subtracted the average across the sky). This is analogous 
to the template-based approach and does not use any special features of the wavelet decomposition. We show the result for the 
average flux calculated in this way for the 3.3 GeV energy bin in Figure~\ref{fig:bubblelatfluxebin2} (\textit{left}), and for other energy bins in 
Figure~\ref{fig:bubblelatfluxebin2} (\textit{right}) using the same latitude bins. The sharp increase in the gamma-ray residual emission for $|b|<5^{\circ}$ 
in all energy bins is partially attributed to the GCE. The GCE also may contribute to the next latitude bins ({\it i.e.}~$5^{\circ} < |b| < 15^{\circ}$). 
We study these latitudes in the next section. 

\begin{figure}[t]
\begin{minipage}{0.45\linewidth}
\includegraphics[width=\linewidth]{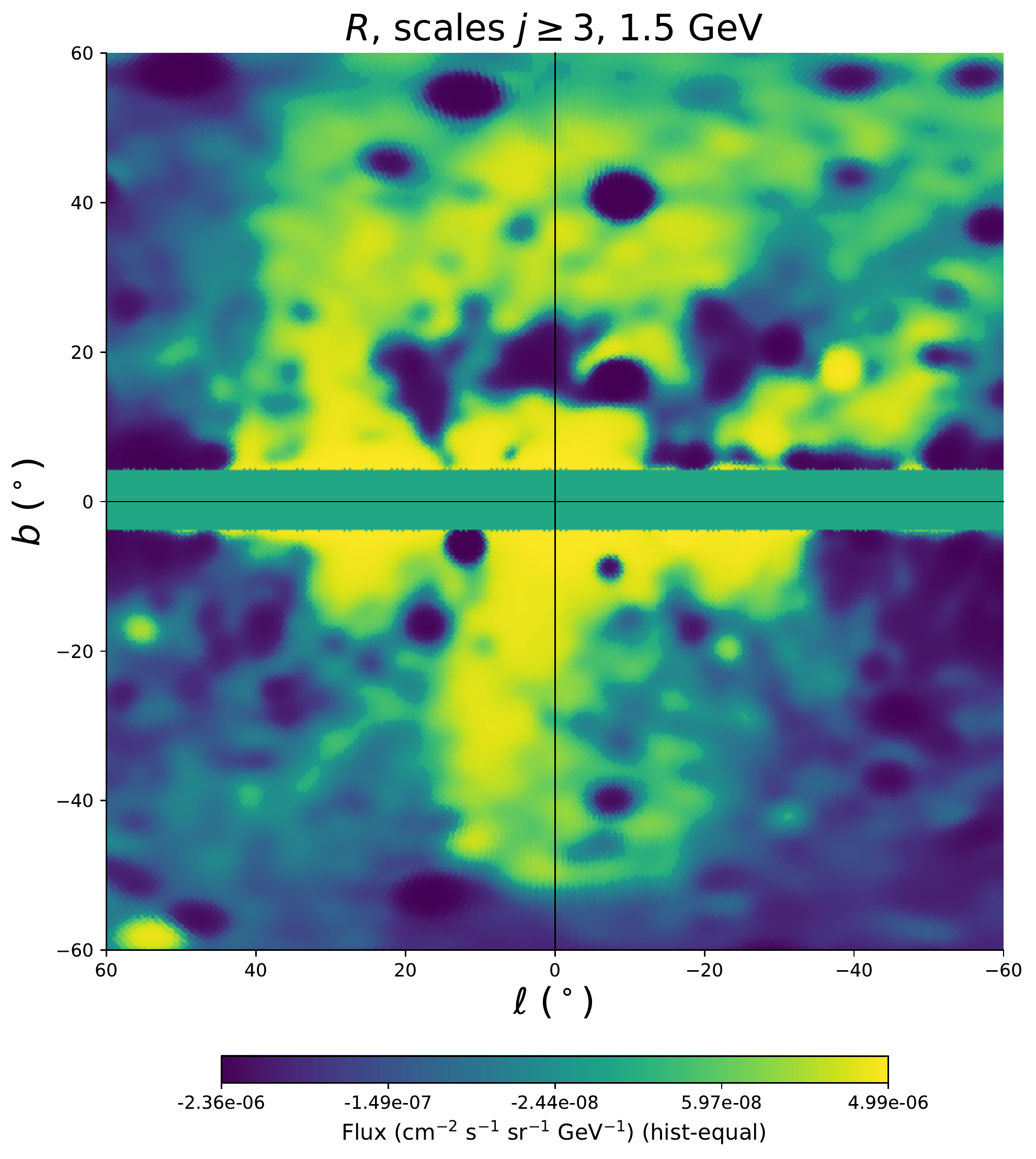}
\end{minipage}
\begin{minipage}{0.45\linewidth}
\includegraphics[width=\linewidth]{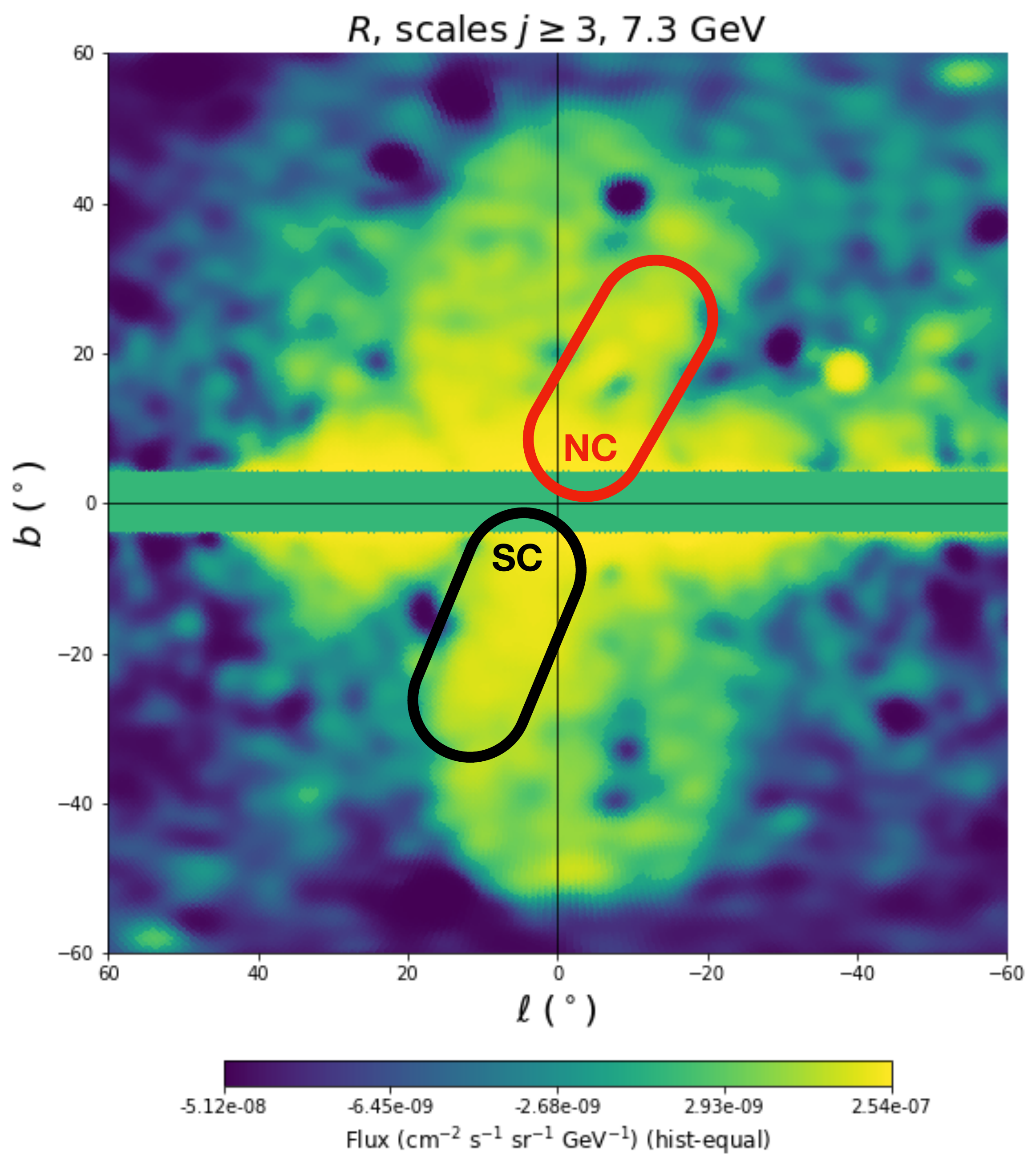}
\end{minipage} \\
\begin{minipage}{0.45\linewidth}
\includegraphics[width=\linewidth]{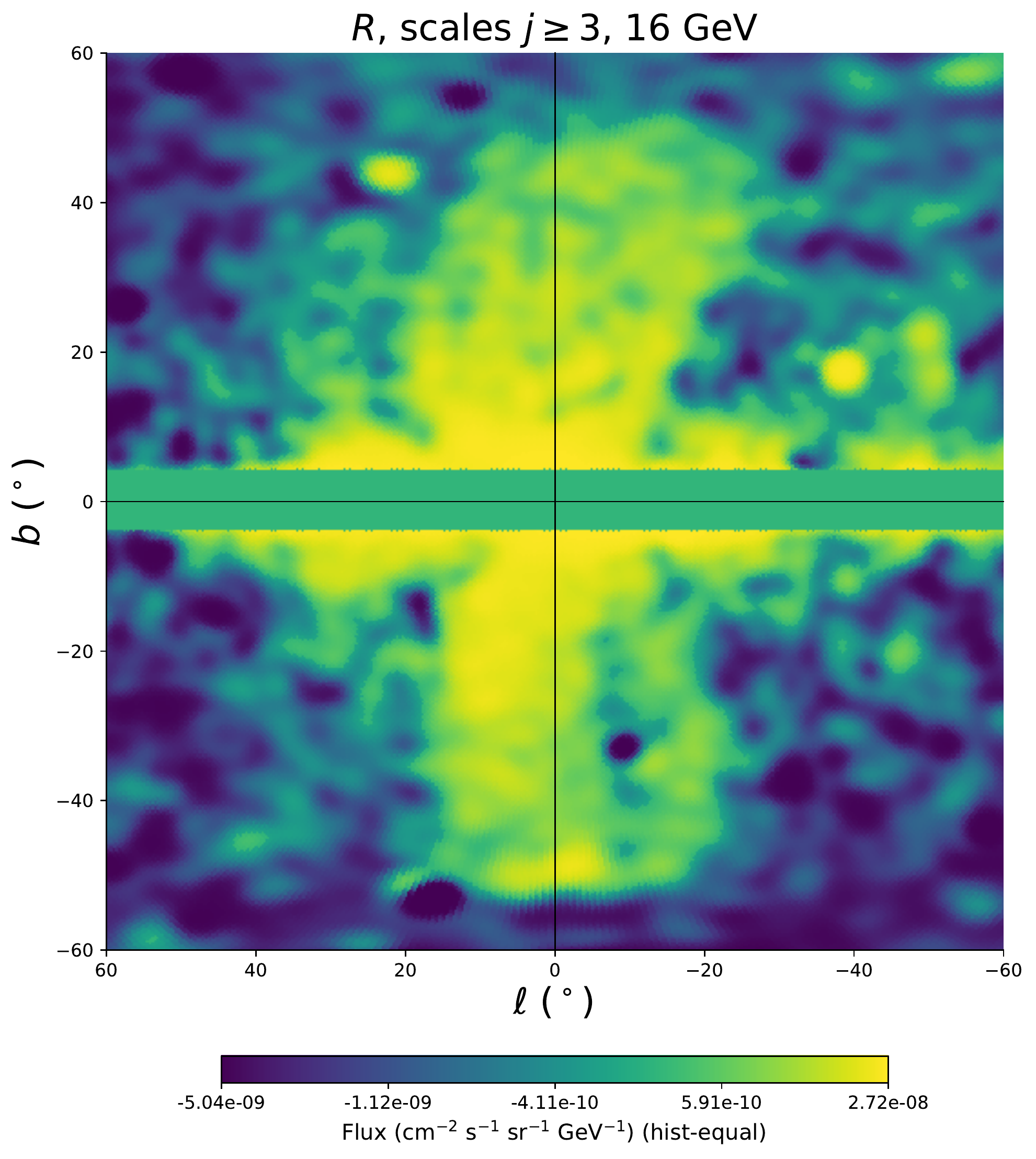}
\end{minipage}
\begin{minipage}{0.45\linewidth}
\includegraphics[width=\linewidth]{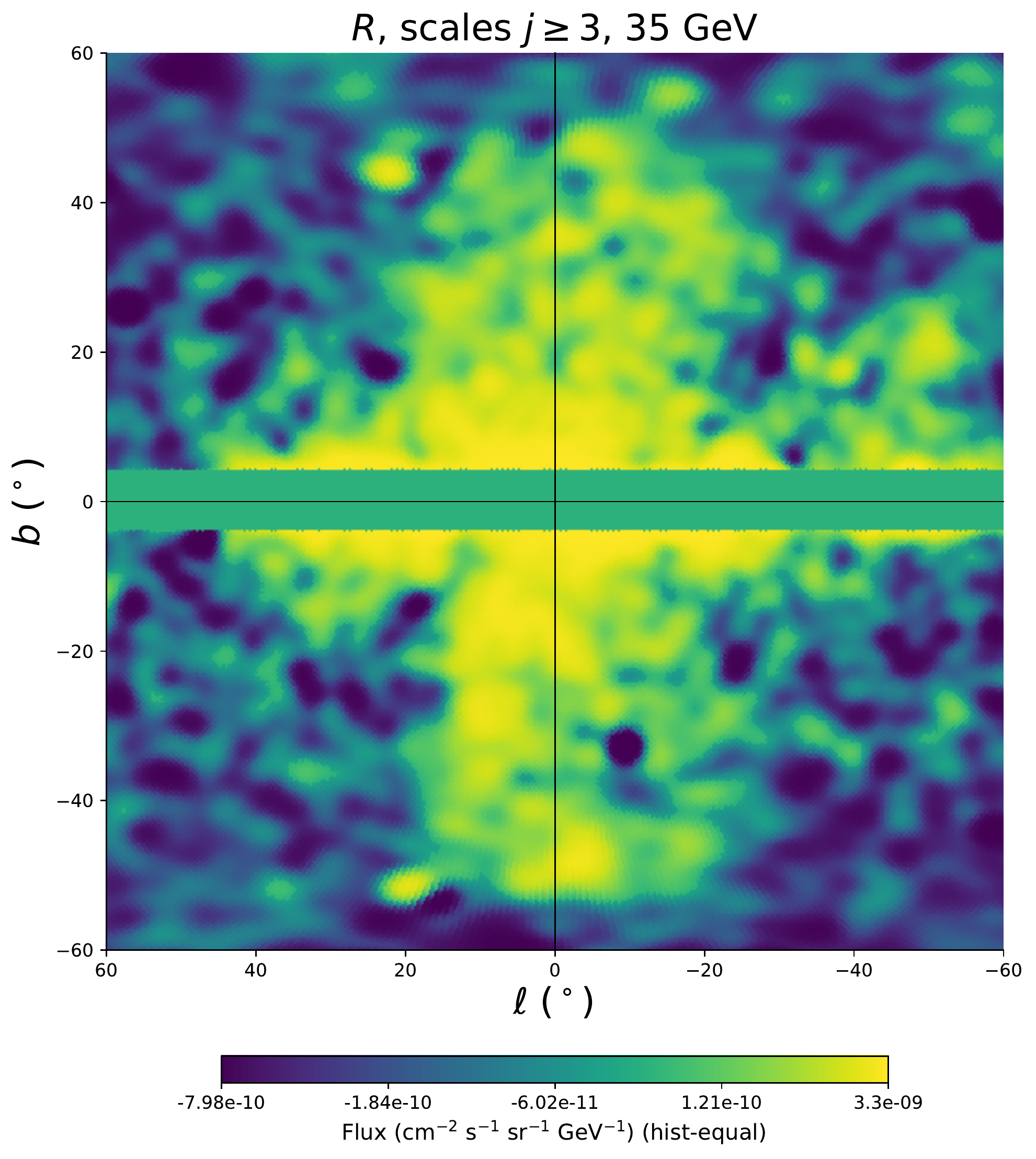}
\end{minipage}
\caption{Residual emission in different energy bins, with the disk masked. \textit{Top left}, 1.5 GeV, \textit{top right}, 7.3 GeV, \textit{bottom left}, 16 GeV
and \textit{bottom right}, 35 GeV. A region around the disk extending to 2$^\circ$ in latitude 
has been masked out. On the 7.3 GeV map we have drawn for reference the Southern Cocoon "SC" and the Northern Cocoon "NC".}
\label{fig:finalbubbles4Ebins}
\end{figure}

Away from the central three bins, the flux is relatively flat until $|b|\approx 50^\circ$ at which point the flux becomes negative, 
denoted in the plot by the "$\times$" symbol. This marks the end of the Bubbles. This same behavior is replicated in all energy 
bins.  This is demonstrated also in Figure~\ref{fig:finalbubbles4Ebins} where we present the residual emission for wavelet 
levels of $j\geq 3$ at the energy bins of 1.5, 3.3, 7.3, and 35 GeV. The \textit{Fermi} Bubbles are clearly observed at these energies. 

In addition, in Figure~\ref{fig:finalbubbles4Ebins} we clearly identify a ``cocoon'' in the Southern Bubble at positive longitudes, 
extending to $b\approx -30^\circ$ in agreement with \cite{Su:2010qj, Fermi-LAT:2014sfa}. Moreover, there is an indication of a Cocoon 
in the Northern Bubble at negative longitudes, extending to $b\approx +30^\circ$. The Northern Cocoon is most evident at the energies 
of 7.3 and 16 GeV. It appears along roughly the same axis as the southern one, but is less bright. This part of the Northern sky has a
larger column density of ISM gas along our line of sight, which results in a brighter background associated with $\pi^{0}$ and bremsstrahlung 
emission and correspondingly larger systematic uncertainties. Wavelets, which reduce susceptibility to systematic uncertainties on small 
angular scales, allow us to find this indication of a Northern Cocoon. At 7.3 GeV, where the Northern Cocoon is easiest to see, the 
associated flux is $\sim 10^{-8}$ GeV$^{-1}$cm$^{-2}$s$^{-1}$sr$^{-1}$, which is $\sim 30 \%$ dimmer than the Southern Cocoon at the 
same energy. These Cocoons may be an indication of jet emission along an axis that has a projected angle $\approx30^{\circ}$ off the 
perpendicular to the disk. We leave further discussion on the interpretation of this result for section~\ref{sec:PrevWork}.

\begin{figure*}[t]
\centering
\includegraphics[width=.45\textwidth]{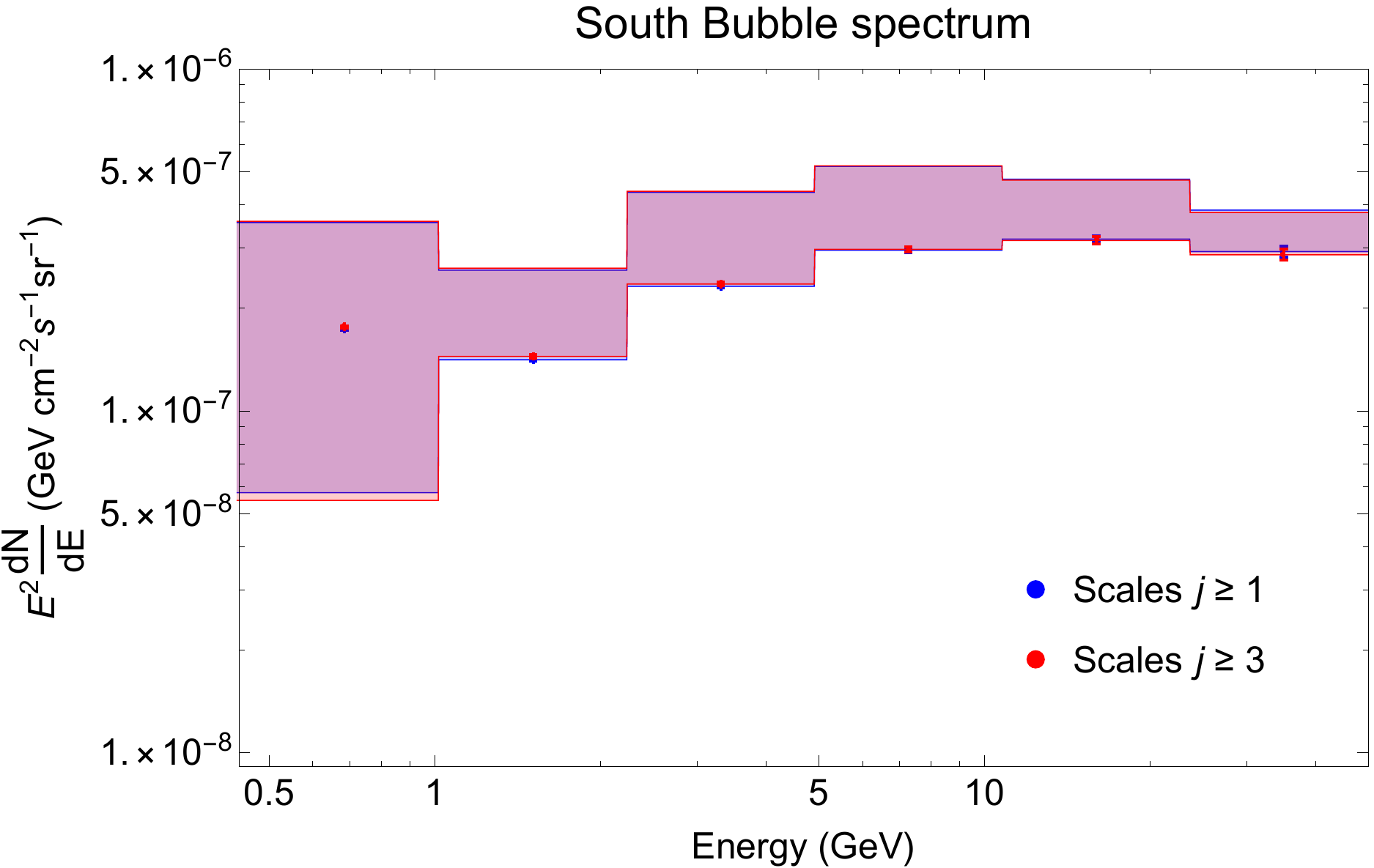}\qquad~~
\includegraphics[width=.45\textwidth]{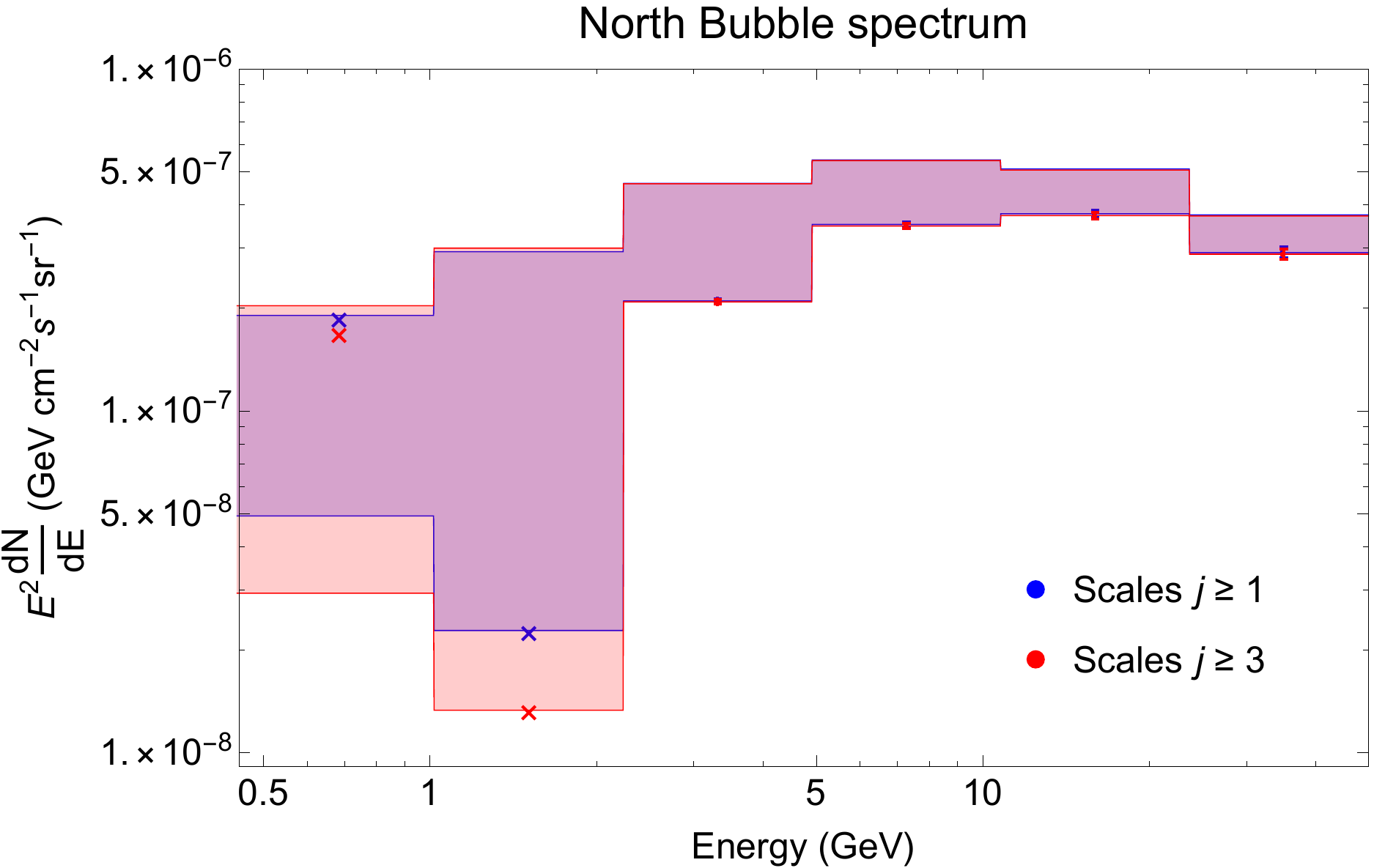}
\caption{The flux in the Southern Bubble (\textit{left}) and Northern Bubble (\text{right}) as a function of energy, restricted to scales larger than 
$\sim$0.7$^\circ$ (blue) and $\sim$3$^\circ$ (red). Systematic uncertainties are shown by bands. The two spectra
are nearly identical, providing compelling proof that the southern bubble is diffuse.}
\label{fig:sbubblespece2w3}
\end{figure*}

As mentioned earlier, Figure~\ref{fig:bubblelatfluxebin2} demonstrates that the Bubbles have 
approximately constant brightness for $-15^{\circ} > b > -45^{\circ}$. We use this range of $b$ and $| \ell | < 20^{\circ}$ to 
determine the energy spectrum of the Southern Bubble. We also calculate the spectrum excluding the first two wavelet levels, 
{\it i.e.}~eliminating power at scales smaller than $2.8^\circ$. We compare the spectra in Figure~\ref{fig:sbubblespece2w3} and 
find that they are nearly identical. This demonstrates that the Southern Bubble is truly diffuse in its nature and not the accumulated 
emission from many small angular-scale sources, as would be expected from gamma-ray point sources or bright filaments ({\it e.g.}, 
from emission correlated with gas). This is a test that can be performed optimally with wavelets, which are \textit{designed} for 
such a type of analysis. The fractional difference in the flux is less than 2.5\% at all energy levels.  By using wavelets, we are thus 
able to make a claim about the details of the Bubble structure on a purely morphological basis.

We produce a similar spectrum for the Northern Bubble in Figure~\ref{fig:sbubblespece2w3}. Above energies of 3 GeV the Northern 
Bubble appears almost entirely diffuse. For energies lower than 3 GeV, the systematics of the Northern Bubble are large, which 
potentially contributes to the power that we notice on the first two wavelet scales. This can be the result of the larger ISM gas column 
density in this part of the sky. The increased ISM gas column density creates large differences between the modeled predictions on 
the diffuse $\pi^{0}$ and bremsstrahlung emissions, which we have to subtract from the original data (see section~\ref{subsec:StartingAss}). 
Any mis-modeling of these emissions leaves enough power at these scales and energies in the residual map $R$ that is picked up
by the wavelet decomposition. Thus, we conclude that the Northern Bubble is mostly diffuse emission, like the Southern Bubble, with 
additional contamination from mismodeled small-scale structure. This evidence that the Bubbles are truly large-scale diffuse emissions 
in their morphology is valuable for understanding their underlying mechanism. We leave this detailed discussion for section~\ref{sec:PrevWork}.

\subsection{The Galactic Center Excess}
\label{subsec:GCE}

\begin{figure}[t]
\centering
\includegraphics[width=0.8\linewidth]{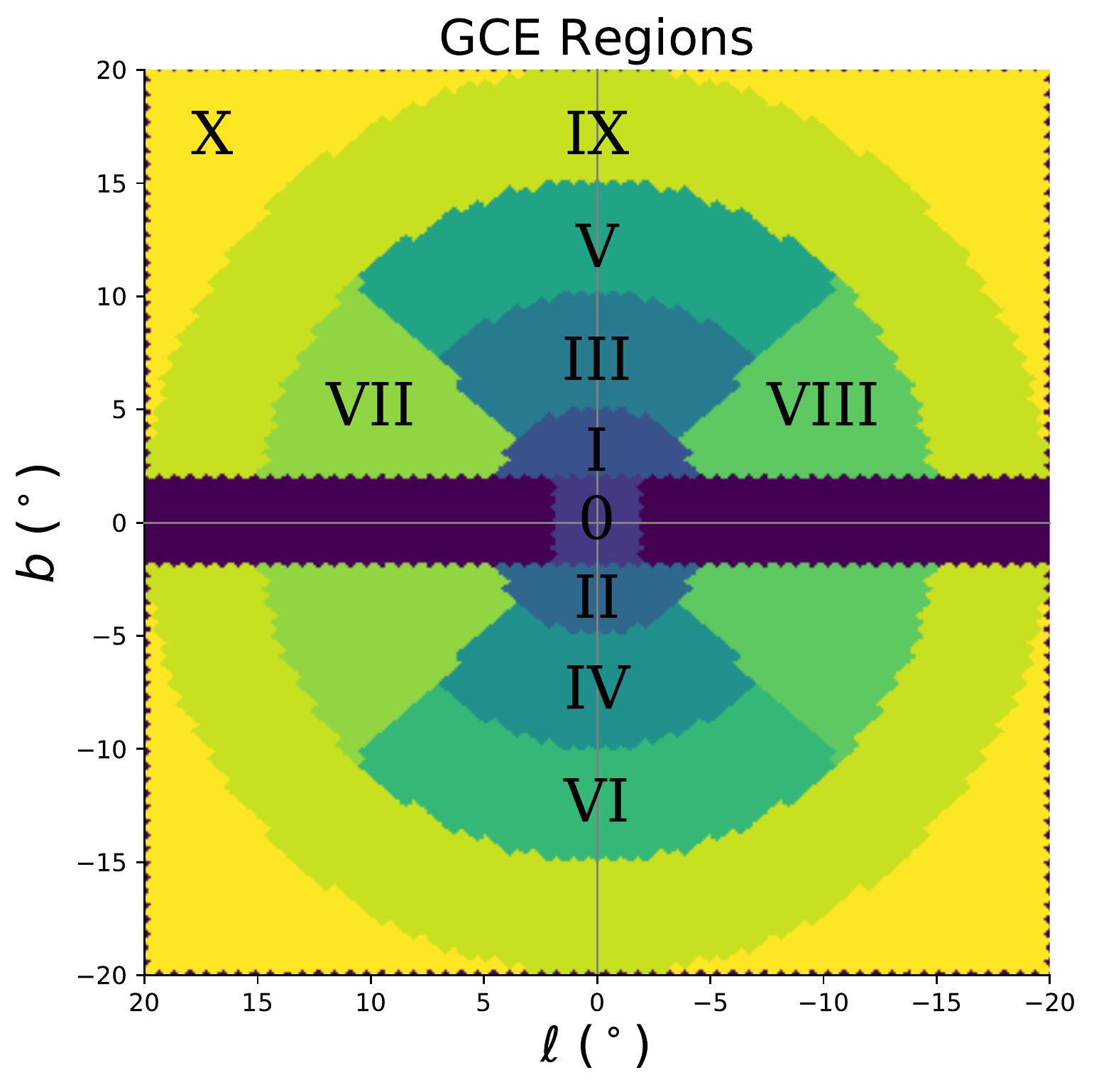}
\caption{The regions used to describe the Galactic Center Excess. We use the same regions as \cite{Calore:2014xka} 
as well as a region centered at the origin, with $|\ell|, |b| < 2$.}
\label{fig:gceregs}
\end{figure}

\begin{figure}[!ht]
\centering
\includegraphics[width=\linewidth]{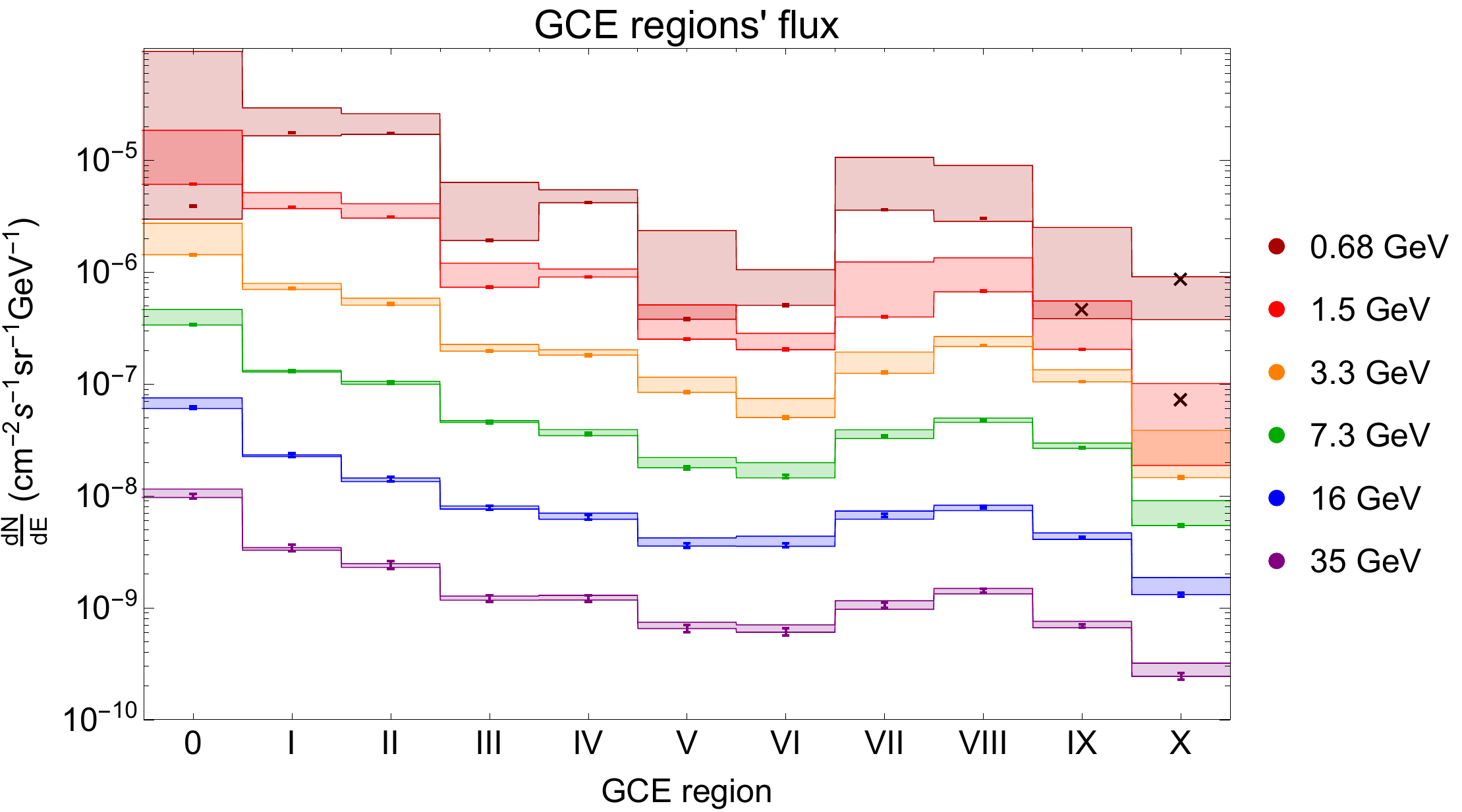}
\caption{The flux around the Galactic center for the regions given in Figure~\ref{fig:gceregs} at all energy levels and with all 
angular scales. The fluxes in region IX in the first energy bin and in region X in the first two energy bins are negative, and we 
do report the absolute value of their uncertainty band.}
\label{fig:gcereg}
\end{figure}

Following \cite{Calore:2014xka}, we calculate the flux around the Galactic center in a number of different regions, as shown in 
Figure~\ref{fig:gceregs}. In addition to those used in \cite{Calore:2014xka}, we also include ``region 0,'' defined by $|\ell|, |b| < 2$. 
In Figure~\ref{fig:gcereg}, we show the residual emission in each region for each of our six energy bins. We find positive emission 
associated with the GCE in all energy bins for regions 0--VIII. Only region IX below 1 GeV and region X below 2 GeV have 
negative residual emission. We remind the reader that for any of these fluxes, the average across the entire sky has been 
removed, so these are fluxes on regions of a map with a zero average. We also point out that the lowest energy bin has the 
largest systematics. 

\begin{figure}[t]
\centering
\includegraphics[width=\linewidth]{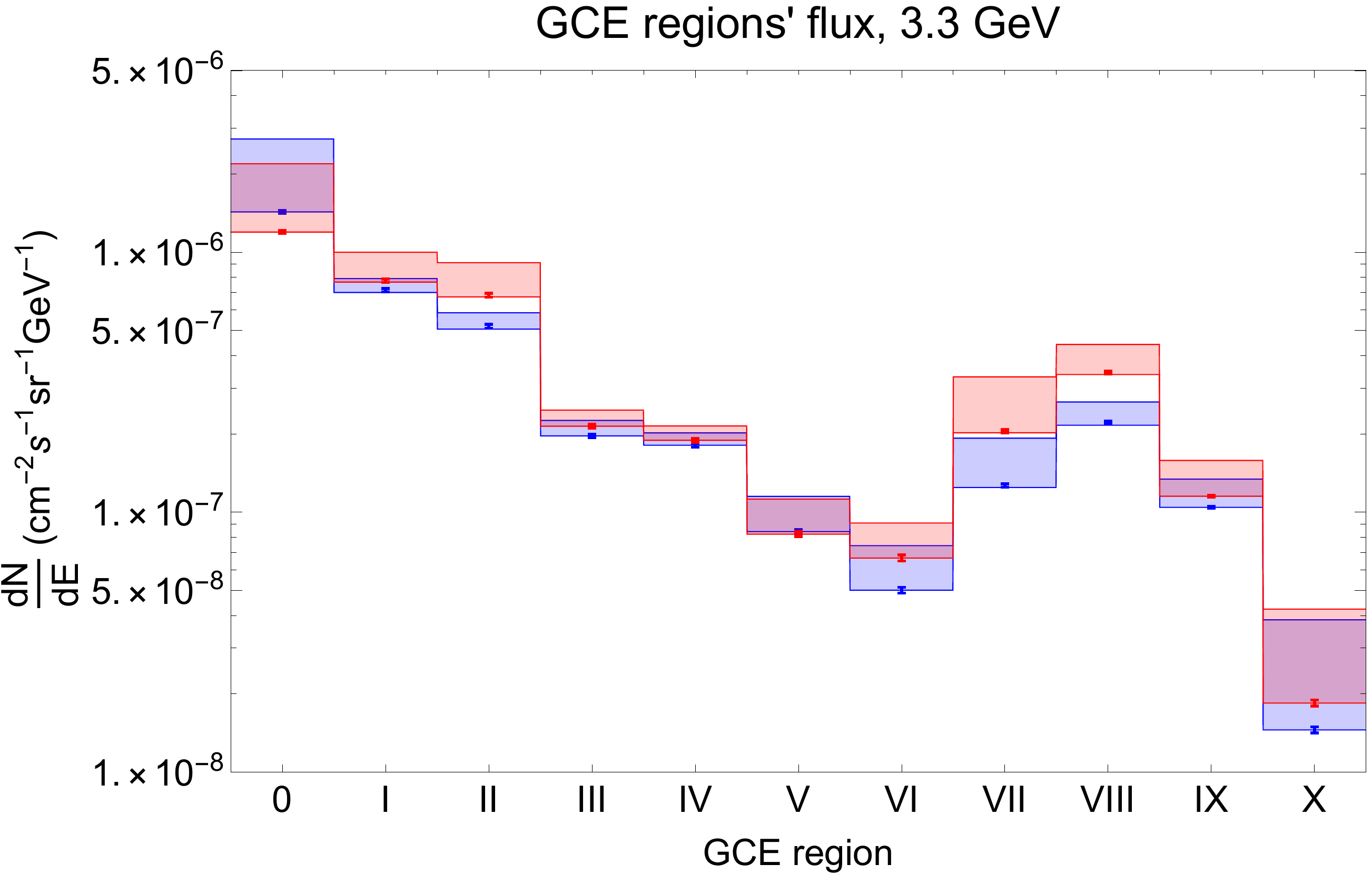}
\caption{The flux around the Galactic center for the regions given in Figure~\ref{fig:gceregs} at $E = 3.3$ GeV. 
The flux is presented as two series, one including power from all scales, $j \geq 1$ (blue), and 
one including only $j \geq 3$ (red). The differences between the two fluxes in regions I, III, IV, V, IX,
and X are $\sim$$10 \%$, while in regions 0, II, VI, VII, and VIII the differences go from $\sim$30 up to $\gtrsim 100 \%$.}
\label{fig:gcerege2}
\end{figure}

As with our analysis of the Bubbles, we wish to compare the small- and large-scale nature of the GCE. This is shown in 
Figure~\ref{fig:gcerege2} for the 3.3 GeV bin. In the blue band of Figure~\ref{fig:gcerege2} we show the total GCE emission, 
or the sum of all 9 scales (as in the orange band of Figure~\ref{fig:gcereg}), while in the red band we show the GCE emission 
only for wavelet scales with $j \geq 3$. In regions I, III--V, IX, and X we find no significant difference with or without $w_1$ 
and $w_2$, indicating little power at small scales. In contrast, the remaining GCE regions, all of which lie close to the Galactic 
midplane, provide evidence that emission with support on smaller angular scales has been mismodeled. Intriguingly, this power 
does not have a fixed sign: in region 0 there is evidence for {\it positive} power in $w_1$ and $w_2$, which can indicate the 
existence of additional point sources in the very inner few degrees around the Galactic Center. However, in regions II, VII, and 
VIII there is significant {\it negative} power at small angular scales: in Figure~\ref{fig:gcerege2} the blue bands lie {\it below} the 
red bands for those regions. This suggests the presence of spurious emission in the diffuse templates along the Galactic disk.

The GCE is complicated in its nature, being neither fully diffuse, nor dominated by emission at the smallest scales. The small-scale 
emission near the Galactic disk is not well accounted for by the templates or the point-source catalog that we have used. We discuss 
the interpretation of this spurious power in more detail in the following section.

\begin{figure}[t]
\centering
\includegraphics[width=\linewidth]{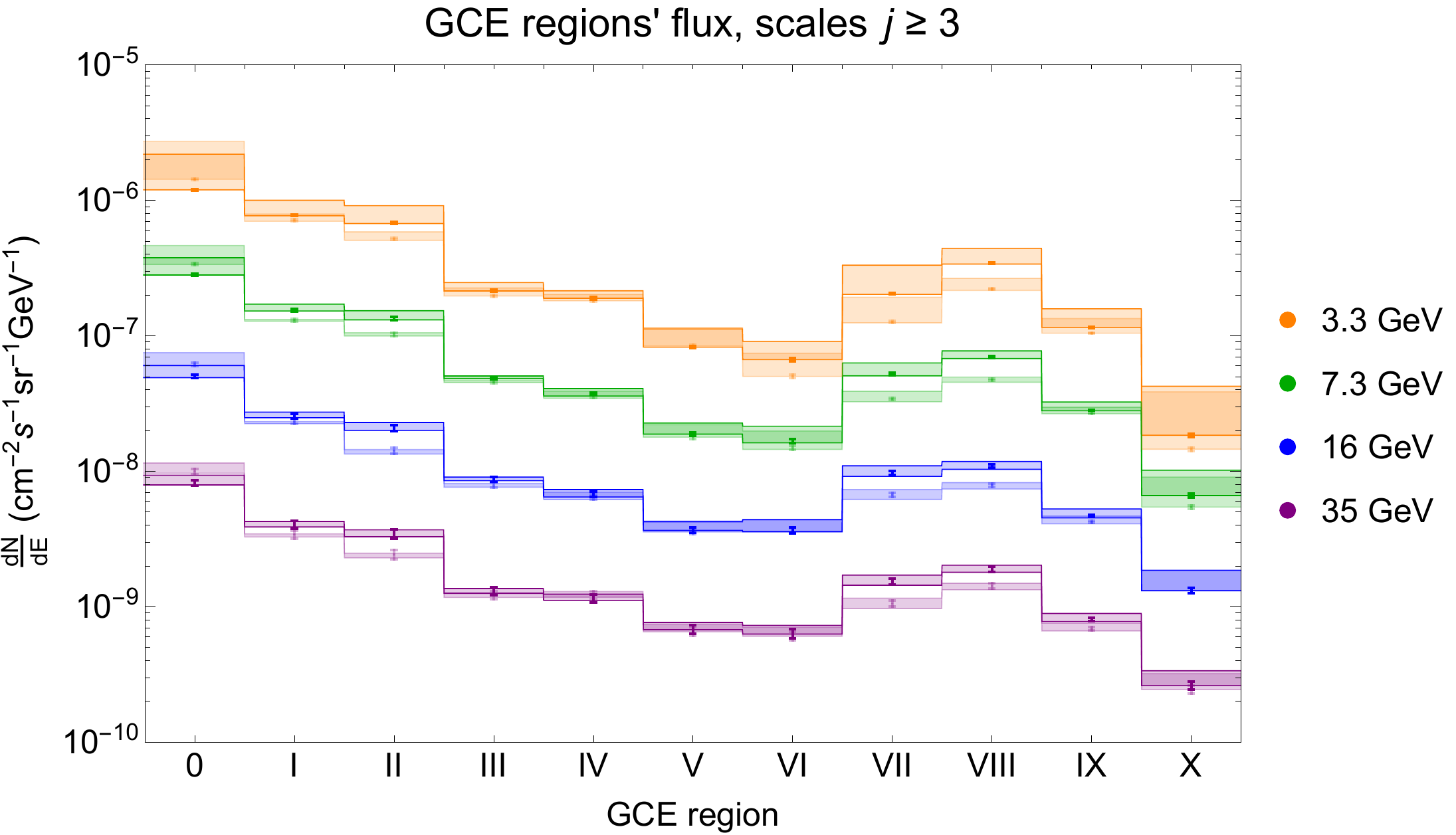}
\caption{The flux around the Galactic center for the regions given in Figure~\ref{fig:gceregs} and at several energies. We show the 
flux from scales $j \geq 3$ in the heavily shaded bands and $j \geq 1$ in the fainter bands.}
\label{fig:gceregAllE}
\end{figure}

In Figure~\ref{fig:gceregAllE} we show the GCE emission associated with scales $j \geq 3$ vs $j \geq 1$ for the energy bins between 
3.3 and 35 GeV. At 0.68 and 1.5 GeV the systematic uncertainties are large enough to prevent any conclusions regarding the nature 
of the GCE emission and are omitted here for clarity (see though Figure~\ref{fig:GCE_w1w3_AllE} of Appendix~\ref{app:15by15}). At 
energies of 7.3, 16, and 35 GeV, our findings are similar to the ones at 3.3 GeV. An insignificant portion of the excess in Regions I, III, 
IV, V, IX, and X can be associated with small angular scales. More power originates at small scales for regions 0, II, VII, and VIII, though 
with the negative sign that we noted above. Region VI is more diffuse in nature at higher energies and becomes similar to its reflection 
across the disk, region V.   As for 3.3 GeV, the sign of these contributions from small scales varies systematically with position relative 
to the Galactic disk. We leave the interpretation of these results for section~\ref{sec:PrevWork} and present additional information in 
Appendix~\ref{app:15by15}. 

\begin{figure*}[t]
\centering
\includegraphics[width=.45\textwidth]{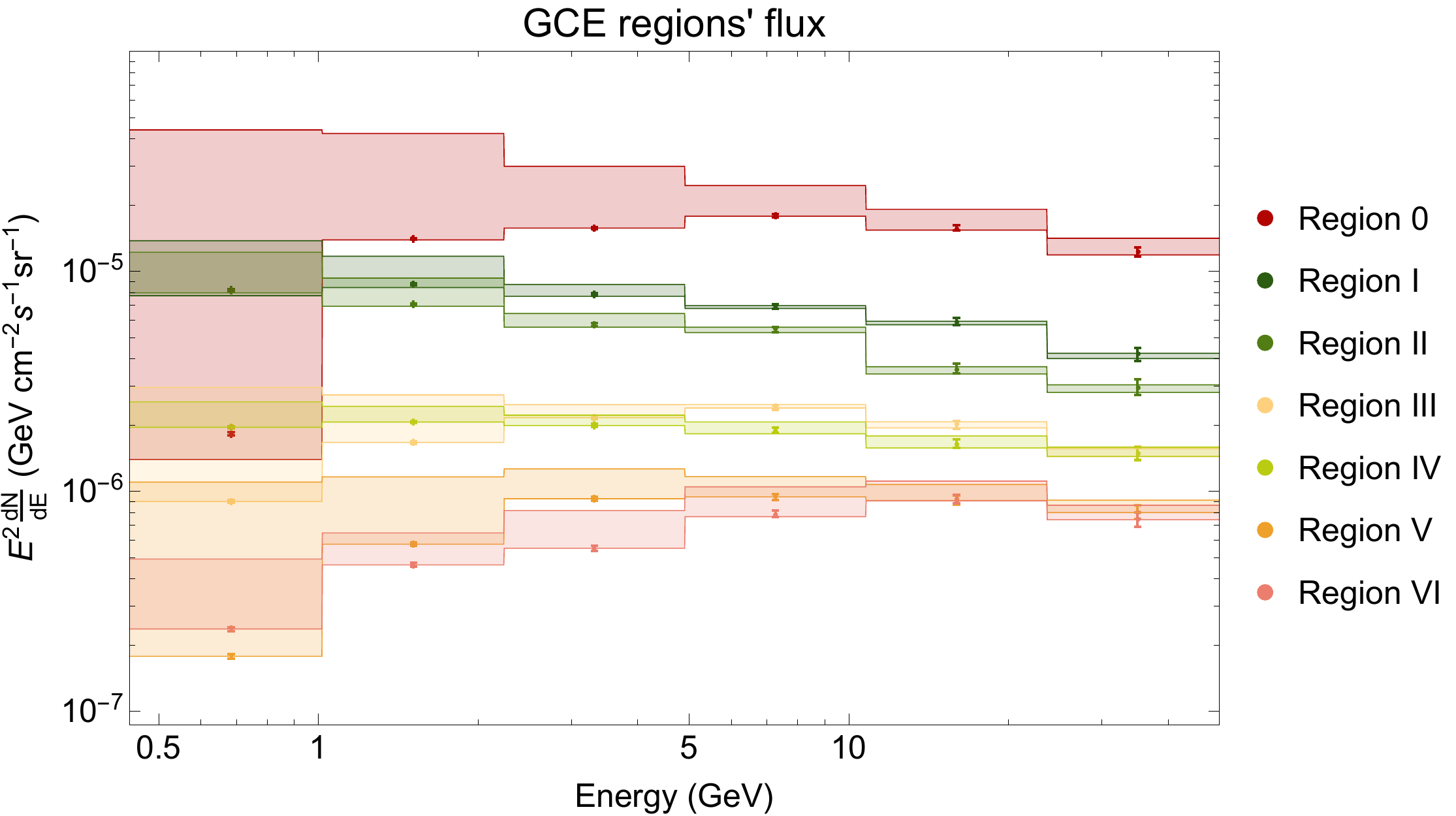}\qquad~~
\includegraphics[width=.45\textwidth]{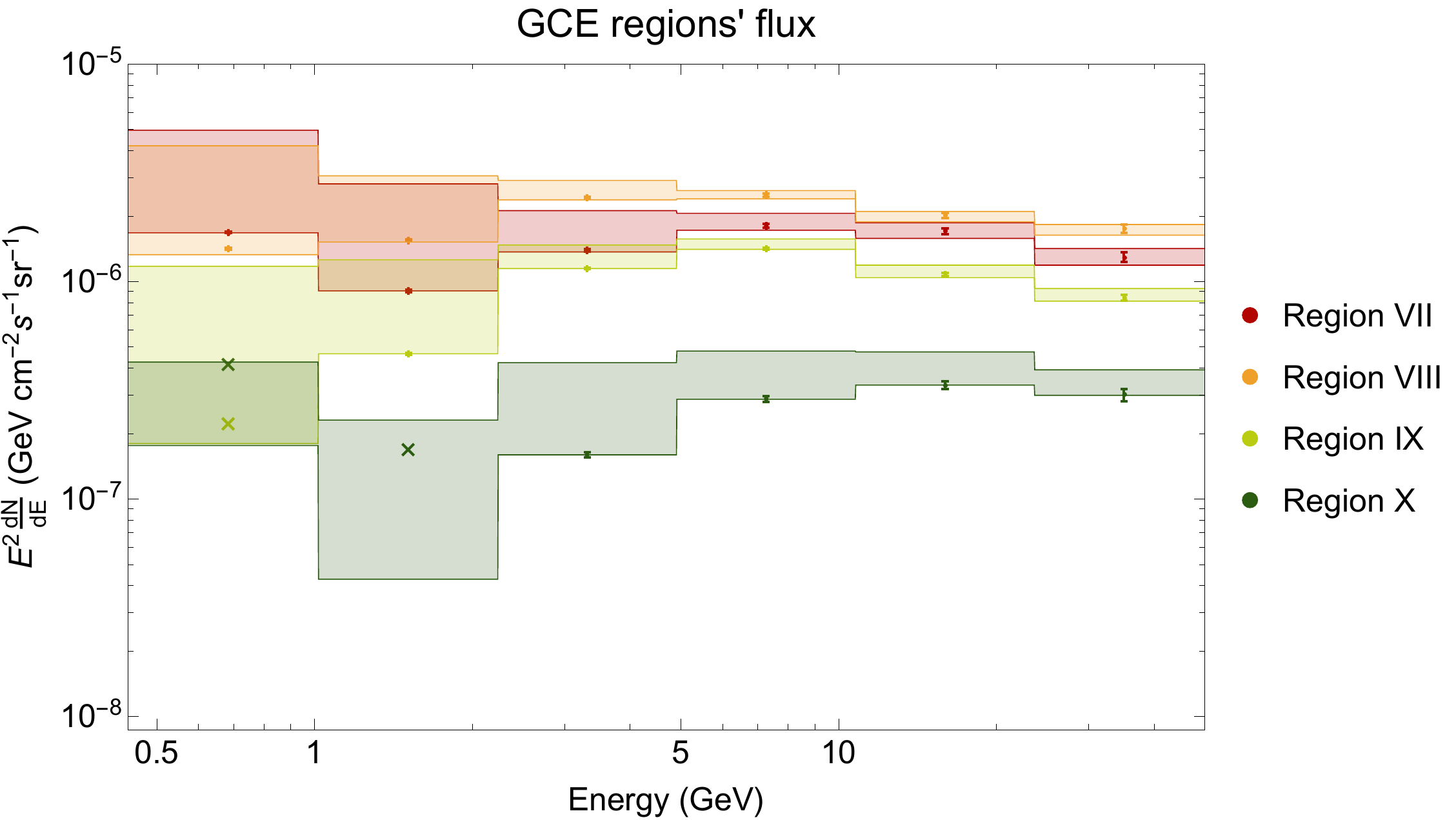}
\caption{Spectra for regions near the Galactic center. \textit{Left}, regions 0, and I-VI. \textit{Right}, regions VII-X. We show the absolute 
value when the flux is negative.}
\label{fig:gceregSpectra}
\end{figure*}
	
In Figure~\ref{fig:gceregSpectra} we show the GCE flux spectra (multiplied by $E^2$) from all wavelet levels for the different regions. The spectra 
from regions 0--VIII are consistently positive, while the flux from region IX is negative in its first energy bin and in region X in its first two energy bins. 
Since the wavelet transform removes the monopole term of the entire sky, negative fluxes are expected in dim regions of the sky that are adjacent 
to an excess.  From Figure~\ref{fig:gceregSpectra} we see that the majority of spectra, when positive, can be fit by a broken power law, 
$dN/dE \propto E^{-\alpha}$, with $\alpha \le 2$ at low energies and $\alpha >2$ above the break. The break in most cases occurs in the 5--10 GeV bin, 
although in regions VI and X it is in the 10--23 GeV bin and in regions I, II, and IV there is little evidence for a break. Since we use a small number 
of bins, we avoid spectral analysis beyond this level. The main goal of this wavelet technique is to identify diffuse emissions, study their morphology 
in a less biased way than a template technique and also discuss the nature of these diffuse emissions in terms of the angular scales where most of 
their power lies. Yet, since our wavelet based technique is carried out on the sky after subtracting an average background and also removes the full 
sky average from the maps, many photons are lost.  Thus, these advantages come at the expense of wider energy bins, and as a result a poorer 
understanding of the spectral properties. 

\begin{figure}[t]
\centering
\includegraphics[width=\linewidth]{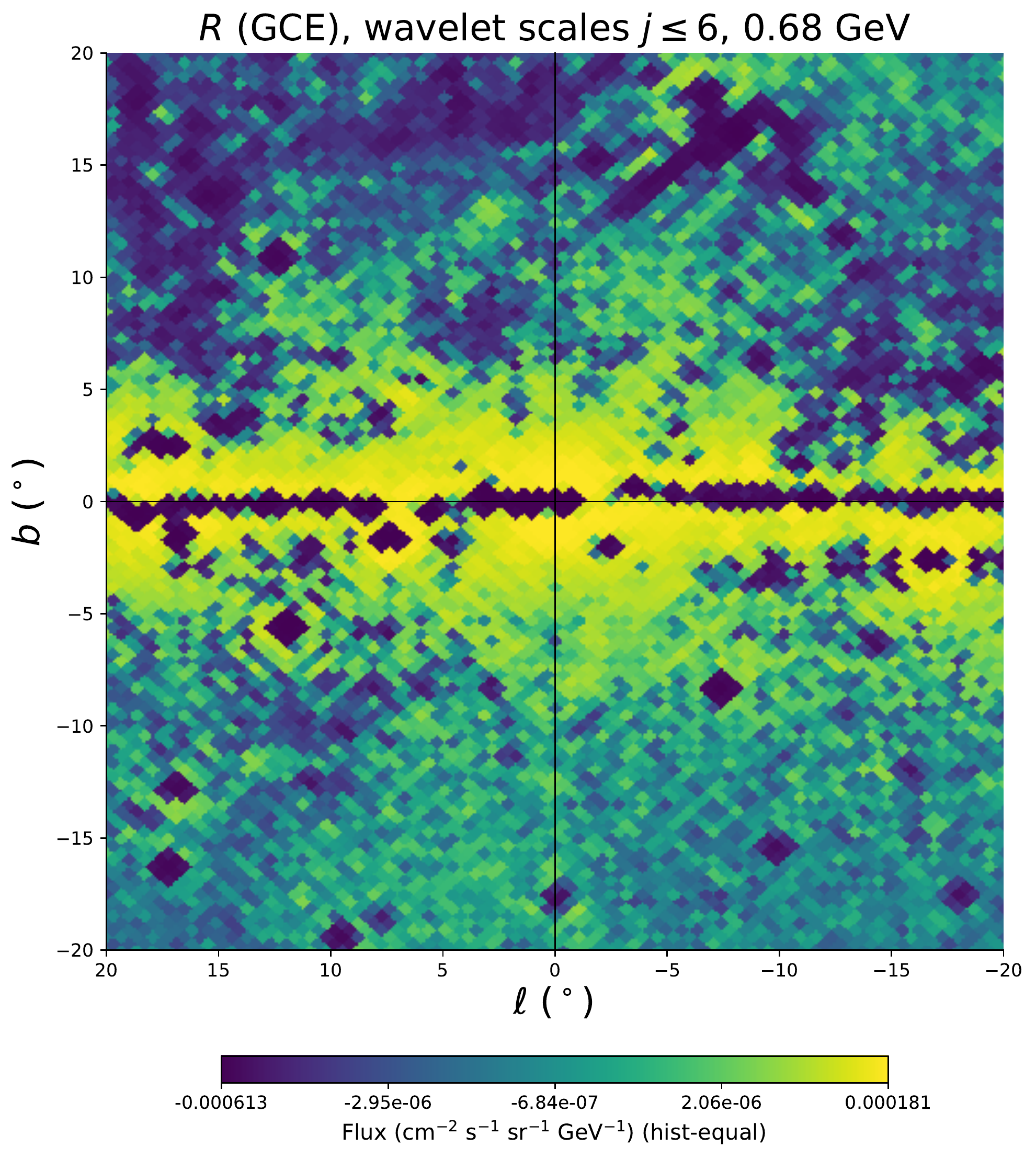}
\caption{The differential flux around the Galactic center from scales $j \leq 6$ at 0.68 GeV. We find 
the center of the emission to be at $(\ell, b) = (-4^{\circ}, 0^{\circ})$, with the emission being slightly
elongated along the disk.}
\label{fig:GCE1to6}
\end{figure}

Our wavelet-based approach can also assess the location of the center of the GCE emission and its extent in longitude and latitude at different energies. 
These important aspects of the GCE are contested by different results (see e.g. \cite{Daylan:2014rsa, Calore:2014xka, Gordon:2013vta, Linden:2016rcf, 
Macias:2016nev, Bartels:2017vsx}) and are crucial for the interpretation of the underlying emission. To initially address these questions, we consider only 
the wavelet scales of 1-6, {\it i.e.}~GCE$^{1-6}$, Equation (\ref{roidef}), which ignores power that is in scales larger than $45^{\circ}$; we do not want the 
\textit{Fermi} Bubbles or other regions of the Milky Way to affect our results. In Figure~\ref{fig:GCE1to6} we show the sum of wavelet levels 1-6  
for a window of $40^{\circ}\times40^{\circ}$ around the Galactic center at 0.68 GeV. GCE$^{1-6}$ appears off-center by $\approx 4^{\circ}$ toward $\ell < 0$.   
We now try to quantify this, using two slightly different approaches. The results are presented in Table~\ref{tab:OffCenter}, along with similar analyses 
carried out on the EDE and the WDE, see Section~\ref{subsec:East_West_Emissions}.

\begin{table*}[!ht]
\begin{center}
\begin{tabular}{c|cccccccc} \hline
region characteristic &$E_1$&$E_2$&$E_3$&$E_4$&$E_5$&$E_6$\\   \hline
GCE$^{1-6}$ (smoothed) Center &$-4^{\circ}$&$-6^{\circ}$&$-6^{\circ}$&$-4^{\circ}$&$-4^{\circ}$&$-4^{\circ}$\\ 
GCE$^{3-6}$ Center & $-4^{\circ}$ & $-6^{\circ}$ & $-6^{\circ}$ & $-4^{\circ}$ & $-3^{\circ}$ & $-3^{\circ}$  \\ 
GCE$^{3-6}$ [FWHM$_\parallel$, FWHM$_\perp$] & $[5^{\circ}, 12^{\circ}]$ & $[8^\circ, 5^{\circ}]$ & $[9^\circ,7^{\circ}]$ & $[7^\circ,5^{\circ}]$ & $[15^\circ,4^{\circ}]$ & $[7^\circ,4^{\circ}]$ \\ 
\hline
WDE$^{1-6}$ (smoothed) Center &$24^{\circ}$ & $26^{\circ}$ & $26^{\circ}$ & $27^{\circ}$ & $27^{\circ}$ & $23^{\circ}$\\ 
WDE$^{3-6}$ Center & $23^{\circ}$ & $24^{\circ}$ & $24^{\circ}$ & $24^{\circ}$ & $26^{\circ}$ & $24^{\circ}$ \\
WDE$^{3-6}$ [FWHM$_\parallel$, FWHM$_\perp$] & $[23^{\circ},10^\circ]$ & $[31^{\circ},6^\circ]$ & $[27^{\circ},4^\circ]$ & $[32^{\circ},6^\circ]$ & $[26^{\circ},4^\circ]$ & $[28^{\circ},4^\circ]$ \\ 
\hline
EDE$^{1-6}$ (smoothed) Center &$-22^{\circ}$&$-19^{\circ}$&$-19^{\circ}$&$-18^{\circ}$&$-19^{\circ}$&$-18^{\circ}$\\
EDE$^{3-6}$ Center & $-18^{\circ}$ & $-22^{\circ}$ & $-22^{\circ}$ & $-21^{\circ}$ & $-23^{\circ}$ & $-20^{\circ}$  \\ 
EDE$^{3-6}$ [FWHM$_\parallel$, FWHM$_\perp$] & $[13^\circ,12^\circ]$ & $[16^\circ,4^\circ]$ & $[16^\circ,5^\circ]$ & $[18^\circ,6^\circ]$ & $[15^\circ,5^\circ]$ & $[17^\circ,4^\circ]$  \\ 
 \hline
\end{tabular}
\end{center}
\caption{For each of the three disk-centric excesses we present their centers calculated using the smoothed approach on wavelet levels 1-6 and 
unsmoothed on levels 3-6.  We also present the full width at half maximum (FWHM) along the disk (FWHM$_\parallel$) and perpendicular to the 
disk (FWHM$_\perp$), using ROI$^{3-6}$.  This is done for all six energy bins.}
\label{tab:OffCenter}
\end{table*}

\begin{figure*}[!ht]
\centering
\includegraphics[width=0.45\linewidth]{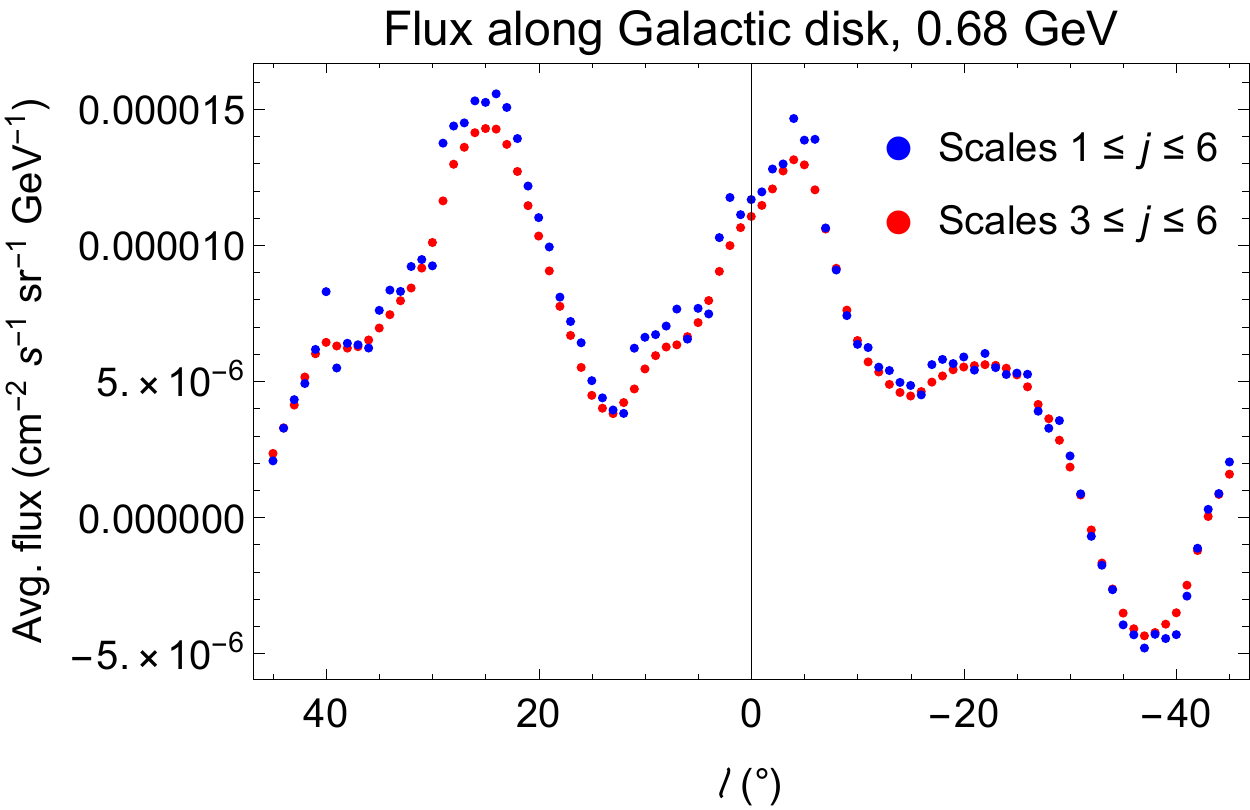}\qquad
\includegraphics[width=0.45\linewidth]{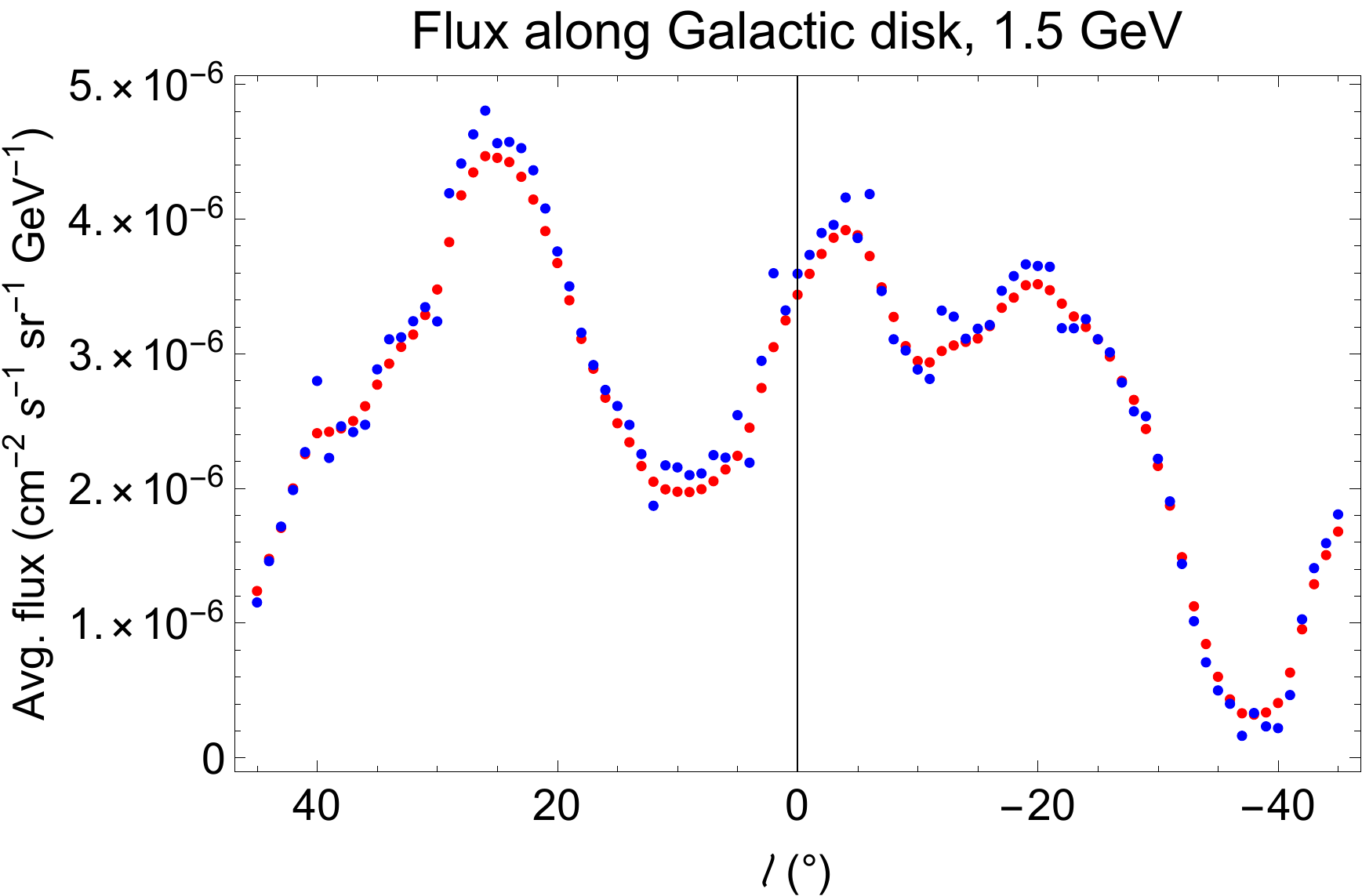} \\
\caption{The averaged flux profile of the Galactic disk for $-45^{\circ} \leq \ell \leq 45^{\circ}$, averaging over a 
$10^{\circ}\times10^{\circ}$ moving box. We show the flux in scales $1 \leq j \leq 6$ (blue points) and 
in scales $3 \leq j \leq 6$ (red points). The GCE, the WDE, and the EDE are all clearly visible 
especially once we remove power from the lowest two wavelet scales.}
\label{fig:WDE_EDE}
\end{figure*}

For our first approach, shown in Figure~\ref{fig:WDE_EDE}, the center of the GCE$^{1-6}$ emission is calculated by translating a $10^{\circ}\times10^{\circ}$ 
window along the disk ($b=0^{\circ}$) and finding the longitude for which the average GCE$^{1-6}$ flux in the window is maximal. This is different from finding 
the pixel with the maximum flux along the disk---for instance, we average over emission from sub-threshold point sources in the $10^{\circ}\times10^{\circ}$ window, 
and are thus less sensitive to individual undiscovered point sources. Although different from finding the brightest pixel, if we had defined the center as the 
brightest pixel of GCE$^{1-6}$ along the disk our results would change by only $\approx 1^{\circ}$.   We see that this approach confirms, across all energy 
bins, what was seen in Figure~\ref{fig:GCE1to6}: the center of GCE$^{1-6}$ is offset from the Galactic center by $\approx -4^\circ$. 

\begin{figure*}[!ht]
\centering
\includegraphics[width=0.49\linewidth]{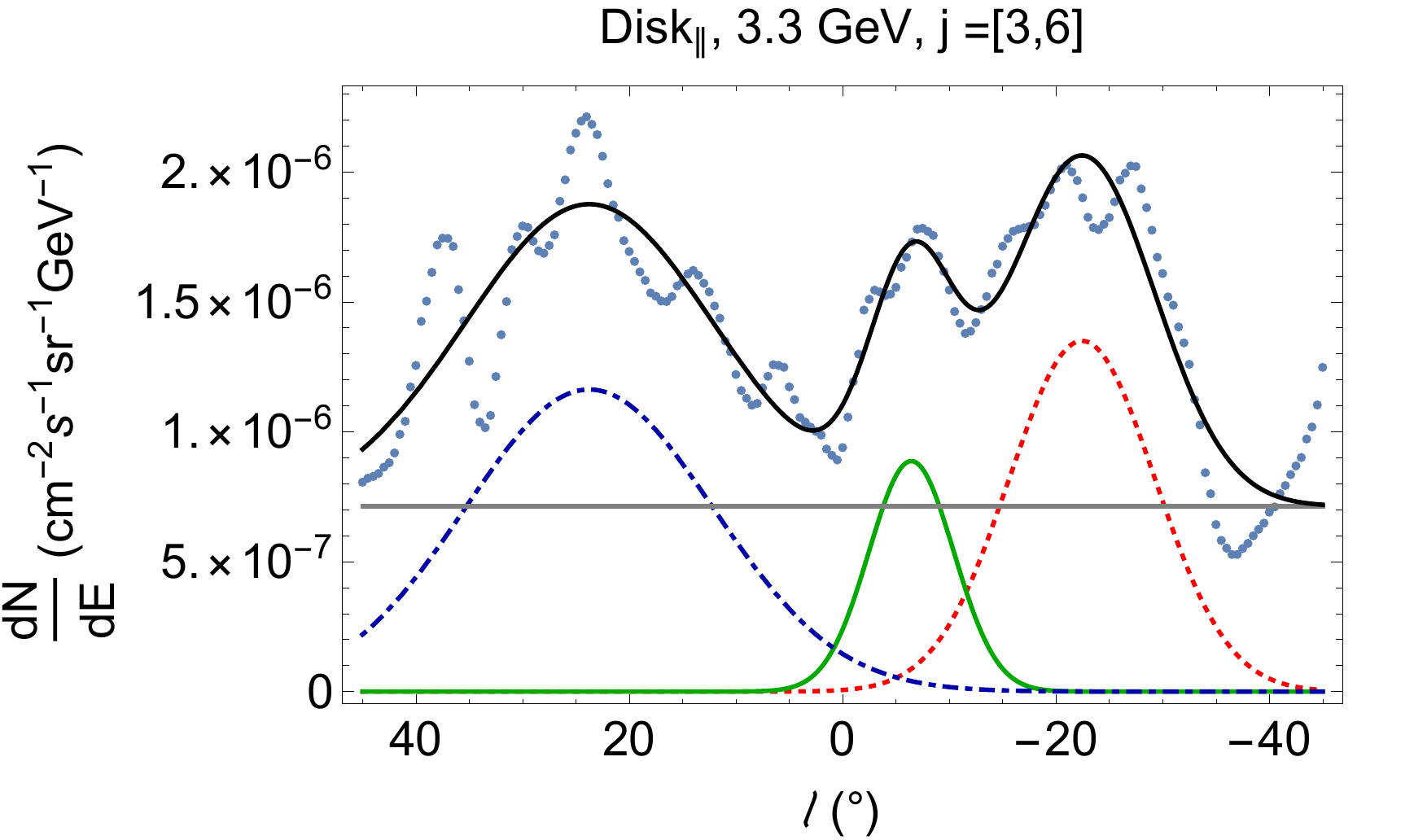}
\includegraphics[width=0.49\linewidth]{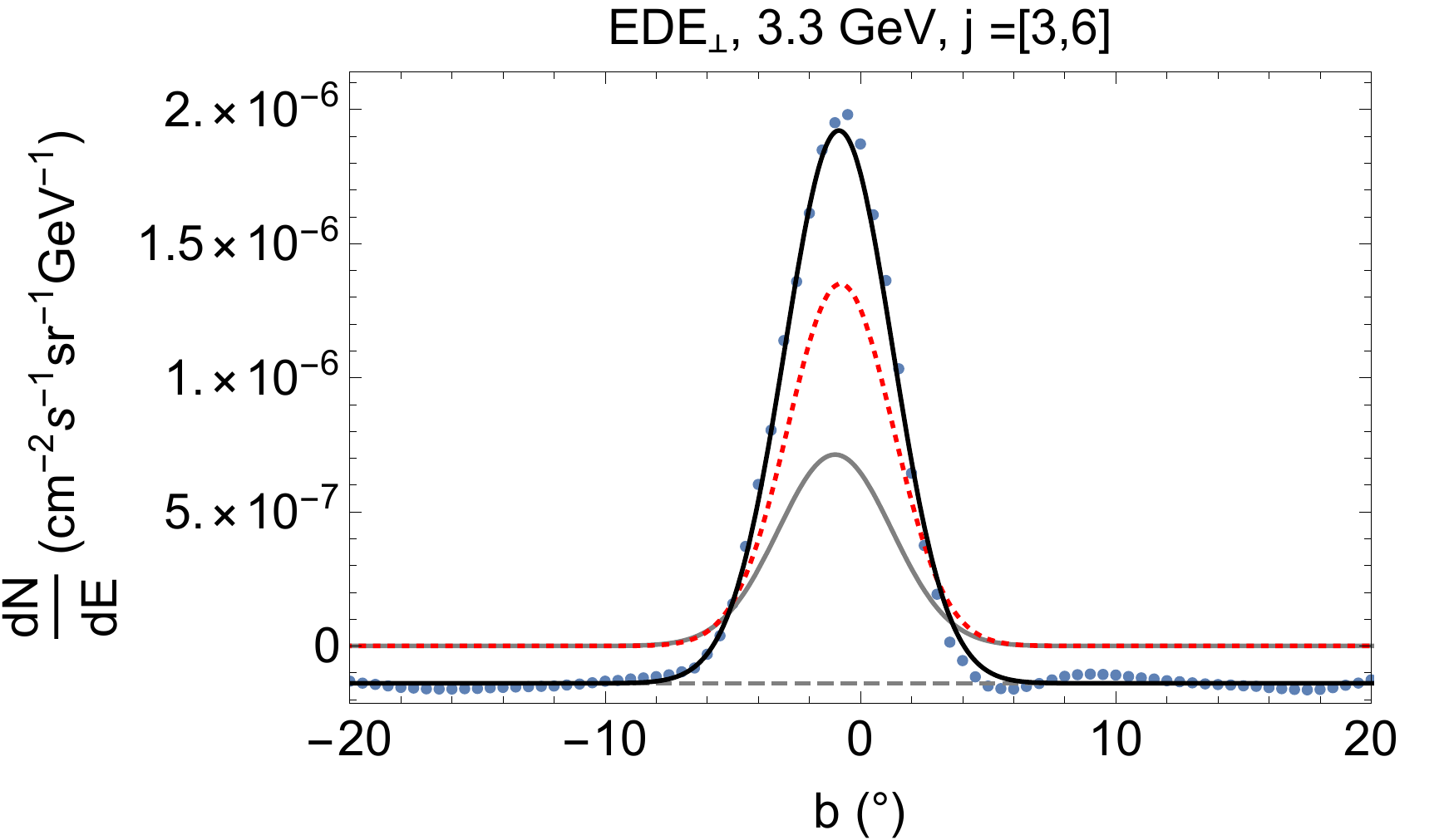}\\
\includegraphics[width=0.49\linewidth]{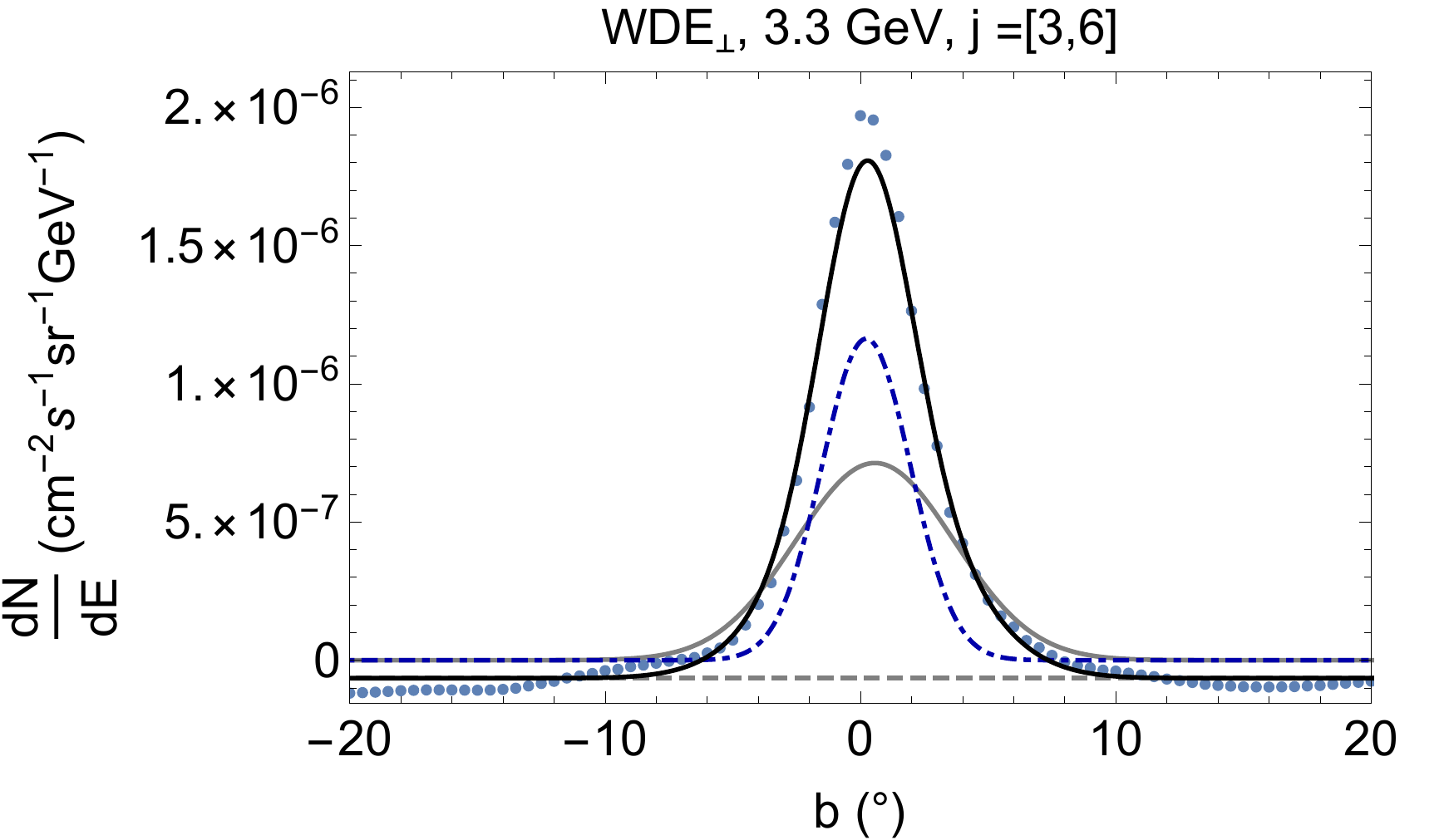}
\includegraphics[width=0.49\linewidth]{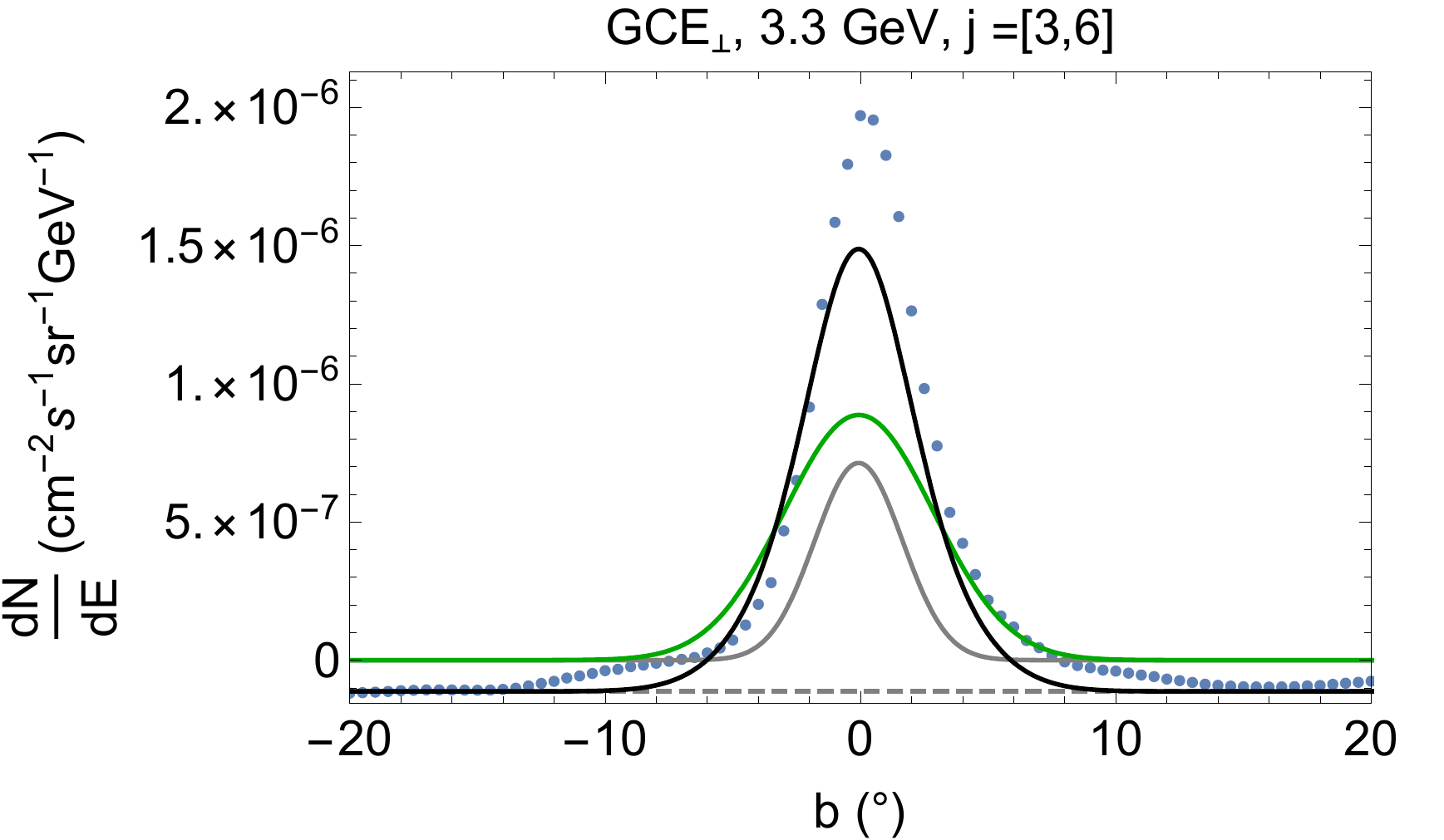}
\caption{{\it Top left:} $R^{3-6}$ for the region within $1^\circ$ of the Galactic midplane, for the 3.3 GeV energy bin.  In each case the wavelet coefficients 
(blue dots) are fit by the sum of a constant and three gaussians, shown in gray, dot-dashed blue, solid green and dashed red to represent the disk, GCE, 
EDE, and WDE, respectively. {\it Remaining panels:} fits to the latitude profiles of the disk, GCE, EDE, and WDE with normalizations fixed from the fit in 
the top left panel.}
\label{fig:waveletgaussianfits}
\end{figure*}

The second approach, shown in Figure~\ref{fig:waveletgaussianfits}, is inherently more wavelet based: we consider the flux within a strip $|b|<0.5^\circ$ 
over the region $|\ell|\le 40^\circ$ and omit small angular scales, analyzing only ${\rm GCE}^{3-6}$.  This ROI contains the GCE as well as both the EDE 
and the WDE.  This choice of levels is motivated by trying to find large diffuse objects, but ignores the wavelet levels $j\ge 7$ at which point the excesses 
merge.  Since the ROI contains three diffuse objects, we fit the flux in each energy bin as the sum of three gaussians and an additional constant; the 
constant is included to represent the ``long wavelength'' modes we have explicitly ignored. The width, central $\ell$ pixel, and normalization of these 
gaussians are allowed to float.  The region of interest made from only the highest of these wavelet levels, ${\rm GCE}^{5-6}$, is well fit by two gaussians 
and ${\rm GCE}^{4-6}$ is well fit by the same solution we find for ${\rm GCE}^{3-6}$ but has less resolved substructure.  We present the results of this 
fit for ${\rm GCE}^{3-6}$ for the 3.3 GeV energy bin in Figure~\ref{fig:waveletgaussianfits} and for the 7.3 GeV bin in Figure~\ref{fig:waveletgaussianfits2}.  
This approach simultaneously finds the centers of the diffuse excesses as well as their extension along the disk.  The results are presented in 
Table~\ref{tab:OffCenter}, and again confirm the shift in the center.  

To find the extent of the excesses perpendicular to the disk we look at the flux along a $1^\circ$ slice in $\ell$ and $-20^\circ<b<20^\circ$. The value of 
$\ell$ is fixed to the center of each excess as found by the first of our wavelet-based fitting procedures.  Since the disk is a sizable background and can 
bias the preferred value of the image component heights, we fit the flux in the perpendicular direction to a sum of two gaussians and a constant, but the 
normalizations of the gaussians are fixed to the best fit values for the constant and relevant gaussian found in the previous step.  The results of this 
procedure are shown Table~\ref{tab:OffCenter}.

We find that the GCE is off-center by $3^{\circ}$--$6^{\circ}$ at all energies. This result is to some extent affected by specific nearby point sources.
Omitting scales 1 and 2, this offset is smaller but still at negative $\ell$.  

\subsection{The Diffuse Disk Emissions Centered at $\ell \approx 25^{\circ}$ and $\ell \approx -20^{\circ}$}
\label{subsec:East_West_Emissions}

Similar to our analysis of the GCE, we study the two additional emissions that we identified along the disk. These diffuse emissions are bright, close 
to the Galactic center, and extended enough to possibly contaminate the GCE at large wavelet scales $j \geq 7$. Template works that extend out 
to these regions will try to fit their emission, potentially biasing the best fit diffuse emission parameters if they are not included in the model. The 
information regarding their centers and widths versus energy is given in Table~\ref{tab:OffCenter}. We find that both the WDE and the EDE are 
significantly more elongated along the Galactic disk than the GCE emission. We tested how these results would have changed if we had used a 
different number of wavelet scales in calculating them, and we found that the WDE and the EDE remain significantly more elongated along the 
Galactic disk than the GCE  emission.

In Figure~\ref{fig:WDE_EDE} we show the data we used to determine centers of the three emissions (red points), as the average flux within a window 
translated along the disk from $\ell = -45^{\circ}$ to $\ell = +45^{\circ}$. In blue are fluxes using scales 1--6 summed; in red are those using only scales 3--6. 
We show our results at all energy bins. The three major diffuse emissions along the disk (GCE, WDE, and EDE) can 
be clearly seen. Furthermore, by comparing the blue to the red points, we quantify the impact of small
scale structures as those coming from collections of unidentified/mis-modeled point sources.

To give the exact decomposition of flux of these emissions, in Table~\ref{tab:Scales} we give the 
associated average flux within a $10^{\circ}\times10^{\circ}$ window centered at the WDE and EDE for each energy 
and wavelet scale. These are \textit{partial} fluxes and inform us of the gamma-ray emission at different scales. We also 
give for comparison the results of the GCE$^{1-6}$. Again for the sake of comparison, we do not use a different center at each energy for a given emission; instead, we choose the emission's center at the third energy bin, $-6\deg$ for the GCE$^{1-6}$, $26\deg$ for the WDE, and $-19\deg$ for the EDE.

\begin{table*}[t]
\begin{center}
\begin{tabular}{c|cccccccc} \hline
region scale &$f(E_1)$&$f(E_2)$&$f(E_3)$&$f(E_4)$&$f(E_5)$&$f(E_6)$\\   \hline
GCE$^{1-6}$ &$1.4 \times 10^{-5}$&$4.2 \times 10^{-6}$&$8.9 \times 10^{-7}$&$1.7 \times 10^{-7}$&$2.5 \times 10^{-8}$&$4.2 \times 10^{-9}$\\ 
GCE$^{1}$ &$8.3 \times 10^{-7}$&$2.5 \times 10^{-7}$&$5.5 \times 10^{-8}$&$6.0 \times 10^{-9}$&$2.9 \times 10^{-10}$&$2.6 \times 10^{-11}$\\ 
GCE$^{2}$ &$1.0 \times 10^{-6}$&$2.1 \times 10^{-7}$&$4.5 \times 10^{-8}$&$6.2 \times 10^{-9}$&$7.1 \times 10^{-10}$&$1.0 \times 10^{-10}$\\ 
GCE$^{3}$ &$2.6 \times 10^{-6}$&$5.6 \times 10^{-7}$&$1.1 \times 10^{-7}$&$2.0 \times 10^{-8}$&$3.0 \times 10^{-9}$&$5.1 \times 10^{-10}$\\ 
GCE$^{4}$ &$3.9 \times 10^{-6}$&$1.2 \times 10^{-6}$&$2.6 \times 10^{-7}$&$4.9 \times 10^{-8}$&$7.7 \times 10^{-9}$&$1.3 \times 10^{-9}$\\ 
GCE$^{5}$ &$3.7 \times 10^{-6}$&$1.2 \times 10^{-6}$&$2.6 \times 10^{-7}$&$5.1 \times 10^{-8}$&$7.8 \times 10^{-9}$&$1.3 \times 10^{-9}$\\ 
GCE$^{6}$ &$1.8 \times 10^{-6}$&$7.2 \times 10^{-7}$&$1.7 \times 10^{-7}$&$3.6 \times 10^{-8}$&$5.8 \times 10^{-9}$&$9.5 \times 10^{-10}$\\ 
WDE$^{1-6}$ &$1.5 \times 10^{-5}$&$4.8 \times 10^{-6}$&$1.0 \times 10^{-6}$&$1.7 \times 10^{-7}$&$2.6 \times 10^{-8}$&$4.5 \times 10^{-9}$\\ 
WDE$^{1}$ &$4.1 \times 10^{-7}$&$1.3 \times 10^{-7}$&$2.1 \times 10^{-8}$&$1.7 \times 10^{-9}$&$-5.9 \times 10^{-11}$&$5.9 \times 10^{-12}$\\ 
WDE$^{2}$ &$7.6 \times 10^{-7}$&$2.1 \times 10^{-7}$&$3.6 \times 10^{-8}$&$5.2 \times 10^{-9}$&$4.1 \times 10^{-10}$&$9.3 \times 10^{-11}$\\ 
WDE$^{3}$ &$3.0 \times 10^{-6}$&$7.4 \times 10^{-7}$&$1.4 \times 10^{-7}$&$2.2 \times 10^{-8}$&$3.1 \times 10^{-9}$&$5.2 \times 10^{-10}$\\ 
WDE$^{4}$ &$4.6 \times 10^{-6}$&$1.6 \times 10^{-6}$&$3.3 \times 10^{-7}$&$5.8 \times 10^{-8}$&$8.9 \times 10^{-9}$&$1.6 \times 10^{-9}$\\ 
WDE$^{5}$ &$4.3 \times 10^{-6}$&$1.4 \times 10^{-6}$&$3.1 \times 10^{-7}$&$5.6 \times 10^{-8}$&$8.5 \times 10^{-9}$&$1.5 \times 10^{-9}$\\ 
WDE$^{6}$ &$2.2 \times 10^{-6}$&$7.5 \times 10^{-7}$&$1.7 \times 10^{-7}$&$3.1 \times 10^{-8}$&$4.9 \times 10^{-9}$&$8.1 \times 10^{-10}$\\ 
EDE$^{1-6}$ &$5.7 \times 10^{-6}$&$3.7 \times 10^{-6}$&$9.8 \times 10^{-7}$&$1.9 \times 10^{-7}$&$3.0 \times 10^{-8}$&$4.7 \times 10^{-9}$\\ 
EDE$^{1}$ &$2.1 \times 10^{-8}$&$3.6 \times 10^{-8}$&$8.7 \times 10^{-9}$&$-1.7 \times 10^{-10}$&$3.9 \times 10^{-10}$&$4.0 \times 10^{-11}$\\ 
EDE$^{2}$ &$2.1 \times 10^{-7}$&$1.2 \times 10^{-7}$&$3.0 \times 10^{-8}$&$5.7 \times 10^{-9}$&$1.2 \times 10^{-9}$&$2.2 \times 10^{-10}$\\ 
EDE$^{3}$ &$9.6 \times 10^{-7}$&$5.4 \times 10^{-7}$&$1.5 \times 10^{-7}$&$3.0 \times 10^{-8}$&$4.4 \times 10^{-9}$&$6.9 \times 10^{-10}$\\ 
EDE$^{4}$ &$1.7 \times 10^{-6}$&$1.2 \times 10^{-6}$&$3.3 \times 10^{-7}$&$6.5 \times 10^{-8}$&$9.9 \times 10^{-9}$&$1.6 \times 10^{-9}$\\ 
EDE$^{5}$ &$1.7 \times 10^{-6}$&$1.1 \times 10^{-6}$&$3.0 \times 10^{-7}$&$5.7 \times 10^{-8}$&$8.9 \times 10^{-9}$&$1.4 \times 10^{-9}$\\ 
EDE$^{6}$ &$1.1 \times 10^{-6}$&$6.5 \times 10^{-7}$&$1.7 \times 10^{-7}$&$3.3 \times 10^{-8}$&$5.0 \times 10^{-9}$&$8.1 \times 10^{-10}$\\ 
 \hline
\end{tabular}
\end{center}
\caption{The average total and \textit{partial} differential fluxes $f(E_{i})$ in units of cm$^{-2}$s$^{-1}$sr$^{-1}$GeV$^{-1}$ 
within $10^{\circ}\times10^{\circ}$ centered at the GCE$^{1-6}$, the WDE, and the EDE for each energy bin and wavelet 
scale up to $j \leq 6$. While the centers of these emissions are energy dependent, for simplicity we use the centers at $E_{3}$ 
as given in Table~\ref{tab:OffCenter}.}
\label{tab:Scales}
\end{table*}

We find that within these windows the WDE, the GCE$^{1-6}$, and the EDE have between 85\% and 
97\% of their averaged emission at scales $j \geq 3$ at all energies. We discuss the interpretation of these results in 
section~\ref{sec:PrevWork}.
	
\section{Connection with previous works and interpretations}
\label{sec:PrevWork}

The \textit{Fermi} Bubbles' morphology has been studied in several works \cite{Dobler:2009xz, Su:2010qj, Fermi-LAT:2014sfa, 
Carretti:2013sc, Su:2012gu}. Several ideas have been proposed for their origin. If the charged particles that produce the Bubbles 
are CR electrons, a few underlying physical mechanisms could be at work. A regular injection of plasma causing 1$^{\mathrm{st}}$ 
order Fermi acceleration in the central part of the Galaxy  \cite{Cheng:2011xd, Lacki:2013zsa}, 2$^{\mathrm{nd}}$ order Fermi 
acceleration in turbulent regions of the Bubbles \cite{Mertsch:2011es}, or anisotropic CR diffusion, preferentially in the direction 
perpendicular to the Galactic plane \cite{Dobler:2011mk, Yang:2012fy}, could cause the Bubbles. Alternatively, they could be evidence 
of jet emission from the supermassive black hole in the center of our Galaxy occurring a few Myrs ago \cite{Guo:2011eg, 
Guo:2011ip, Zubovas:2012bn, Mou:2014pea}. If instead the charged particles are protons, then these protons would be the result 
of star formation activity over a period of Gyrs transferred away from the disk by strong Galactic winds \cite{Crocker:2010dg, 
Crocker:2014fla}.If the Bubbles are sourced by CR protons, the gamma-ray emission that constitutes the Bubbles would acquire 
a morphology that is similar to the filamentary target gas. Thus, there would be small scale structure within the Bubbles which would 
be manifestly evident in our wavelet-based analysis.

In Figure~\ref{fig:bubblelatfluxebin2} we give the Bubbles' latitude profile, while in Figures~\ref{fig:finalbubblese3} 
and~\ref{fig:finalbubbles4Ebins} we show their morphology on the sky. Our results for $|b| \geq 15^{\circ}$ are in agreement with 
previous works regarding their flatness and extent. More importantly, and as a direct result of the wavelet analysis, we are able to observe 
that the flux derives almost entirely from scales $j \geq 3$ (Figures~\ref{fig:sbubblespece2w3}), which favors the leptonic origin of the 
Bubbles. In no part of the \textit{Fermi} Bubbles spectrum for $|b| \geq 15^{\circ}$ do we find an indication for emission in small angular 
scales. The lack of power on low wavelet scales disfavors an explanation of the {\it Fermi} Bubbles that originates in collections of 
point sources or brightening of gas filaments.  This points towards a leptonic CR origin for the Bubbles.

Interestingly, we find evidence not only for a Southern Cocoon, but also for a similar emission in the Northern hemisphere, along 
the same axis as the southern one, dimmer by a factor of $\sim$$30 \%$. Previous works have found the Southern Cocoon 
\cite{Su:2012gu, Fermi-LAT:2014sfa}, with no clear consensus on the emission at the Northern hemisphere. The brighter overall emission 
in the Northern hemisphere and the dimmer emission from the Northern Cocoon make its detection more challenging. These Cocoons, 
as with the Bubbles, can be the result of episodic CR outflows either originating directly from accretion by the super-massive black 
hole in the center of the Milky Way, or from conditions in the surrounding environment. The axis of the Cocoons has an apparent 
$\sim30^{\circ}$ inclination to the perpendicular to the disk. The actual inclination angle may be different, with the Southern Cocoon 
being directed toward us, such that its relative brightness may be due to projection effects. Finally, from observations of radio galaxies 
where CR electrons are injected from the central black hole, we know that the brightness of the radio jets and Bubbles are not the same 
\cite{Becker:1995ei, Kapinska:2017bce}. The fact that the Northern Cocoon is only about a factor $\sim30\%$ dimmer than the southern 
one indicates that these effects are mild in the case of our Galaxy.

\begin{figure}[!ht]
\centering
\includegraphics[width=\linewidth]{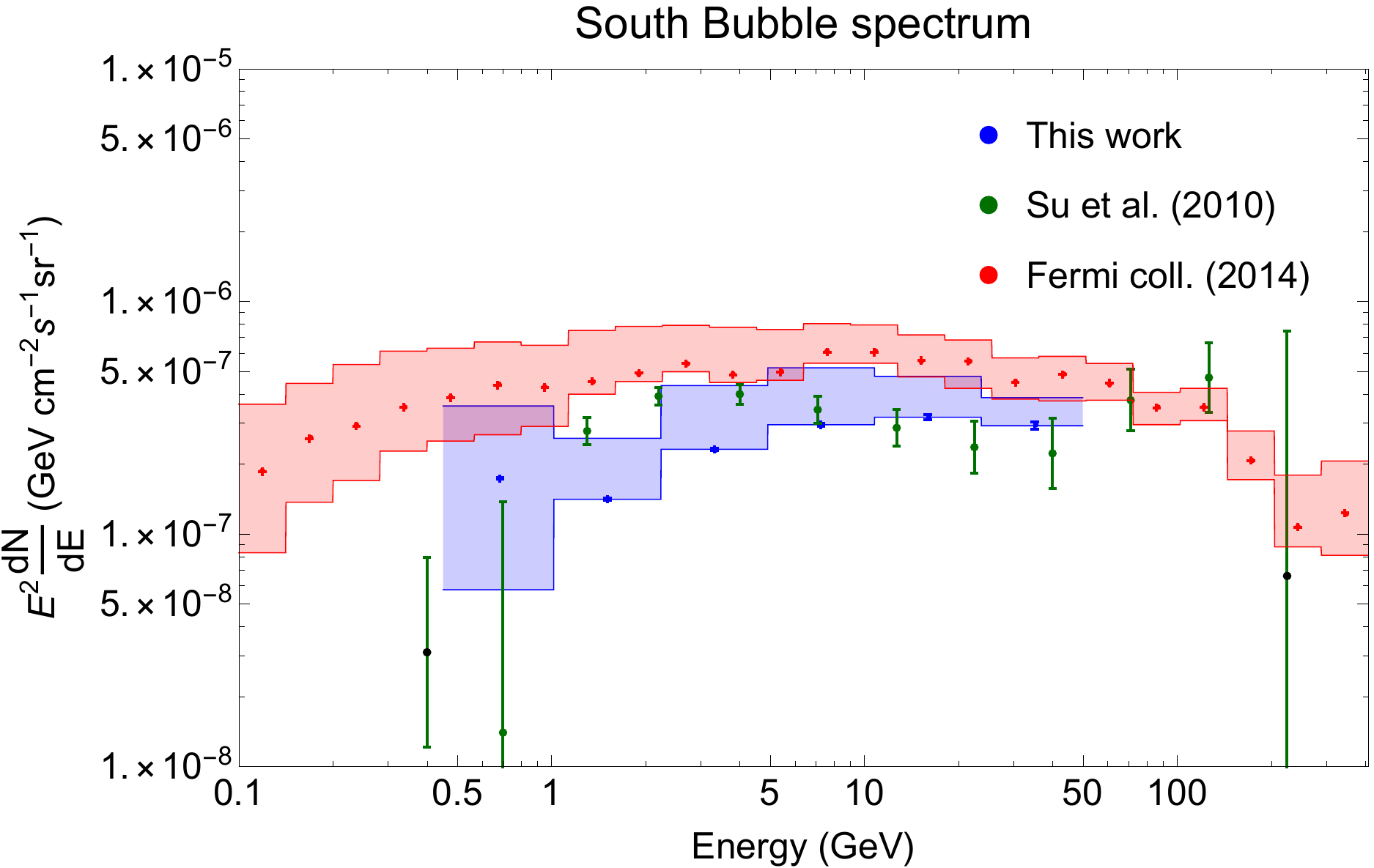}
\caption{The Southern Bubble spectrum, at $ -45^{\circ} \leq b \leq -15^{\circ}$, $|\ell | \leq 20^{\circ}$ from this work, compared to template 
works \cite{Su:2010qj, Fermi-LAT:2014sfa}.}
\label{fig:BubblesComparison}
\end{figure}

In Figure~\ref{fig:BubblesComparison}, we compare the Southern \textit{Fermi} Bubble spectrum at wavelet levels 1-9 from 
Figure~\ref{fig:sbubblespece2w3} to earlier results \cite{Su:2010qj,Fermi-LAT:2014sfa}.The Southern hemisphere is cleaner than the 
Northern hemisphere, making the comparison easier. We find that while the flux from the Bubbles is smaller than that of  \cite{Fermi-LAT:2014sfa} 
it agrees well with \cite{Su:2010qj}, but within the uncertainties our results agree with both \cite{Su:2010qj, Fermi-LAT:2014sfa}.

Now we discuss the residual emission at the lower latitudes of the gamma-ray sky; eventually we will discuss possible connections to the Bubbles.
The inner galaxy is inherently more complicated than the high latitudes where we identify the Bubbles. The expected background emission is 
comprised of several stellar components, including the stellar bulge, stellar halo, and clusters of stars. Moreover, the preliminary results indicated 
above in Figure~\ref{fig:WDE_EDE} illustrate that there are multiple interesting emission regions of similar extent, separation, and brightness 
at low latitude. Finally, the {\it a priori} expectation at lower latitudes is that there are contributions of unknown size and extent from point sources 
which, with uncertain fidelity, follow the stellar distribution. Most point sources from the inner galaxy will be below the \textit{Fermi} detection 
threshold, which is highest in that direction \cite{Abdo:2010ru, Acero:2015hja}. Yet, how much of the total gamma-ray emission comes from 
detectable point sources depends on the exact luminosity distribution of the underlying population. All of these confounding factors make the 
low Galactic latitudes a rich environment for study, and the different characteristic angular scales of the emission components makes the wavelet 
decomposition a promising tool for investigating this region.

Millisecond pulsars (MSPs) may potentially play a role in the GCE. The exact assumptions regarding their luminosity distribution, spectra, and 
progenitor history have a strong impact on whether they are a viable option for the GCE \cite{Hooper:2013nhl, Calore:2014oga, Cholis:2014lta, 
Hooper:2015jlu, Hooper:2016rap}, or if they are indeed responsible for it \cite{Abazajian:2012pn, Yuan:2014rca, Petrovic:2014xra, 
Bartels:2015aea, Brandt:2015ula}. In the case where MSPs haven't been excluded, we expect to find many more such sources in the next years 
with \textit{Fermi} data or through observations at different energy \cite{Cholis:2014lta, Lee:2014mza, Yuan:2014yda, Calore:2015oya, Gaskins:2016cha}.
In a wavelet-based approach, one might expect that point sources will show up as additional power at small angular scales \cite{McDermott:2015ydv}.
 
For these reasons, the results in Figures~\ref{fig:gcerege2}, \ref{fig:gceregAllE}, and \ref{fig:GCE_w1w3_AllE} are relevant, since they give us a first 
indication of whether the observed GCE emission has power in the lowest angular scales. We point out that the comparison between diffuse and 
point-source emission is not so straightforward in region 0, because this region is only $4^{\circ} \times 4^{\circ}$ in size and the radial distribution of 
the gamma-ray sources may have different structure here than elsewhere.  In this region, even an inherently diffuse emission that peaks at the 
Galactic center, {\it e.g.}~GeV-scale annihilating dark matter, would predict significant power at the lowest wavelet scales, and conversely, inherently 
point-source-like emission would become so crowded that it would have substantial power above the lowest wavelet scales. Thus, results in the regions 
further from the Galactic center are a more interesting test of the contribution from small angular scales to the GCE.

Regions VII and VIII are the easiest to understand and compare to, since they are removed from the center, far from the Bubbles, and in these parts 
of the sky point sources from the Galactic disk are expected to be relatively most dominant. At 1.5 GeV and above, in these two regions we find that 
 $\sim$30--50\% of the total ($1 \leq j \leq 9$) emission is in the first two wavelet scales, and moreover the first two wavelet scales contribute 
 \emph{negatively}. There are 1.2 3FGL  point sources per deg$^{2}$ on average in these two windows. This is still higher than the average of 1.02 
 3FGL point sources per deg$^{2}$ along the two stripes of $2^{\circ}\leq |b| \leq 5^{\circ}$ extending at all longitudes: Regions VII and VIII are rich in 
 detected point sources. Only Regions II and VI have a similar $\sim 30 \%$ of their emission in the first two wavelet scales, which is also negative. 
 The magnitude and the sign of this small scale contribution is intriguing. The negative sign in the first two wavelet levels for the regions near the Galactic 
 center and Galactic disk means that unphysical flux has been imparted to the templates on small angular scales at intermediate angular distances from 
 the Galactic center. This is suggestive either of mismodelled bremsstrahlung and pion emission or the inclusion of spurious point sources near the 
 galactic center. We note that Region 0 does not suffer from a similarly large negative contribution at small angular scales. This may be an indication of 
 the large positive contribution from the GCE, or an issue with the procedure to determine the point-source  maps.

At higher energies, we find that region VI shows less power at the first two wavelet scales, while for region II the significance of the low scales remains 
relevant. We show our results at all energies in
 Appendix~\ref{app:15by15}. In region I we find that there is less power in small scales compared to region 
 II, but still more than what we find for regions III, IV, V, IX, and X (and at all energies). These regions along
 with VI show a robust excess emission that is diffuse in its nature, {\it i.e.}~contained on scales $w_j$ with $j \geq 3$. This is a significant
 result and together with the derived spectra of the GCE for these regions (given in Figure~\ref{fig:gceregSpectra})
 confirms results from template analyses in these latitudes \cite{Hooper:2013rwa, Huang:2013pda, Calore:2014xka}.
The GCE does indeed extend above 5$^{\circ}$ in latitude and can be observed up to $\sim$$20^{\circ}$. For
 latitudes between $5^{\circ}$ and $15^{\circ}$ there is a hard spectrum with a break at $\sim$5 GeV. Still, there may be an 
 underlying connection between the GCE and the \textit{Fermi} Bubbles, as discussed in \cite{Petrovic:2014uda, Cholis:2015dea}. Our 
analysis can not fully disentangle these two emissions but does find clear evidence of an excess diffuse gamma-ray 
emission in these latitudes, indicating that either the Bubbles extend down to $\sim$$5^{\circ}$, getting brighter at low 
latitudes, or that the GCE extends to higher than 5$^{\circ}$.

Interestingly, regions I and II, where point sources may contribute to the spectra at low energies, show no clear change from a smooth power-law.  
The combination of spectral and morphological results in Figures~\ref{fig:gcereg} through~\ref{fig:gceregSpectra} gives indications that a different type of gamma-ray emission 
mechanism may be operating, as compared to the emission at higher latitudes. Any such emission mechanism also dominates the point-source 
rich regions VII and VIII. It is our opinion that  these results show point source contribution at latitudes 
$|b| \leq 5^{\circ}$. We believe that this point alone is a matter worthy of a follow up analysis \cite{US2018spring}.

Our results are qualitatively unchanged upon using different point source data. 
Identifying a point source toward the Galactic center and accurately characterizing its spectrum is not a trivial 
task. There is strong diffuse emission from the Galactic disk as well as other point  sources that are at an angular 
separation below the point spread function of the instrument at low energies where, most photons are observed. 
Models of the diffuse disk emission have been shown to impact not only the spectra of the point sources 
but also their identification \cite{TheFermi-LAT:2015kwa}. The Fermi-Collaboration has produced different point 
source catalogs towards the inner galaxy \cite{Acero:2015hja, TheFermi-LAT:2015kwa, Fermi-LAT:2017yoi}. 
In our analysis we remove the point source emission using the spectra from the 3FGL catalog 
\cite{Acero:2015hja} for the Galactic sky at angles outside of a $15^{\circ} \times 15^{\circ}$ window centered at the 
Galactic center. For the inner galaxy window we use the 1FIG point source catalog \cite{TheFermi-LAT:2015kwa}. 
In Appendix~\ref{app:15by15} we show how the results for regions 0 and I-IV are affected by changing the point
source catalog that we use. As we include more flux from point sources, the total residual 
GCE emission decreases, with---as expected---the emission in the lowest two wavelet scales $j = 1, 2$ being  
affected the most. Removing the emission from more point sources impacts almost entirely the emission from the 
lower wavelet scales. Changing the point-source catalog in the inner $15^{\circ} \times 15^{\circ}$ window does
not affect our main conclusions regarding the diffuse nature of regions III and IV, which are affected by about 
$\sim 10 \%$ from our reference choices. Regions V, VI, and further out are entirely unaffected. For more 
discussion we direct the reader to Appendix~\ref{app:15by15}. 

Recent works \cite{Macias:2016nev, Bartels:2017vsx} have suggested that there are indications that the GCE is 
off the center of the Milky Way and has deviations from spherical symmetry. Earlier works \cite{Daylan:2014rsa, 
Calore:2014xka, Gordon:2013vta, Linden:2016rcf} had found the GCE to be spherical and only with weak statistical 
indication of being off-center. In this work we find that the amount by which the GCE is off center is mildly affected by 
the wavelet scales used. Using all scales we find the offset to be at $\ell = -4^{\circ}$ to $-6^{\circ}$, while if we omit the 
lowest two wavelet scales we find it to be at $\ell = -3^{\circ}$ to $-6^{\circ}$; for details, see Table~\ref{tab:OffCenter} 
or Figure~\ref{fig:WDE_EDE} and~\ref{fig:waveletgaussianfits}. Masking out the bright disk contribution within $|b| \leq 2^{\circ}$, 
we find the offset to be reduced to $\ell = -1^{\circ}$ for energies between 1 and 10 GeV. Above 10 GeV the offset 
remains at $\ell = -4^{\circ}$, $-5^{\circ}$, while below 1 GeV the offset becomes positive, $\ell = +3^{\circ}$. This strong 
dependence of the offset on the latitude cut imposed is suggestive of missing or mis-modeled point sources in that region. 

It has been recently suggested that the GCE is elongated along the disk \cite{Bartels:2017vsx}. We support claims that the 
GCE is elongated along the disk. Still, this emission is more spherical than the WDE and the EDE. 
Point sources along the disk at low energies or emission from the Bubbles at high latitudes
and high energies could potentially contaminate the GCE at the level seen here.

\begin{figure}[!ht]
\centering
\includegraphics[width=\linewidth]{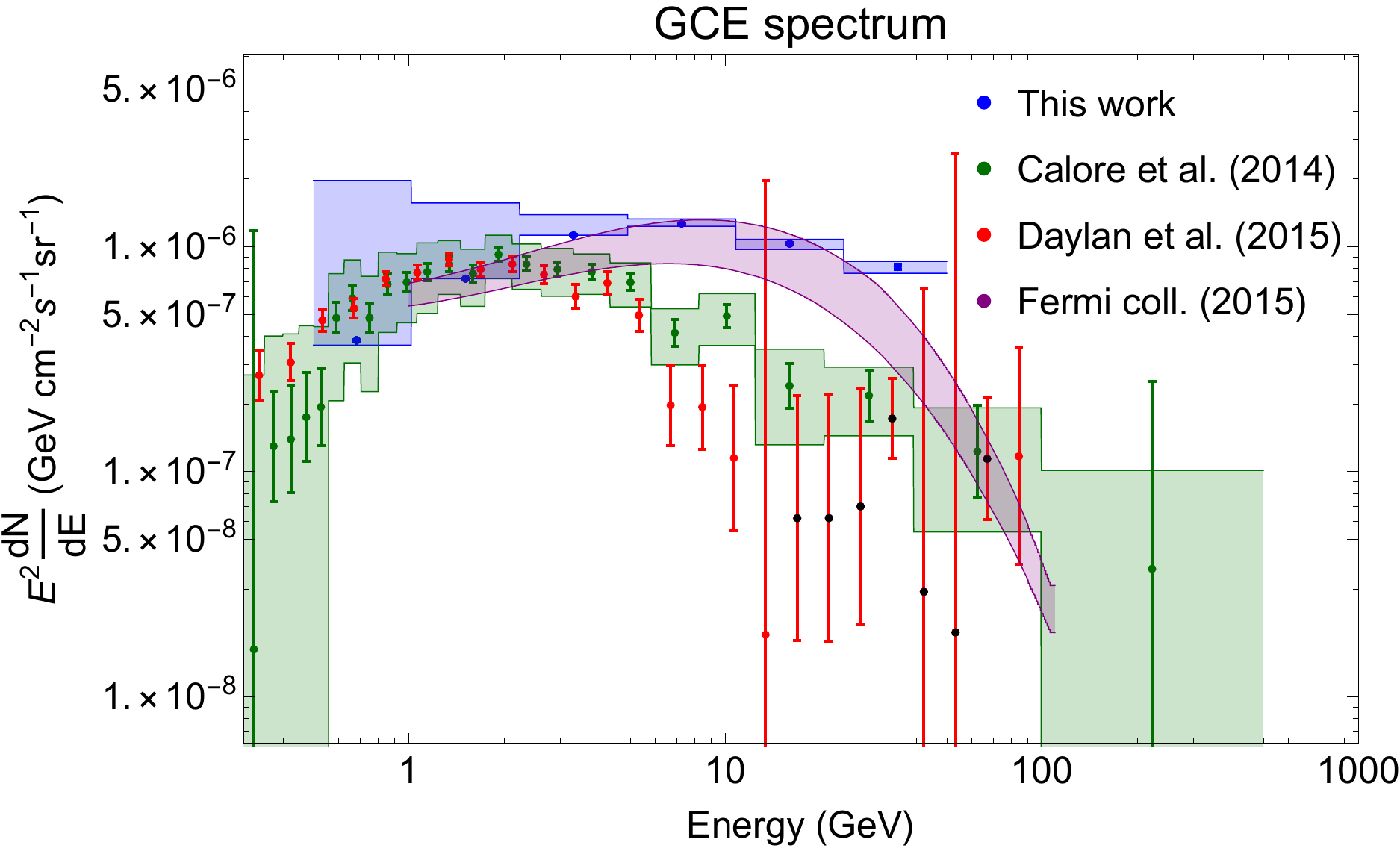}
\caption{The GCE spectrum, at $2^{\circ} \leq |b| \leq 20^{\circ}$, $|\ell| \leq 20^{\circ}$ from this work and its comparison 
to the template works of \cite{Daylan:2014rsa} and \cite{Calore:2014xka}. Since in those works the \textit{Fermi} Bubbles 
were removed from the GCE region emission, we subtract the Southern Bubble spectrum, properly weighting its 
coefficient based on the overlap between the Bubbles template and the region defined above.}
\label{fig:GCEComparison}
\end{figure}

Finally, in Figure~\ref{fig:GCEComparison}, we compare the energy spectrum of the GCE to the works of \cite{Calore:2014xka, 
TheFermi-LAT:2015kwa} in the same region and the extrapolated spectrum in the same region from the work of \cite{Daylan:2014rsa}. 
These works modeled the Bubbles as a separate template component, so we show the difference in our GCE spectra
in the window of $2^{\circ} \leq |b| \leq 20^{\circ}$, $\ell \leq 20^{\circ}$ from the Southern Bubble spectrum of Figure 
~\ref{fig:sbubblespece2w3}. Our spectrum agrees with \cite{Calore:2014xka, Daylan:2014rsa} within the quoted uncertainties
up to 5 GeV. Above 5 GeV we find a significantly harder spectrum, leading to a higher flux, closer to \cite{TheFermi-LAT:2015kwa} 
(see also \cite{TheFermi-LAT:2017vmf}). This may indicate either that the GCE is indeed brighter at these latitudes and energies 
than previous works have suggested, or we may be seeing contamination from the Bubbles, which might be brighter at 
$|b| \leq 20^{\circ}$ than further away from the disk.

\section{Discussion and Conclusions}
\label{sec:Conclusions}	
	
In this work we have used wavelets to analyze the \textit{Fermi} data from August of 2008 to November 2017 for energies from 0.48
to 52 GeV, following the techniques developed first in \cite{McDermott:2015ydv}.  We identify the \textit{Fermi} 
Bubbles and the Southern Cocoon while finding evidence for a Northern Cocoon that is $\sim$$30 \%$ dimmer than the 
southern one and aligned on the same axis. This axis is at an angle of $\sim$$30^{\circ}$ from the perpendicular to the  
Galactic disk.  The wavelet decomposition naturally separates emission power in different angular scales, and we use 
this to show that the Bubbles are diffuse in nature and not a collective effect of emission in small angular scales 
({\it e.g.},~from point sources or filaments). We also find the spectrum of the Bubbles in our results are in agreement with 
previous analyses. 

Focusing on the Galactic center, we find clear evidence for power up to $|b|$ of $20^{\circ}$, with the emission at 
$|b| \geq 5^{\circ}$ being clearly diffuse, {\it i.e.}~having little power in the smaller angular scales. Instead at $|b| 
\leq 5^{\circ}$ and all along the disk, the GCE has power in smaller angular scales.
This could indicate some point source contamination or mismodeling of the diffuse emission at small angular scales.
Still, more than $50 \%$ of that emission is at angular scales of $j \geq 3$, which translates to diffuse emission power 
on scales of $5^{\circ}$ and larger. Our results strongly indicate that there is a smooth transition between the \textit{Fermi}
Bubbles and the GCE or that one of the following statements hold: 
the GCE extends up to $20^{\circ}$ in $|b|$, or the Bubbles extend down to at least $5^{\circ}$ in $|b|$ and are 
brighter in those latitudes ($5^{\circ} \leq |b| \leq 20^{\circ}$) than at $|b| \geq 20^{\circ}$. Possibly, 
these two emissions are physically connected, as the result of CR outbursts either directly from the supermassive 
black hole or from the surrounding environment \cite{Petrovic:2014uda, Cholis:2015dea}.

We have broken the inner Galaxy into 11 subregions to see how the spectrum changes between different regions. 
We find the GCE spectra at low latitudes to be different from those of regions with $|b| > 5^{\circ}$. The model of point 
sources in the inner part of the Galaxy does not qualitatively affect our results. Yet, the fact that we find negative flux emission 
from the lowest scales in that part of the sky, indicates that we need a better understanding of these point sources 
before clearly deciding on the origin of the GCE emission. We find that the GCE is potentially offset by $\simeq -4^{\circ}$ in $\ell$, 
larger than what has been found by other approaches. Masking out the Galactic disk, the offset becomes $-1^{\circ}$ in $\ell$ at 
energies between 1 and 10 GeV, but is larger at lower and higher energies. This larger offset, and its sensitivity to masking, 
may again indicate contamination from uncertain point-source distributions.

Further out from the Galactic center, we identify two diffuse emission components at  $\ell \simeq  -20^{\circ}$ and 
$\ell \simeq  +25^{\circ}$. Each is significantly more elongated along the disk than the GCE. While physically the 
sources of these emissions are separated by a few kpc from the center of the Galaxy and arise from different underlying 
physics, their angular separation and their extent of about $20^{\circ}$ in $\ell$ may contaminate the GCE emission. 

Looking forward, we consider that further analysis of point sources in the inner Galaxy is necessary to better
understand the GCE, leaving this to future work \cite{US2018spring}. In this paper we have shown the insights that a
wavelet analysis of the \textit{Fermi} sky can provide, in terms of the morphologies of gamma-ray emissions. 
This technique can be used on observations from instruments like \textit{CALET} \cite{2013arXiv1311.4084Y}, 
\textit{DAMPE} \cite{TheDAMPE:2017dtc}, Gamma 400 \cite{Moiseev:2013vfa}, e-ASTROGAM \cite{DeAngelis:2017gra}
at different energies and with different angular resolution. Wavelets work best for observations where the instrument has 
a large field of view, good angular resolution, and good sensitivity. The first two characteristics are important to ensure a 
large dynamic range in wavelet scales, including both small and large angular scale emission. Instrument sensitivity 
ensures a large number of photons, necessary for extending this technique to higher energy. Wavelet analyses provide 
morphological information with less bias, as long as the data set contains sufficient photons. Templates on the other hand 
are efficient even with relatively low statistics, and thus provide valuable information on spectral studies, where one can 
split the data into many energy bins. Gamma-ray analyses of future observations should include the use of wavelets along 
with templates to combine the strengths of both techniques.

\bigskip                  
                  
\textit{Acknowledgements:} We thank Dalya Baron, Dan Hooper, Marc Kamionkowski, Tim Linden, Brice Menard, Simona Murgia, and 
Christoph Weniger for helpful discussions. This work makes use of {\tt HEALPix} \cite{Gorski:2004by} and {\tt healpy}\footnote{https://github.com/healpy/healpy}  and was supported by NASA Grants No NNX15AB18G, NNX17AK38G, the Simons Foundation, the DoE under contract number DE- SC0007859 and Fermilab, operated by Fermi Research Alliance, LLC under contract number DE- AC02-07CH11359 with the United States Department of Energy.
	
\bibliography{Wavelet_Analysis}
\bibliographystyle{apsrev}
	
\begin{appendix}
\section{Point Sources}
\label{app:AuxEff}
	
\subsection{Identifying and correcting for point source J1709.7-4429}
\label{sec:bright-PS}
	
\begin{figure}[!ht]
\centering
\includegraphics[width=0.49\linewidth]{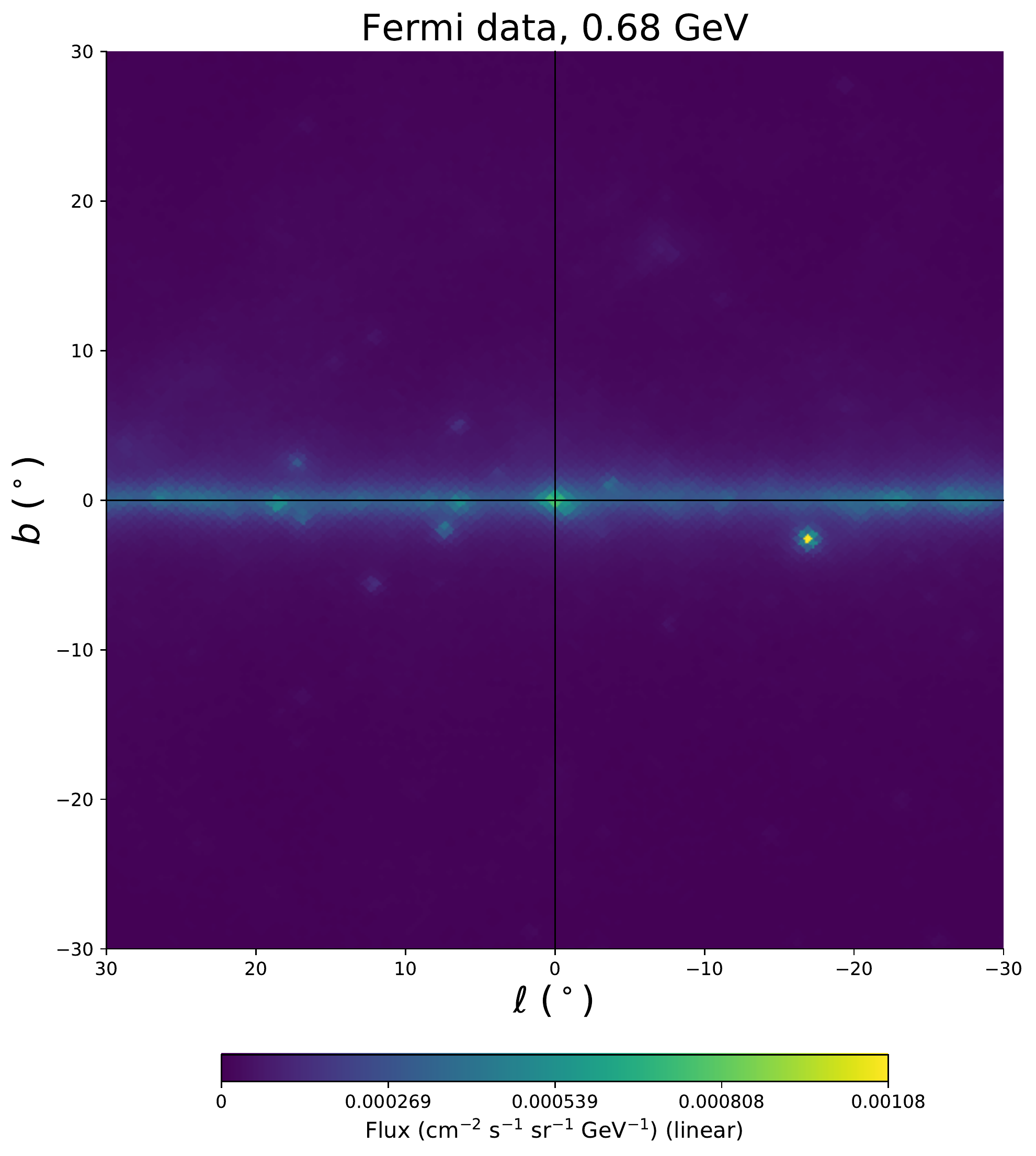}
\includegraphics[width=0.49\linewidth]{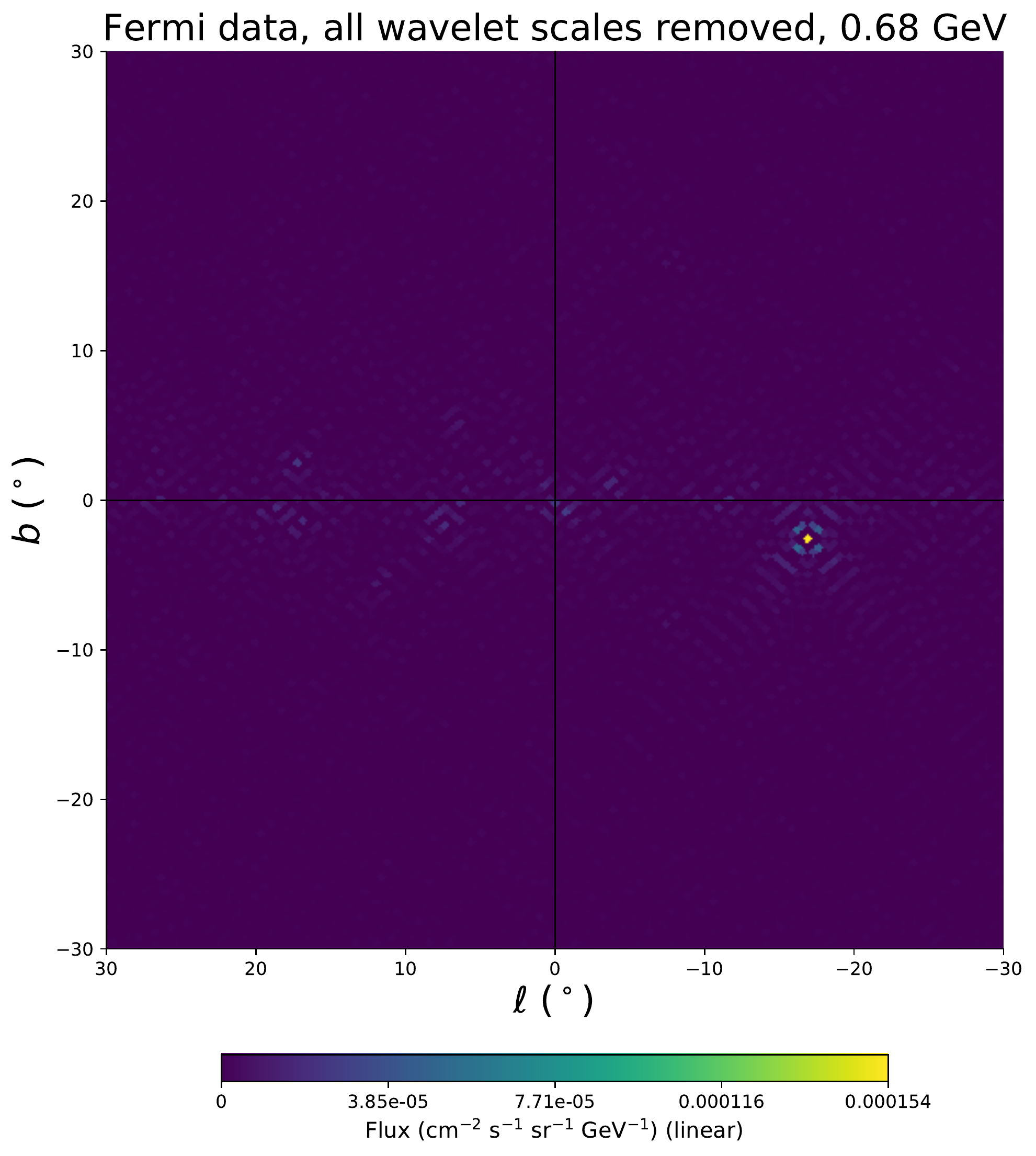}
\includegraphics[width=0.49\linewidth]{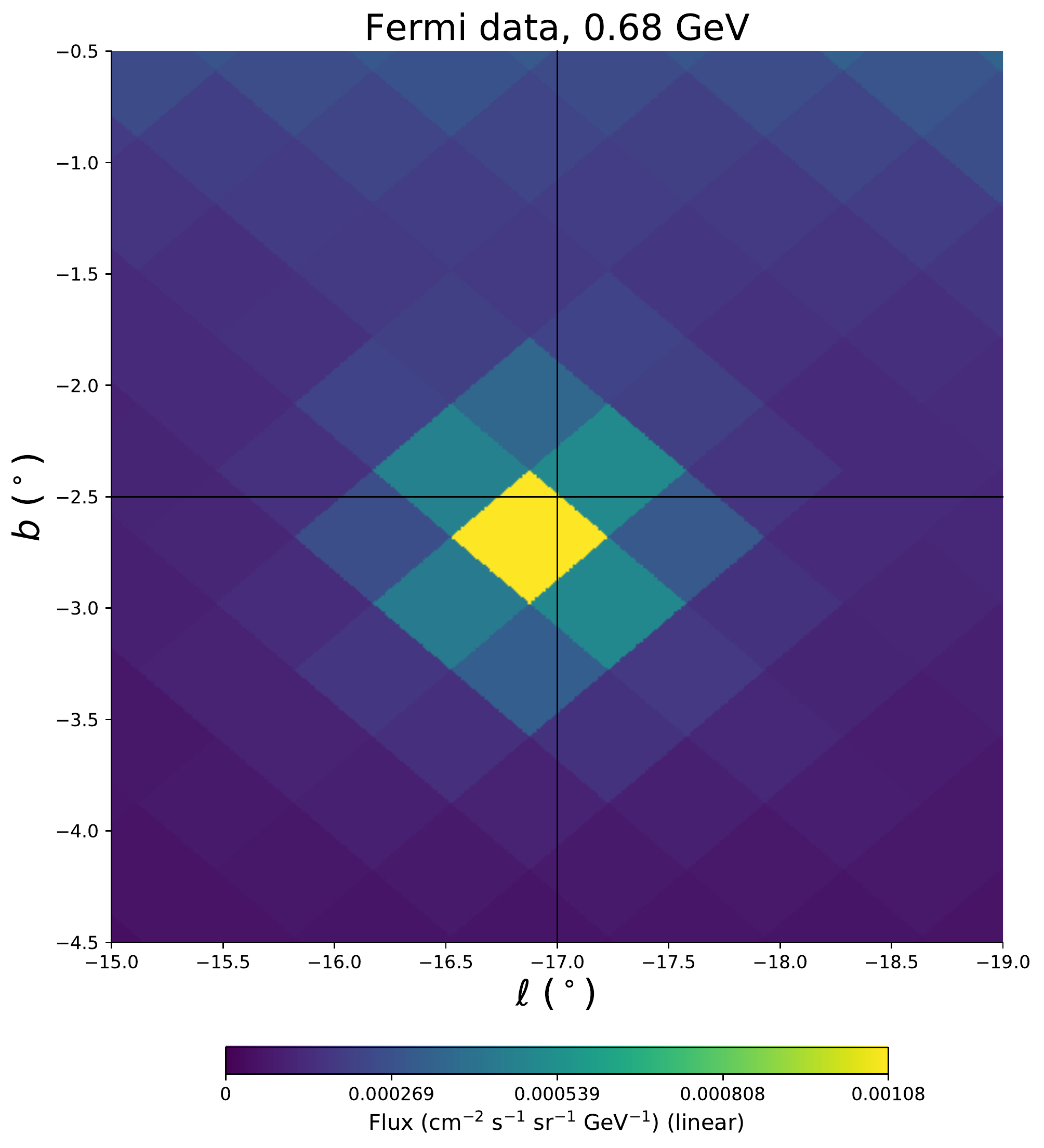}
\includegraphics[width=0.49\linewidth]{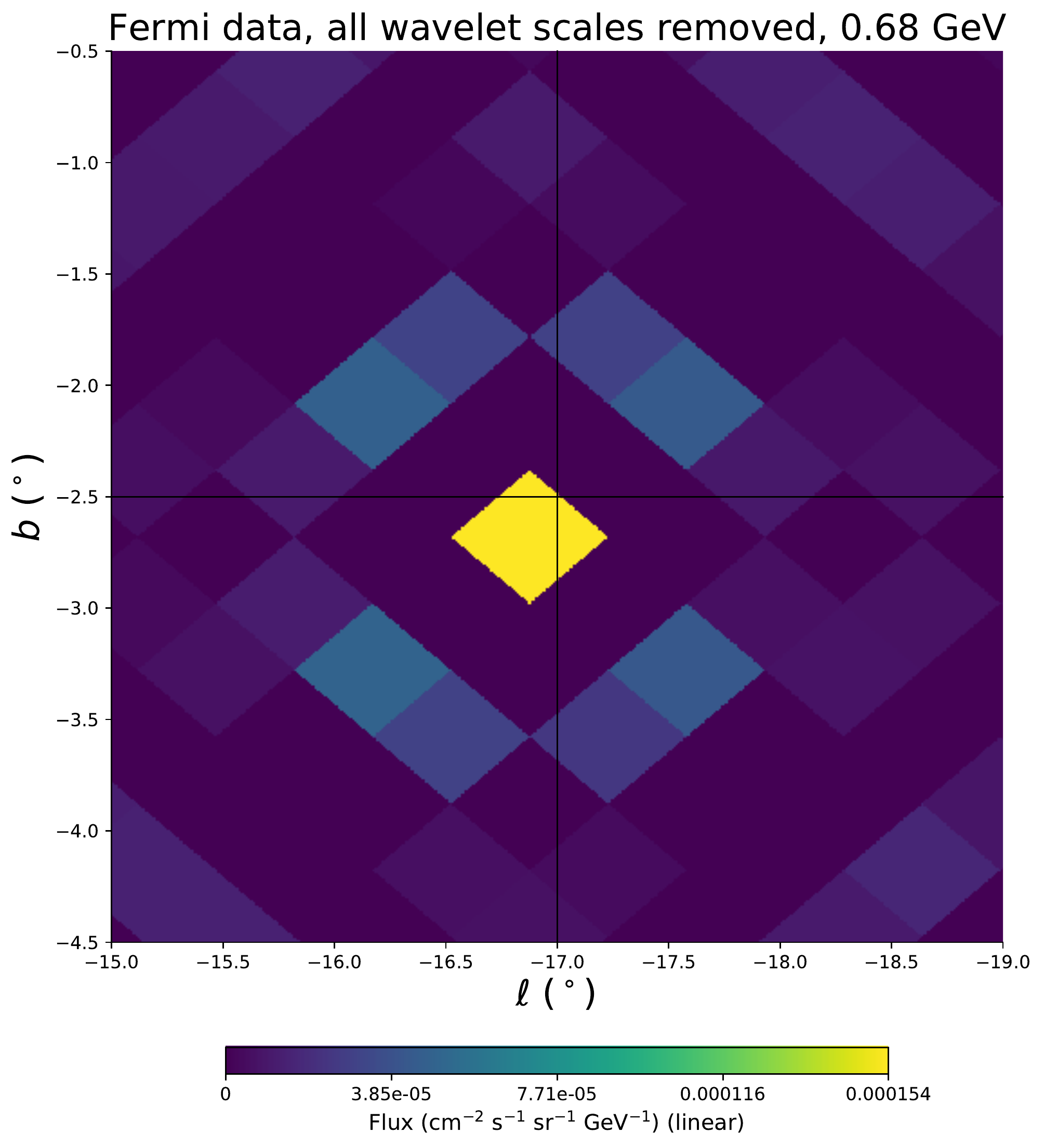}
\caption{Four maps demonstrating the use of wavelets to identify or confirm point sources. 
The first 
map is that of the data, in the region of  $|\ell|, |b| < 30^\circ$. 
The second is the data map, with all 9 wavelet levels (and the monopole) subtracted off. The location 
of  J1709.7-4429 point source becomes obvious. The second pair of maps matches the first but zooms 
in to with $2^\circ$ around J1709.7-4429. In the bottom right residual map, showing remaining flux after retuning the parameters describing the point-source spectrum, $85\%$ of the of the emission of the point source has
been removed.}
\label{ptsrc}
\end{figure}

Using the method described in Section~\ref{sec:method}, we identified a strong point source at 
$(\ell, b) = (-17^\circ, -2.6^\circ)$, corresponding to J1709.7-4429 of the 3FGL catalog \cite{Acero:2015hja}. Due to this large flux in our residual maps we concluded that this point source was under-modeled.  To correct for this we adjusted the parameters (spectral index, cut-off, and normalisation) describing the spectrum of that source in a trial-and-error process, ultimately modeling out $\sim$$85 \%$ of its residual emission.  This process is summarized in Figure~\ref{ptsrc}.
	
\subsection{The Impact of Point Sources in the Inner $15^{\circ} \times 15^{\circ}$ Galactic Sky}	
\label{app:15by15}	

\begin{figure*}
\begin{centering}
\includegraphics[width=2.3in,angle=0]{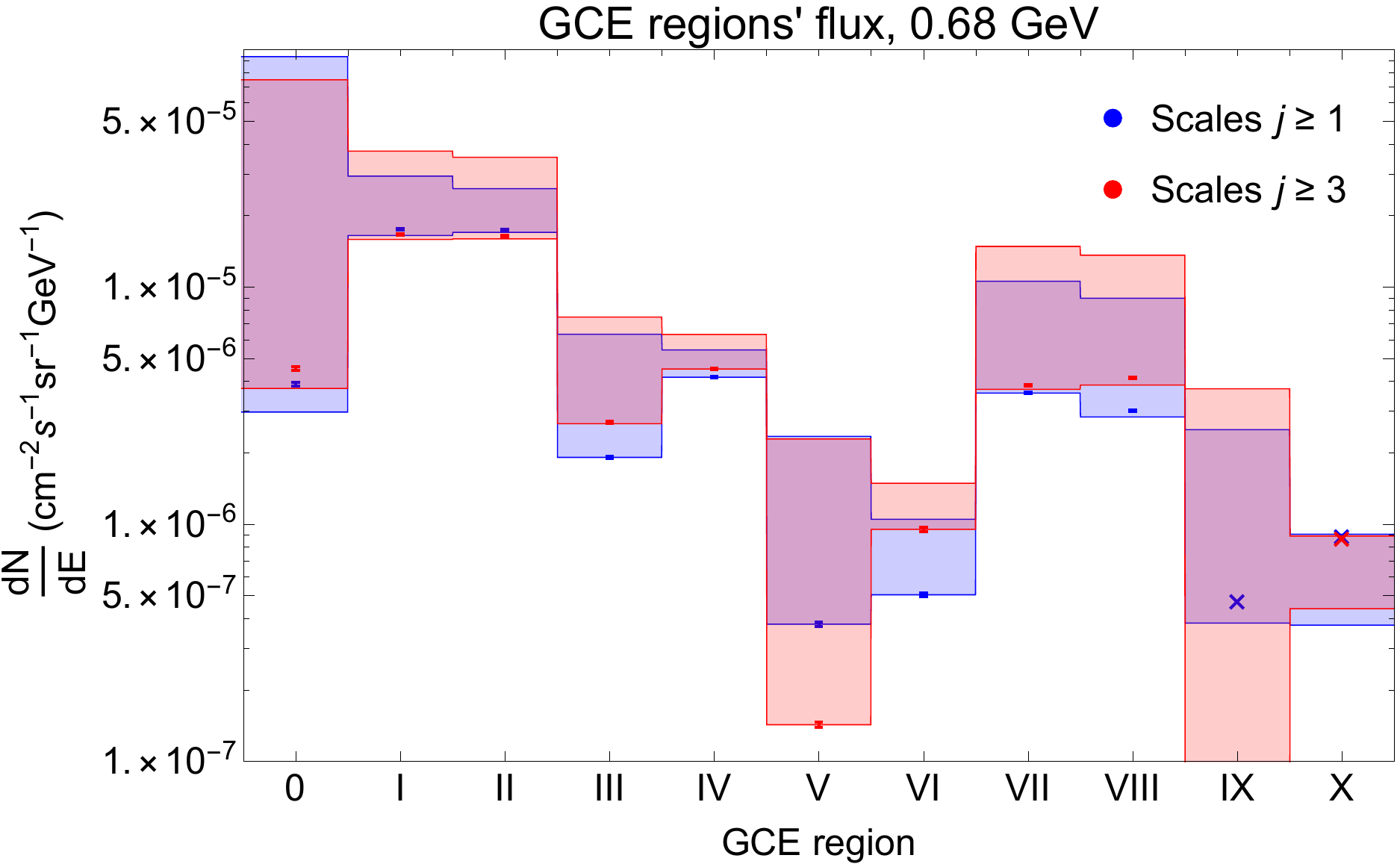}
\includegraphics[width=2.3in,angle=0]{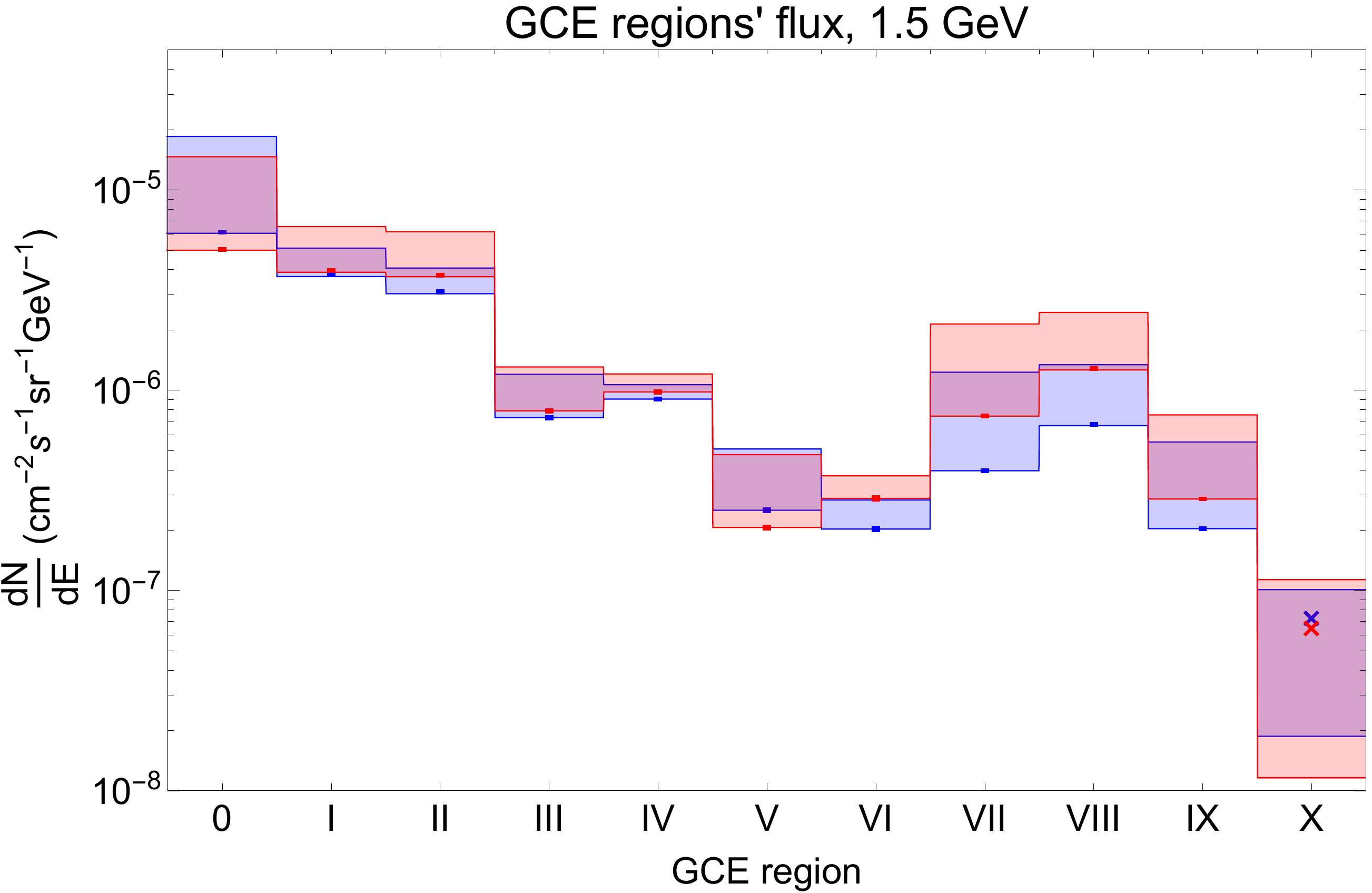}
\includegraphics[width=2.3in,angle=0]{plotsWA/GCEregFlux_w1w3_E3.pdf} \\
\includegraphics[width=2.3in,angle=0]{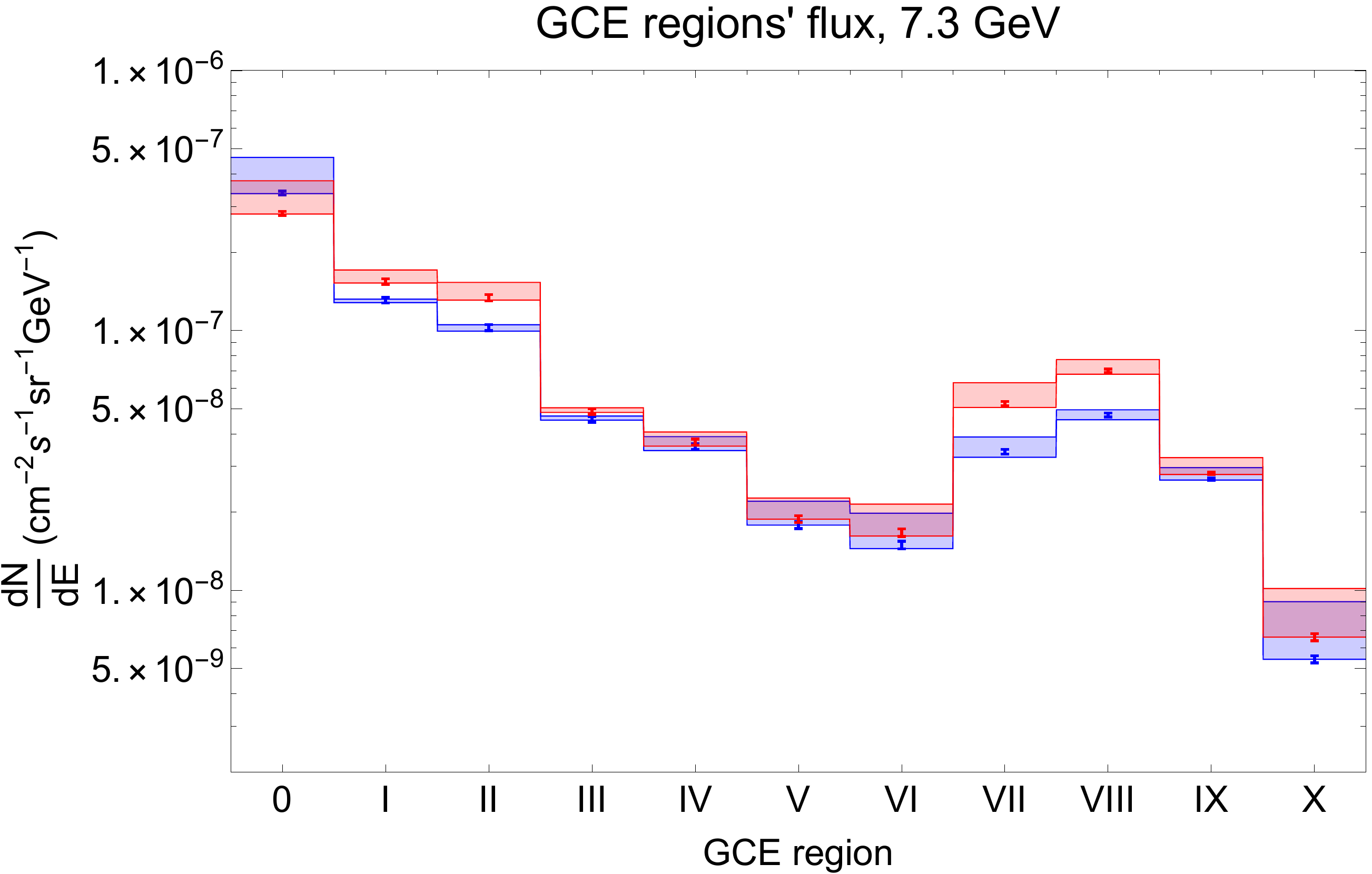}
\includegraphics[width=2.3in,angle=0]{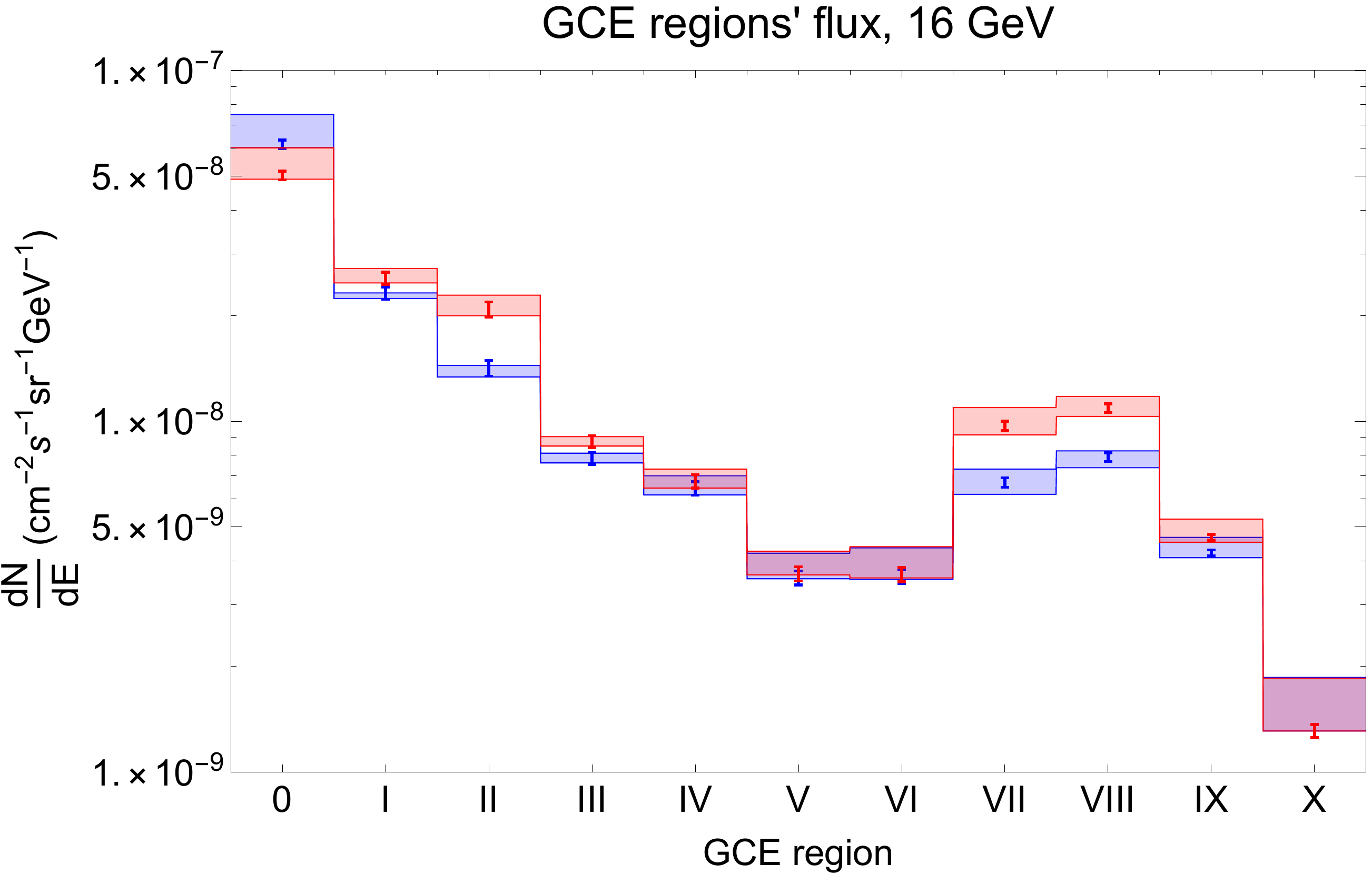}
\includegraphics[width=2.3in,angle=0]{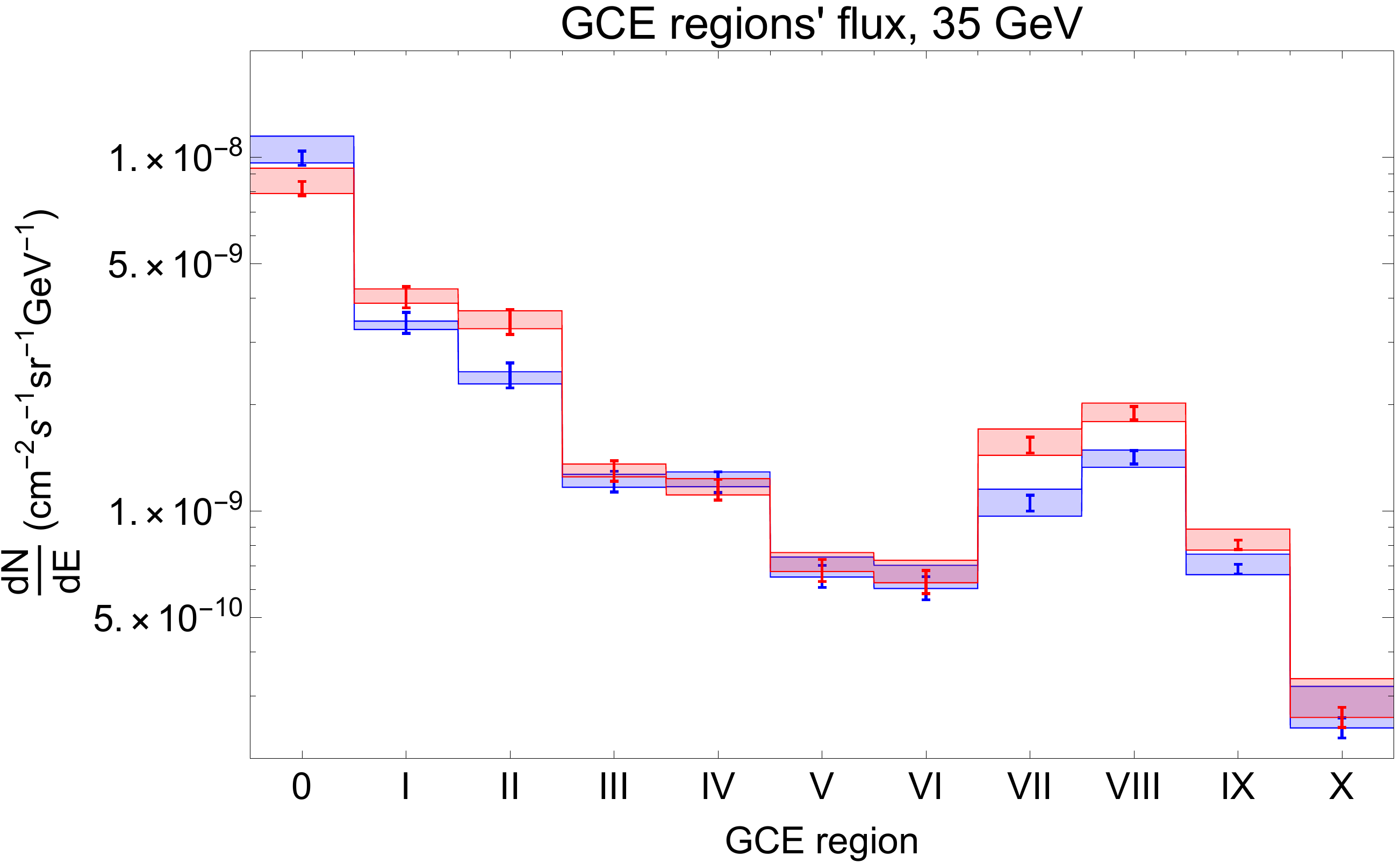}
\end{centering}
\vspace{-0.2cm}
\caption{The flux around the Galactic center for the regions given in Figure~\ref{fig:gceregs} and at each of our energy bins. Fluxes  include power from all scales, $j \geq 1$ (blue), or from only $j \geq 3$ (red). At the lowest energies the 
systematics are large. At energies 3.3 GeV and above we can see the contrast between 
small-scale and large-scale emission. Small-scale emission 
is most important in regions 0, II, VII, and VIII, while regions III, IV, V, and IX are more diffuse. Region I is 
intermediate, and region VI at the higher three energy bins is clearly as diffuse as its 
mirror to the Galactic disk, region V. Compare to Figure~\ref{fig:gcerege2} and Figure~\ref{fig:gceregAllE}.}
\vspace{-0.3cm}
\label{fig:GCE_w1w3_AllE}
\end{figure*}

In Section~\ref{subsec:GCE} we discussed how negative flux in low wavelet scales in the inner Galaxy points towards mismodelling of small scale structure in these regions.  In particular, we showed in Figure~\ref{fig:gcerege2} the GCE emission from all regions at 3.3 GeV both from all scales or from scales $j \geq 3$. Regions I, II, VII, and VIII all showed significant negative power at small angular scales.  We now present the equivalent of Figure~\ref{fig:gcerege2} for all energy bins, in Figure~\ref{fig:GCE_w1w3_AllE}.  This information was contained in Figure~\ref{fig:gcereg} (all scales) and Figure~\ref{fig:gceregAllE}, but we present it again in Figure~\ref{fig:GCE_w1w3_AllE} for ease of comparison.
	
\begin{figure*}
\begin{centering}
\includegraphics[width=2.3in,angle=0]{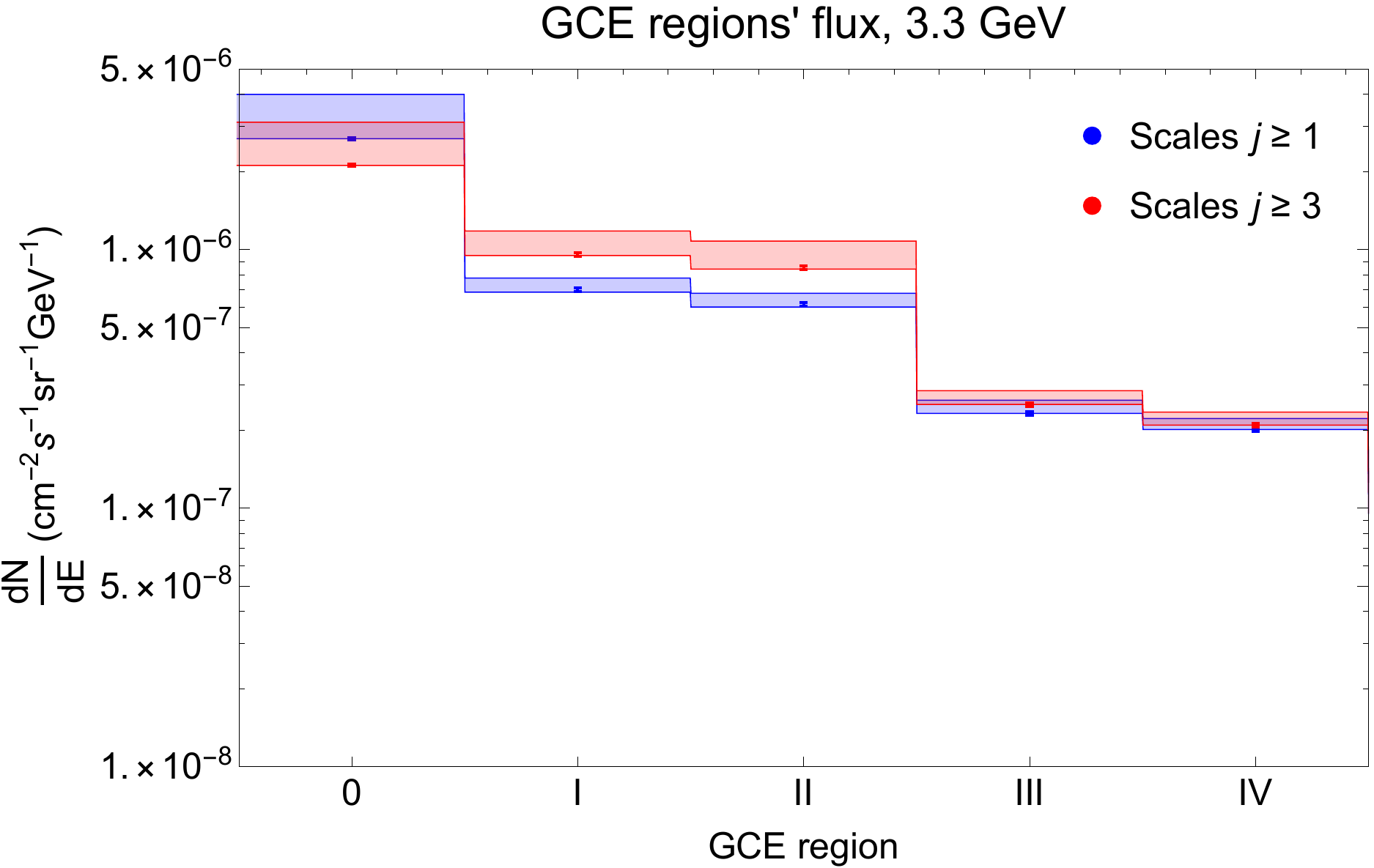}
\includegraphics[width=2.3in,angle=0]{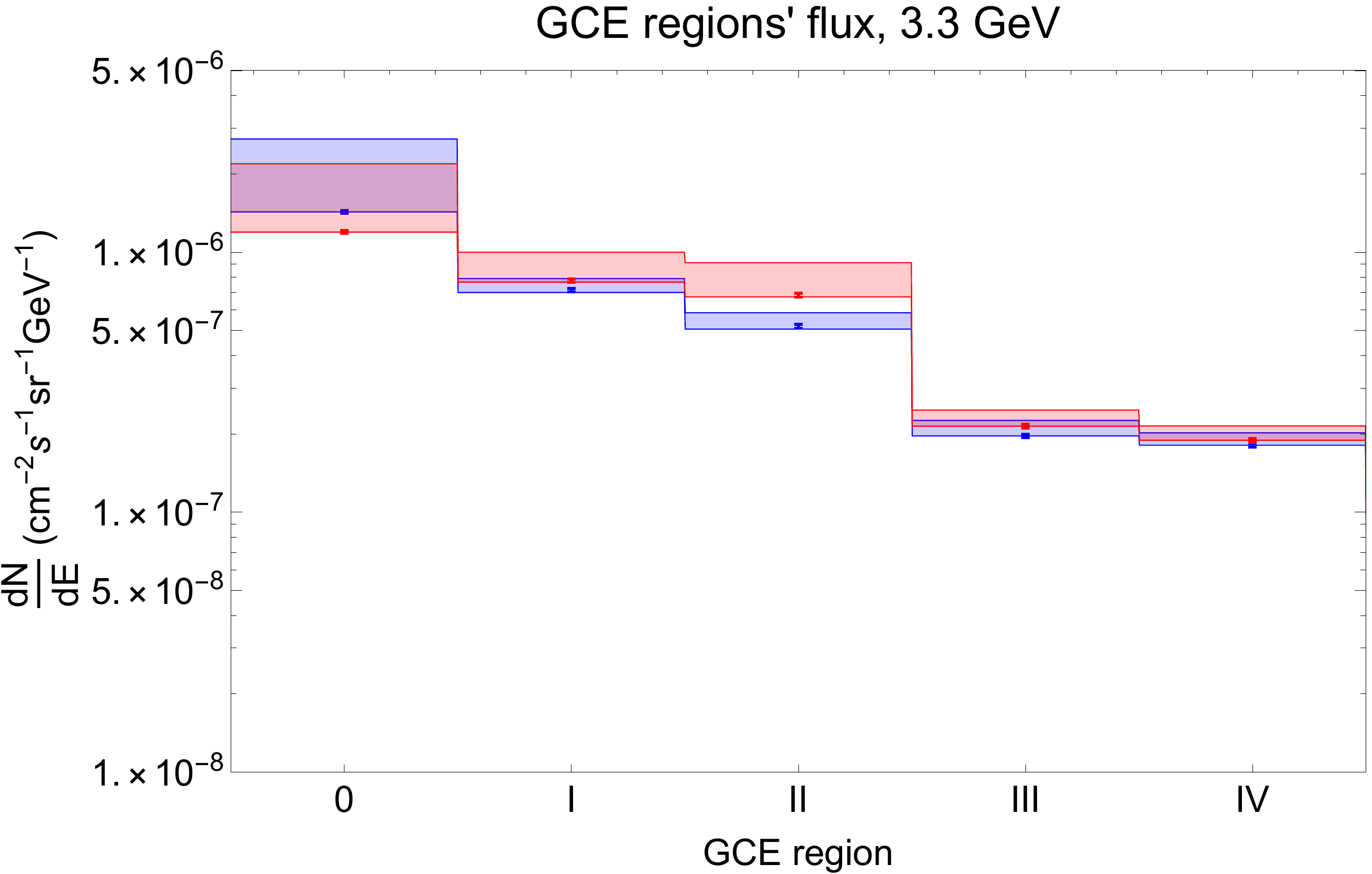}
\includegraphics[width=2.3in,angle=0]{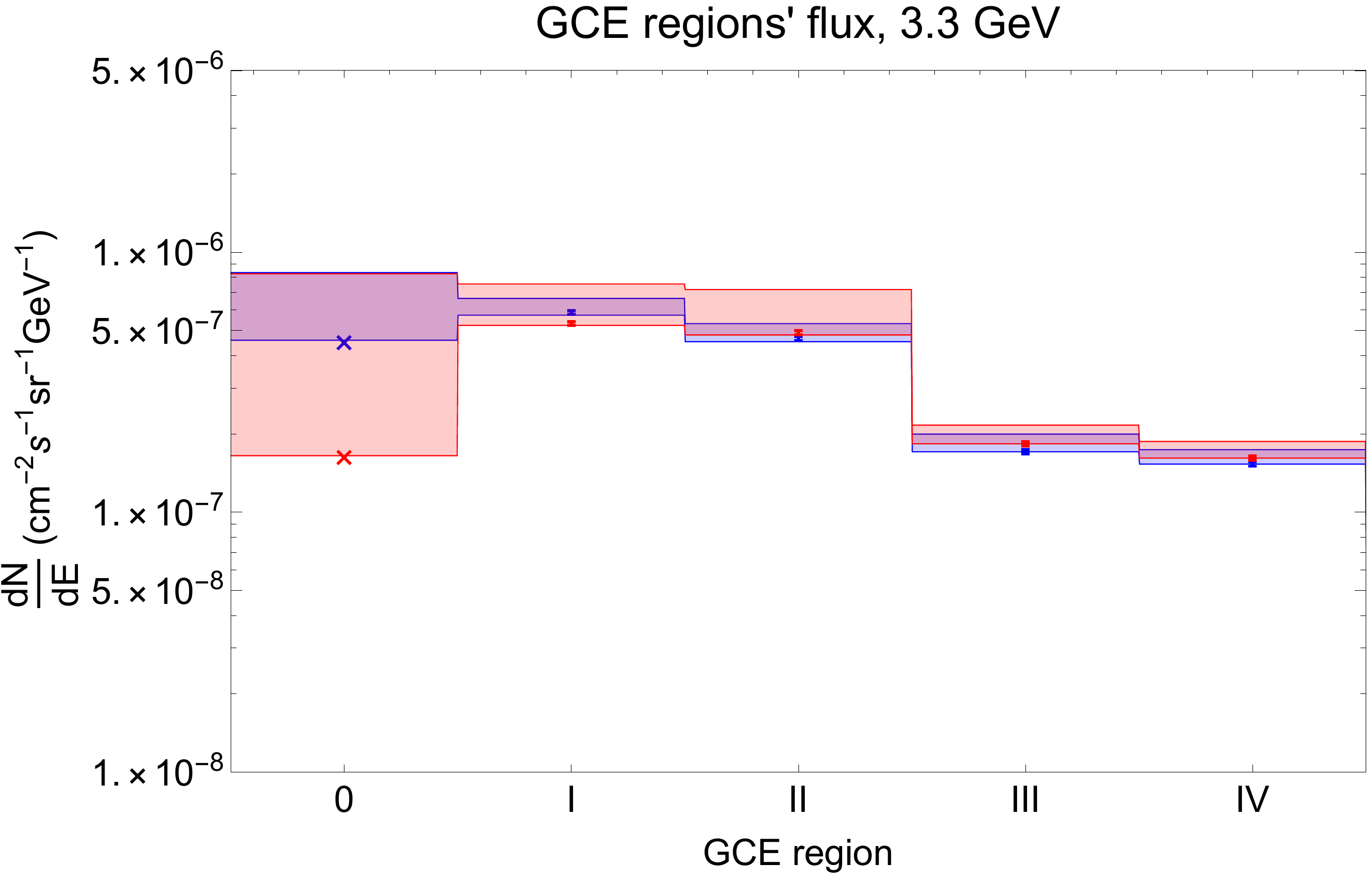}\\
\includegraphics[width=2.3in,angle=0]{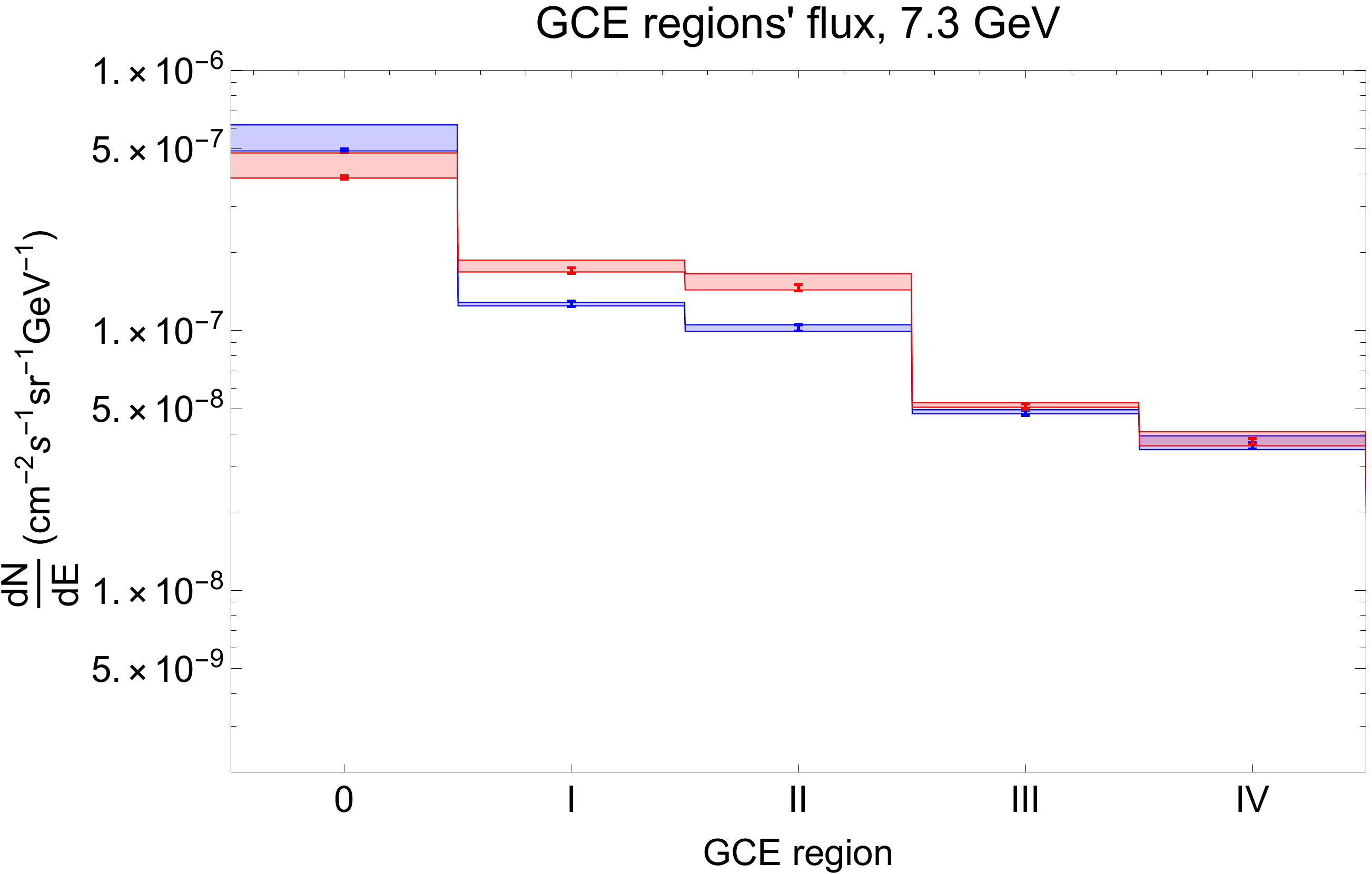}
\includegraphics[width=2.3in,angle=0]{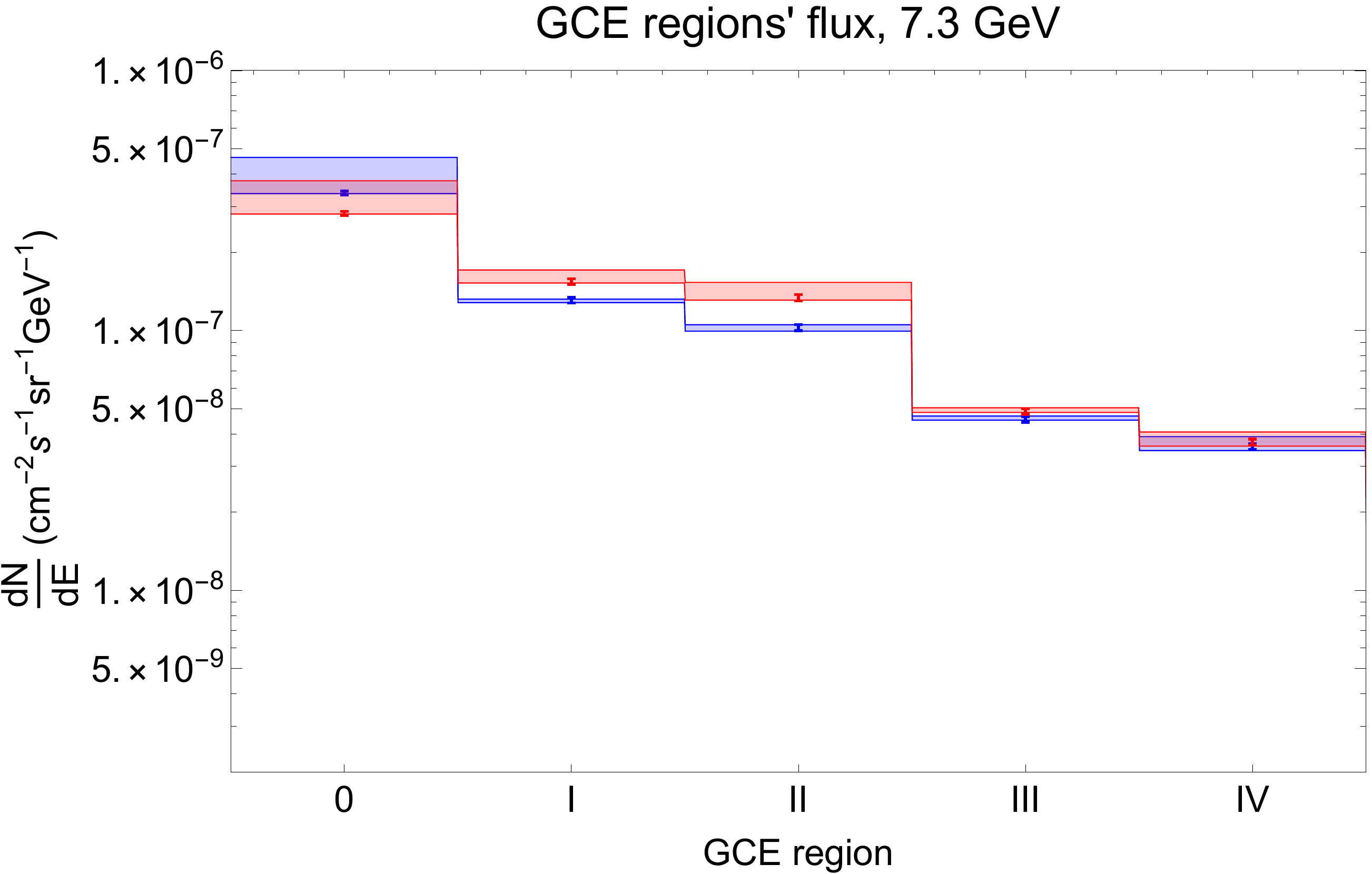}
\includegraphics[width=2.3in,angle=0]{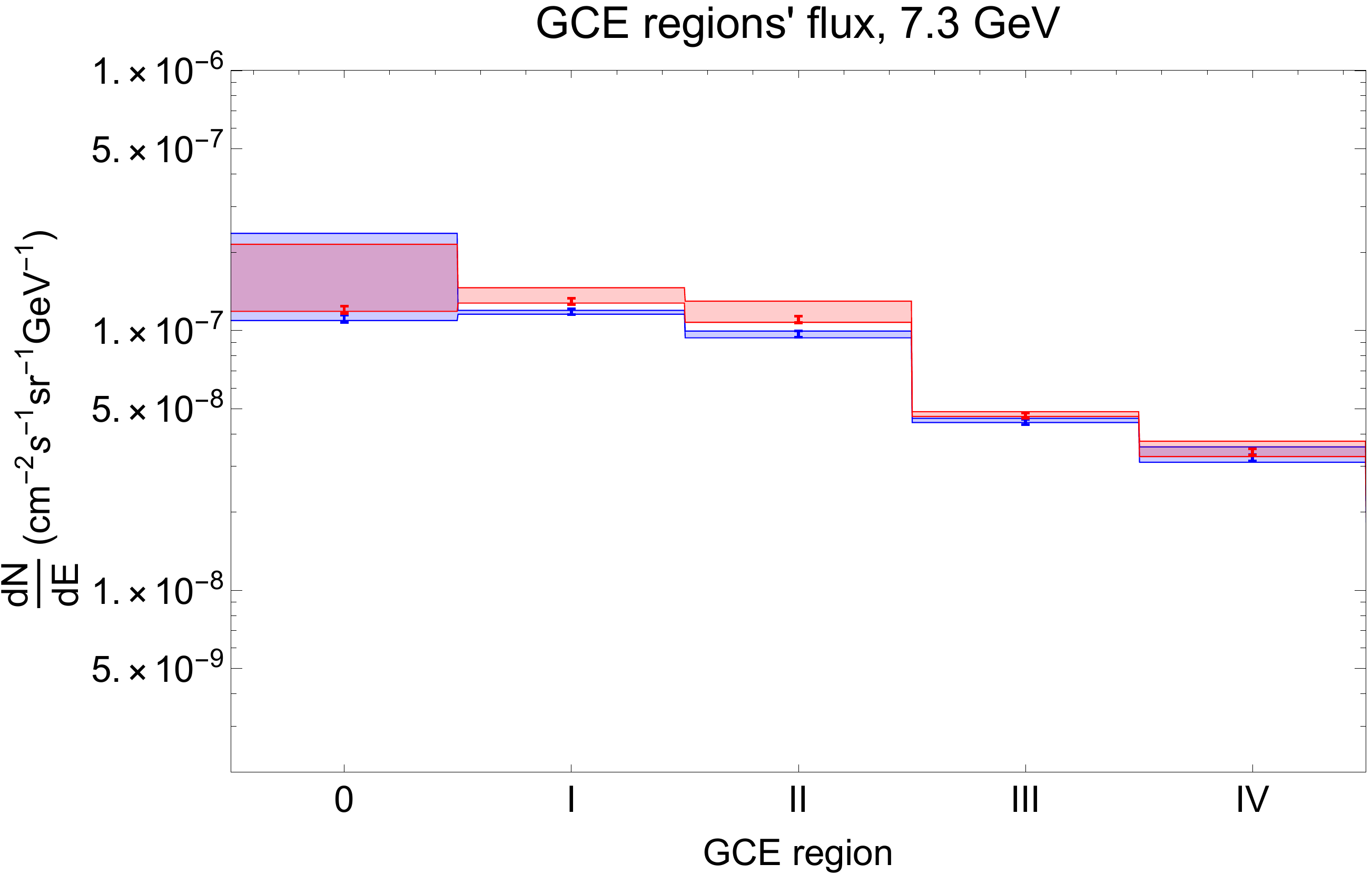}\\
\includegraphics[width=2.3in,angle=0]{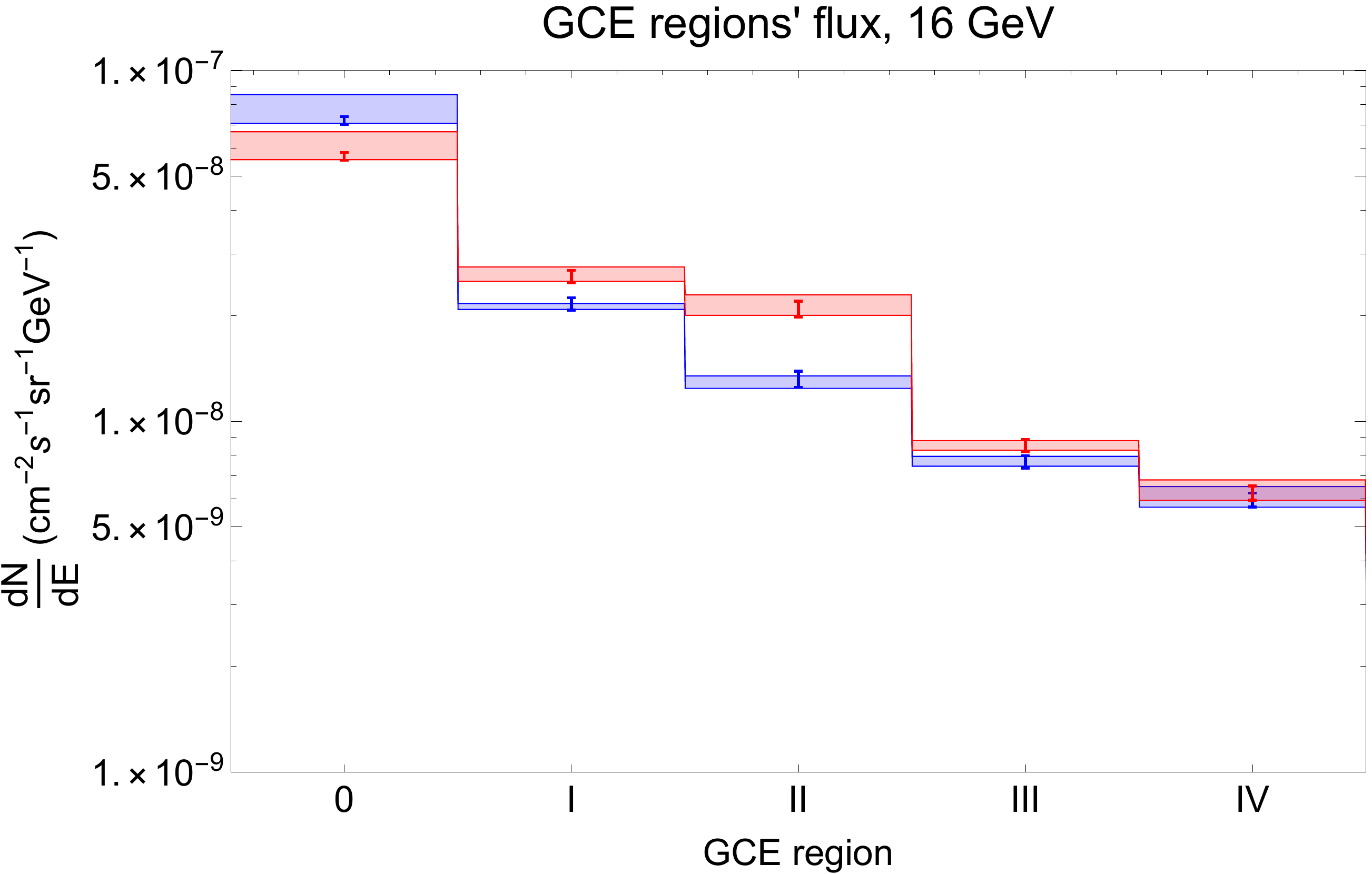}
\includegraphics[width=2.3in,angle=0]{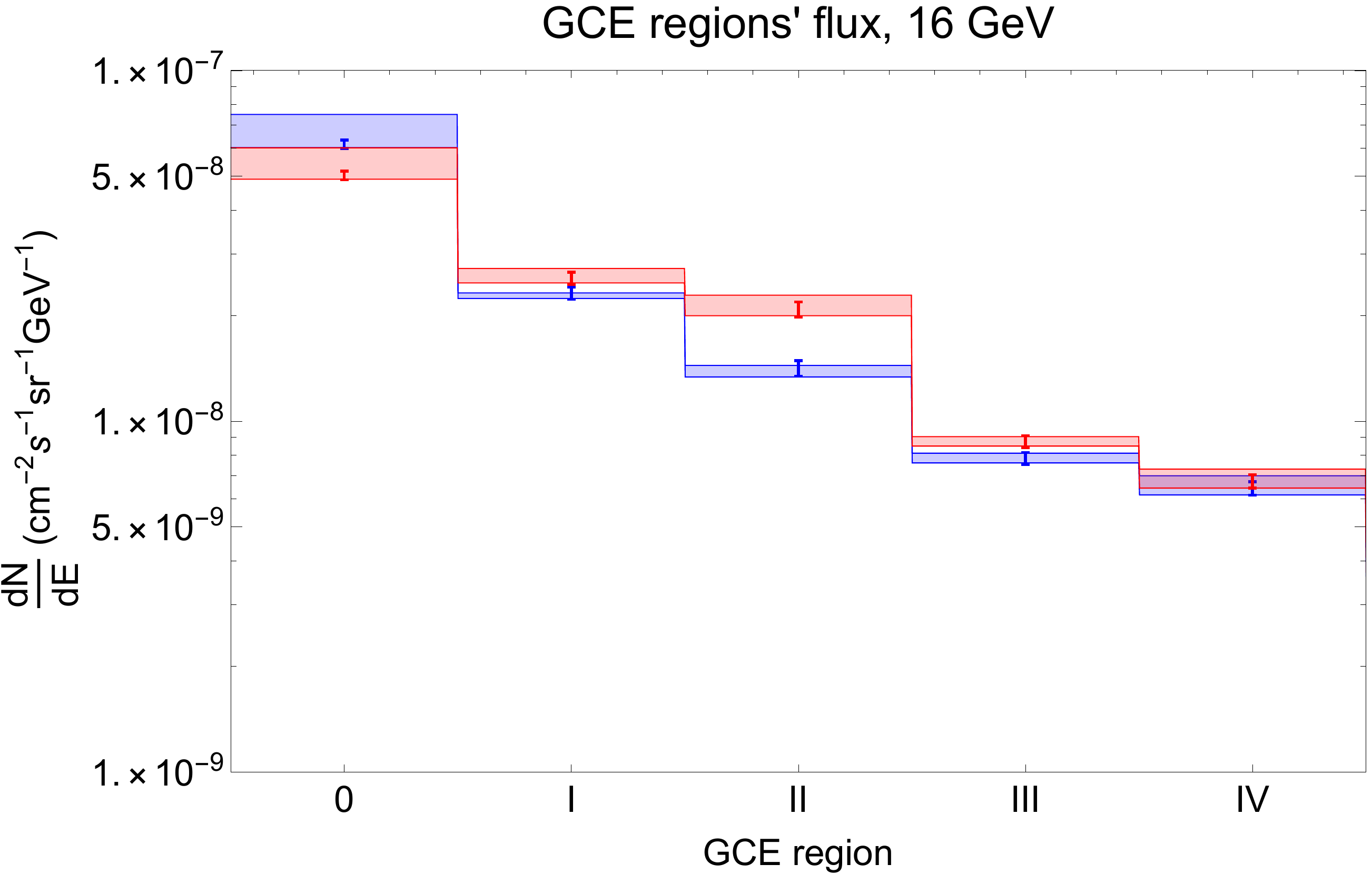}
\includegraphics[width=2.3in,angle=0]{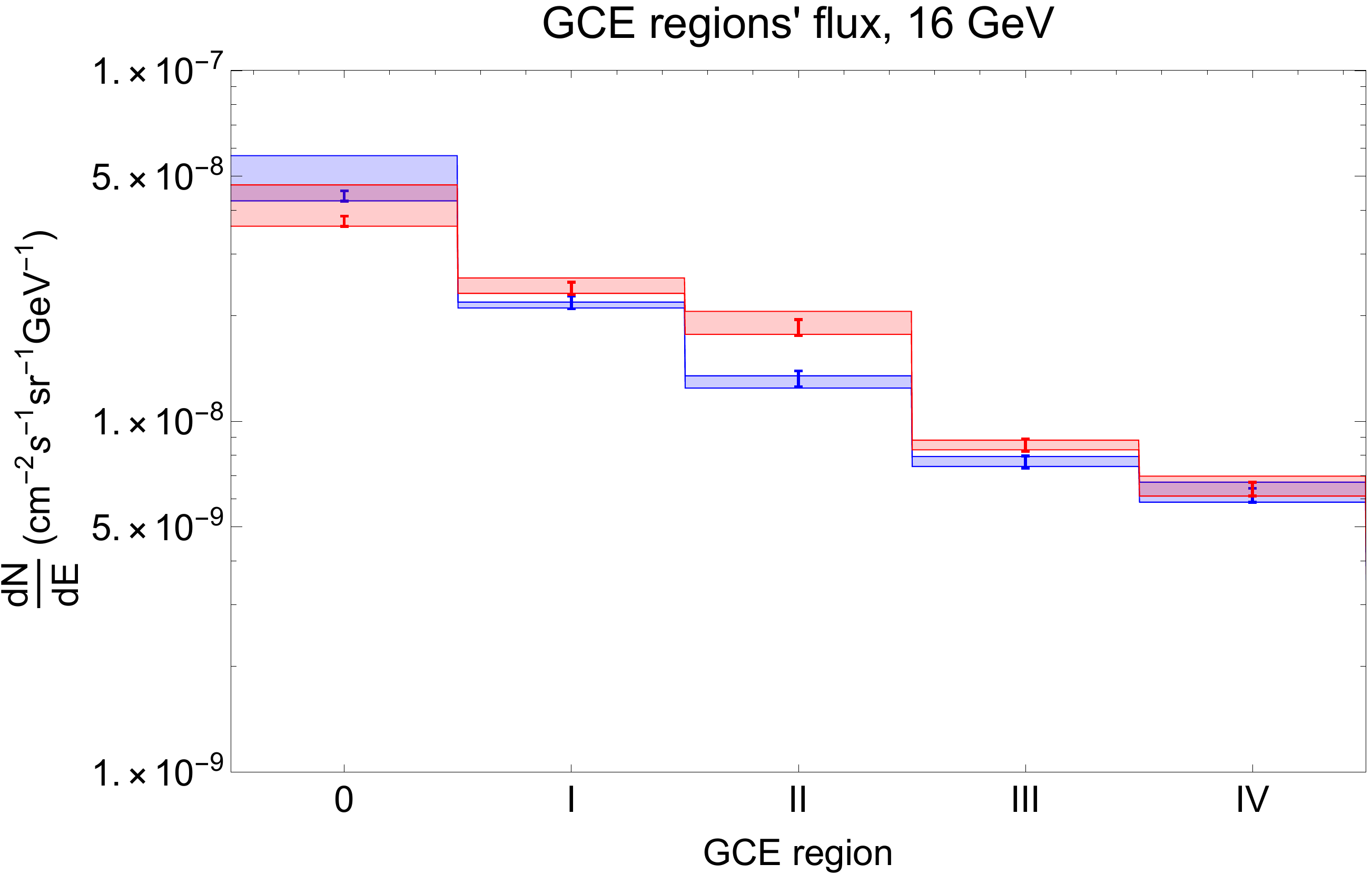}
\end{centering}
\vspace{-0.2cm}
\caption{Similar to Figures~\ref{fig:gcerege2}, \ref{fig:gceregAllE} and~\ref{fig:GCE_w1w3_AllE}, changing the selection of point sources that we model out.
We show subset of the regions given in Figure~\ref{fig:gceregs} for the energy bins at 3.3 (\textit{top row}), 7.3 (\textit{middle row}), and 
16 GeV (\textit{bottom row}). Fluxes include power from all scales, $j \geq 1$ or include only $j \geq 3$. 
In the \textit{left column} we include 3FGL 
point sources  only. In the \textit{middle column} which is our reference choice, we replace the 3FGL point sources with 
the 1FIG point sources in the inner $15^{\circ} \times 15^{\circ}$ box. In the \textit{right column} we include both the 
3FGL and the 1FIG point sources (see text for details). As we move from left to right we increase the flux from point sources. 
(Regions V and beyond are not shown: they are unaffected by these choices since the 1FIG catalog extends out to $7.5^{\circ}$.) The fluxes in region 0 are sensitive to the point sources chosen 
and can drop by up to $\sim$$80\%$. Regions I and II are somewhat sensitive as well, with a maximum flux reduction of $\sim$$40\%$. For regions III and IV the point source selection can affect the GCE fluxes by $\sim$$20\%$ at 3.3 GeV and 
$\lsim$$10\%$ at higher energies.}
\vspace{-0.3cm}
\label{fig:GCE_w1w3_AllE_PS}
\end{figure*}

We also study the impact on the GCE emission from changing the catalog of point sources that 
we model out from the inner $15^{\circ} \times 15^{\circ}$ window. As shown in Figure~\ref{fig:gceregs},
only regions 0 and I-IV can be affected by these choices. In Figure~\ref{fig:GCE_w1w3_AllE_PS}, we 
show the GCE emission in these regions for three different choices of point-source catalog that 
we use and for three different energies; 3.3, 7.3, and 16 GeV. Moving from left to right we increase 
the modeled flux from point sources. In the left column we use only the 3FGL point sources as done
in every other part of the sky. In the middle column we replaced the 3FGL with the 1FIG sources. Going from 3FGL to 1FIG exchanges 21 new 1FIG point 
sources for 38 3FGL sources (which are typically dimmer). In the third column we show our results when we use both 
3FGL and 1FIG point sources (for the 27 sources that appear in both catalogs we use the 
1FIG information).
	
Changing the point sources in the inner $15^{\circ} \times 15^{\circ}$ window affects the results of region 0 by a factor of 2--3,
while for regions III and IV the effect is only 10--20\%, in agreement with our basic conclusion that these regions are diffuse in 
nature.
	
\section{Additional Combined Fit Results for GCE, WDE, and EDE}
\label{app:moregaussianfits}	

\begin{figure*}[ht]
\centering
\includegraphics[width=0.49\linewidth]{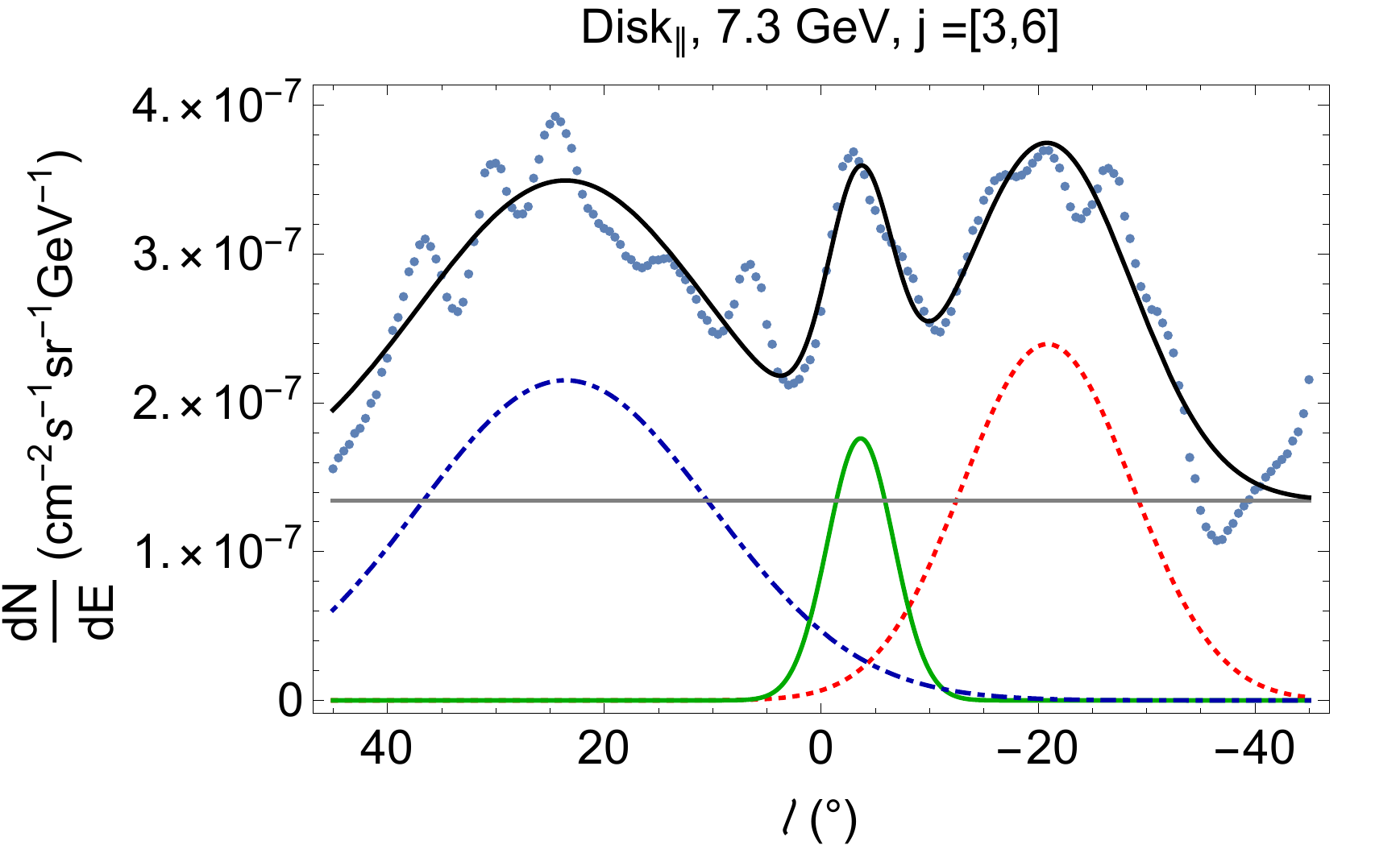}
\includegraphics[width=0.49\linewidth]{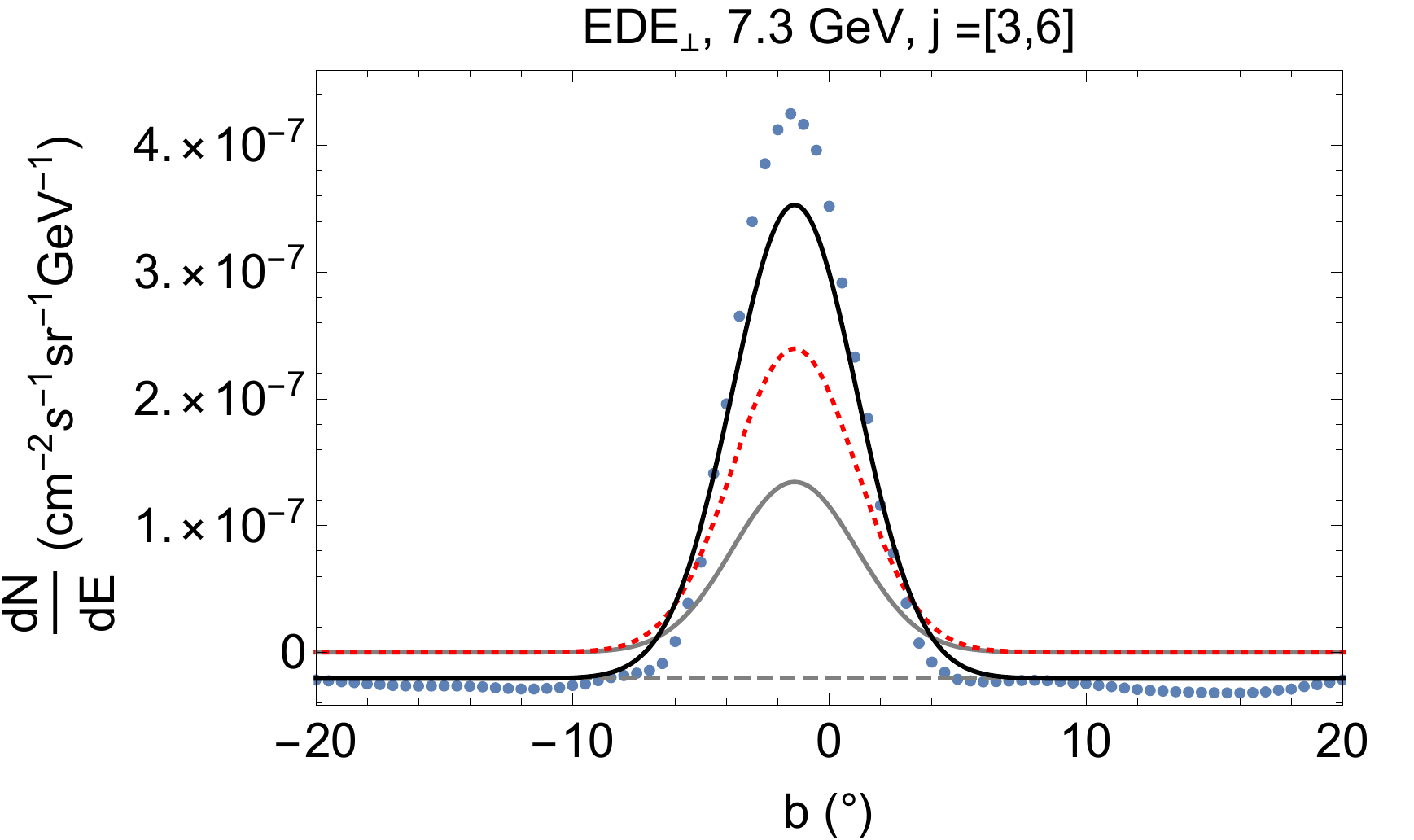}\\
\includegraphics[width=0.49\linewidth]{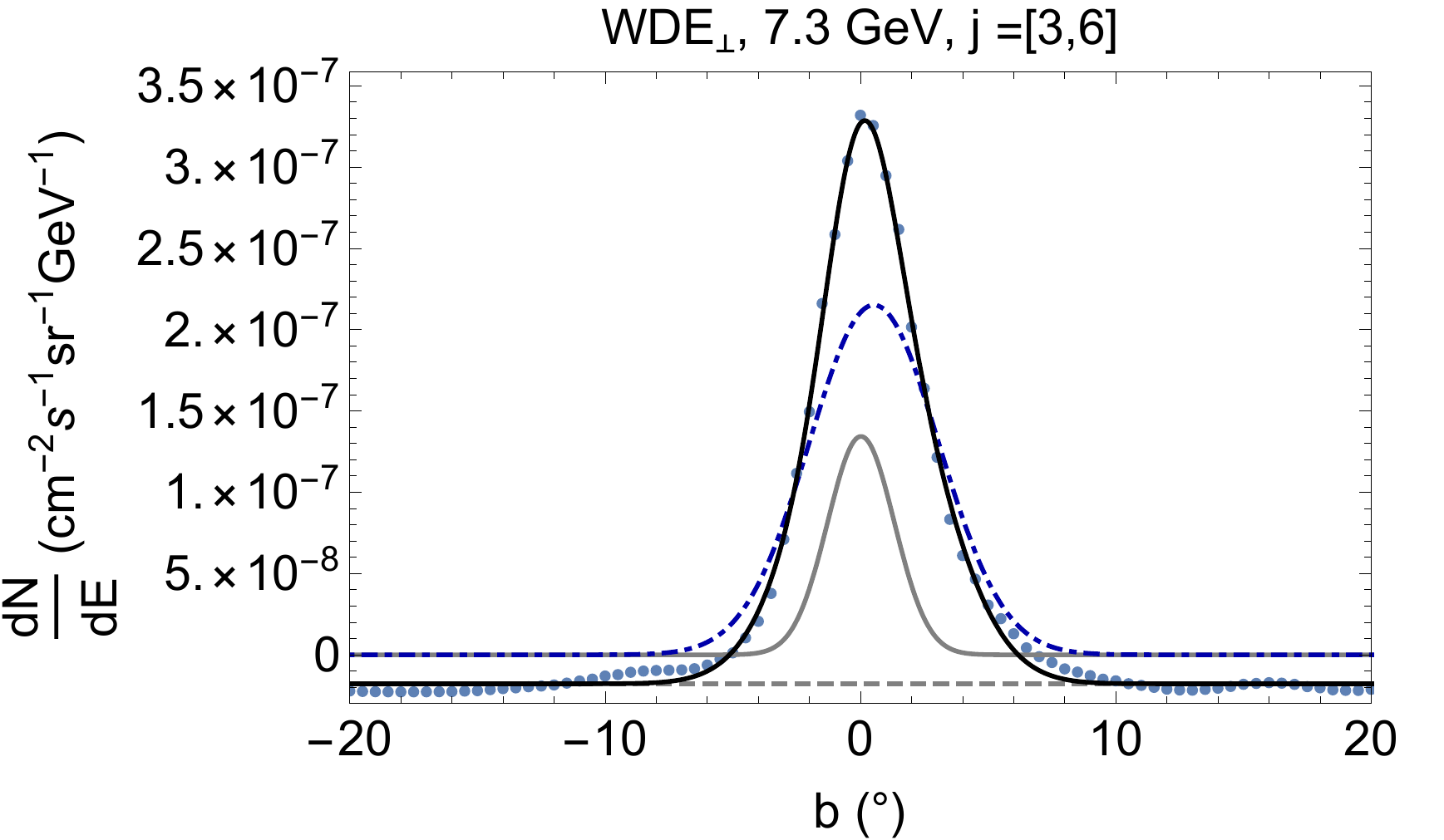}
\includegraphics[width=0.49\linewidth]{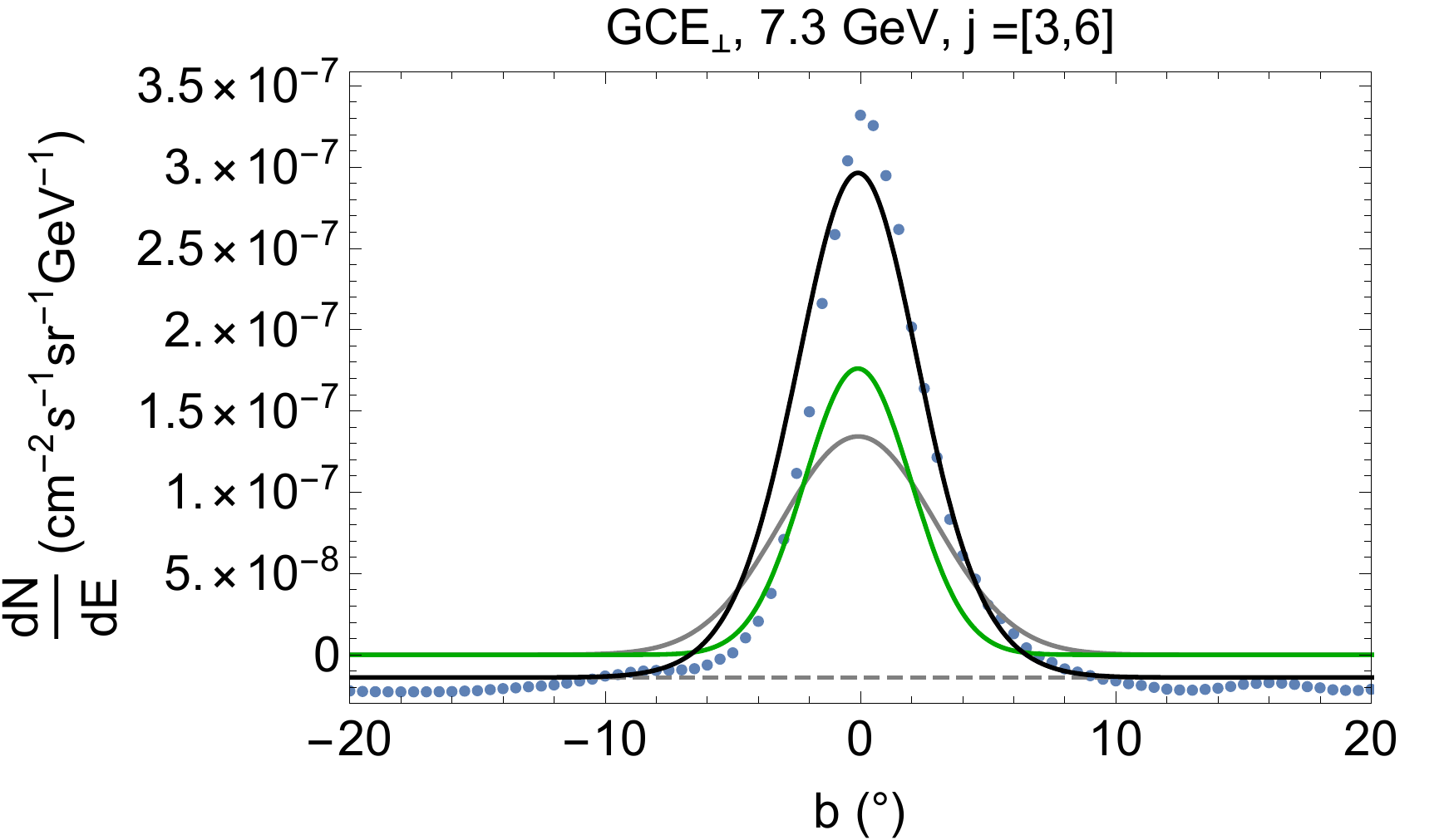}
\caption{Same as Figure~\ref{fig:waveletgaussianfits} but for the fourth energy bin.}
\label{fig:waveletgaussianfits2}
\end{figure*}

\begin{figure*}[ht]
\centering
\includegraphics[width=0.49\linewidth]{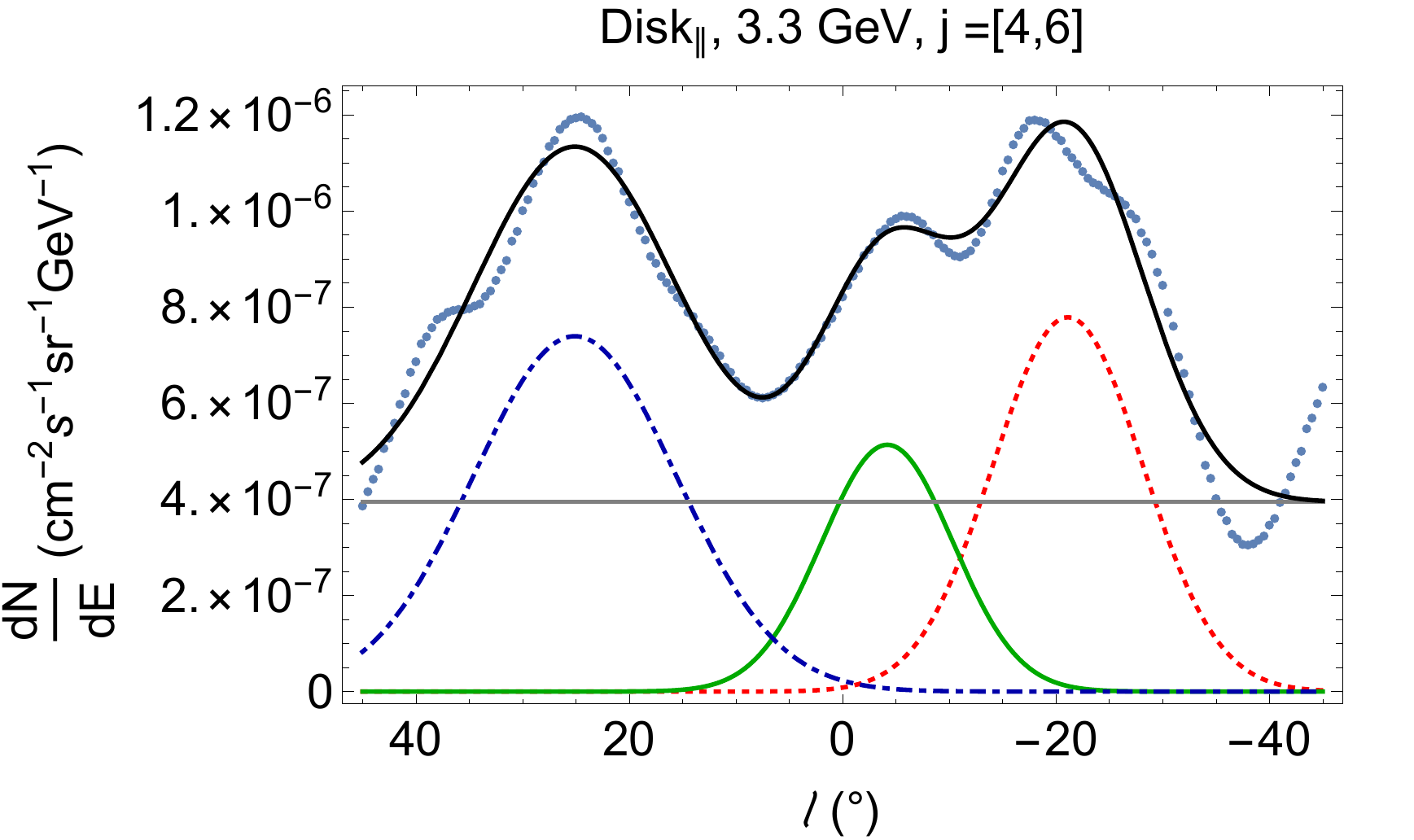}
\includegraphics[width=0.49\linewidth]{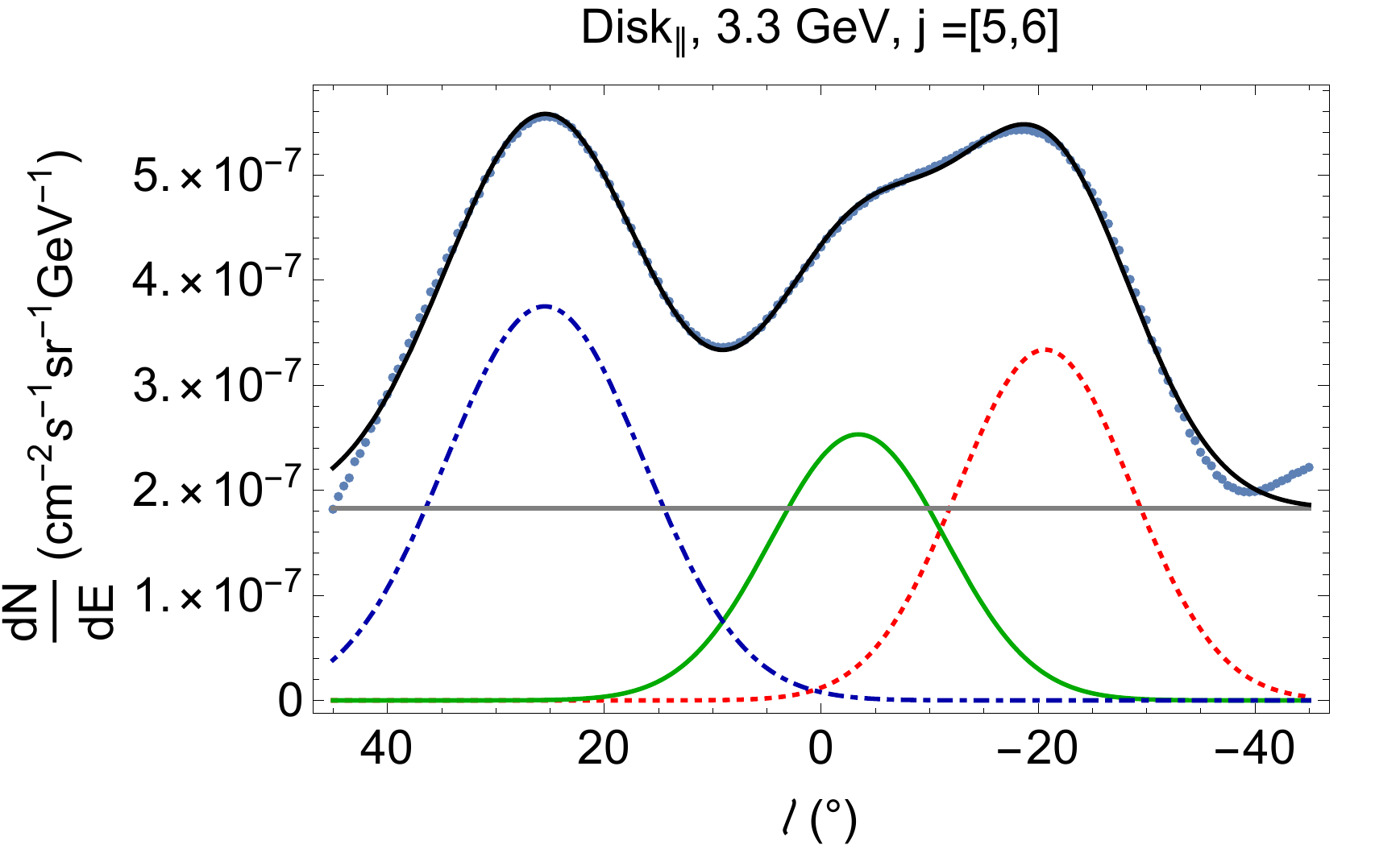}
\caption{Same as the top left of Figure~\ref{fig:waveletgaussianfits} but omitting wavelet levels 3 and 4.}
\label{fig:waveletgaussianfits3}
\end{figure*}

In Figure~\ref{fig:waveletgaussianfits2} we present a fit for the excess emission components that lie along the Galactic disk, following the procedure discussed in Section~\ref{subsec:GCE} for the energy bin centered at 7.3 GeV. These fits reveal structures of essentially the same location and extent as those in the energy bin at 3.3 GeV, which were presented in Figure~\ref{fig:waveletgaussianfits}. We obtain similar results for the remaining energy bins. Furthermore, we find that the same three well-separated residuals are apparent when omitting $w_3$ from the analysis, though the residuals begin to merge when we omit $w_4$. The are shown in Figure~\ref{fig:waveletgaussianfits3}.

\section{Results with PASS 7 Data}
\label{app:Pass7}
\begin{figure*}
\centering
\includegraphics[width=0.49\linewidth]{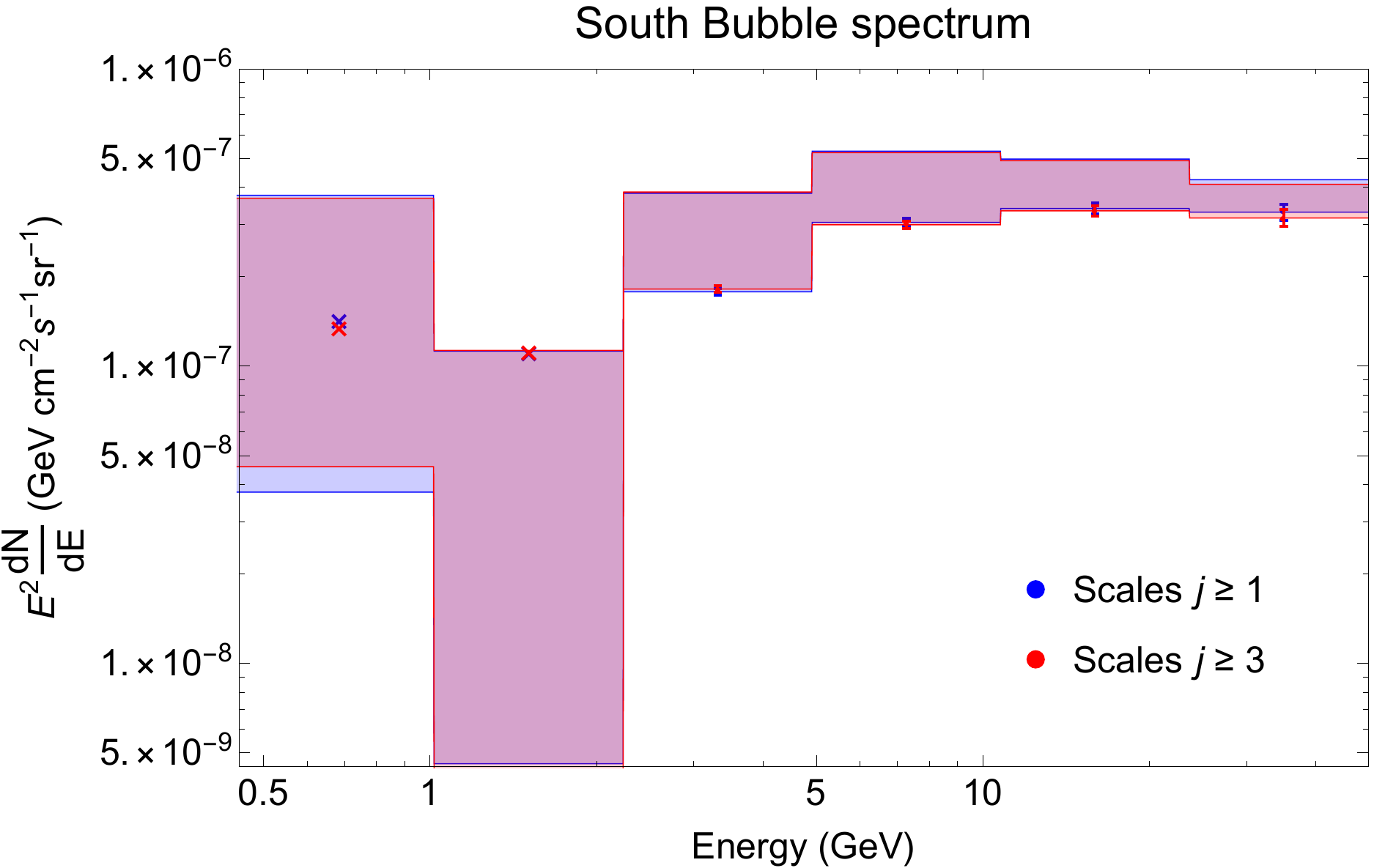}
\includegraphics[width=0.49\linewidth]{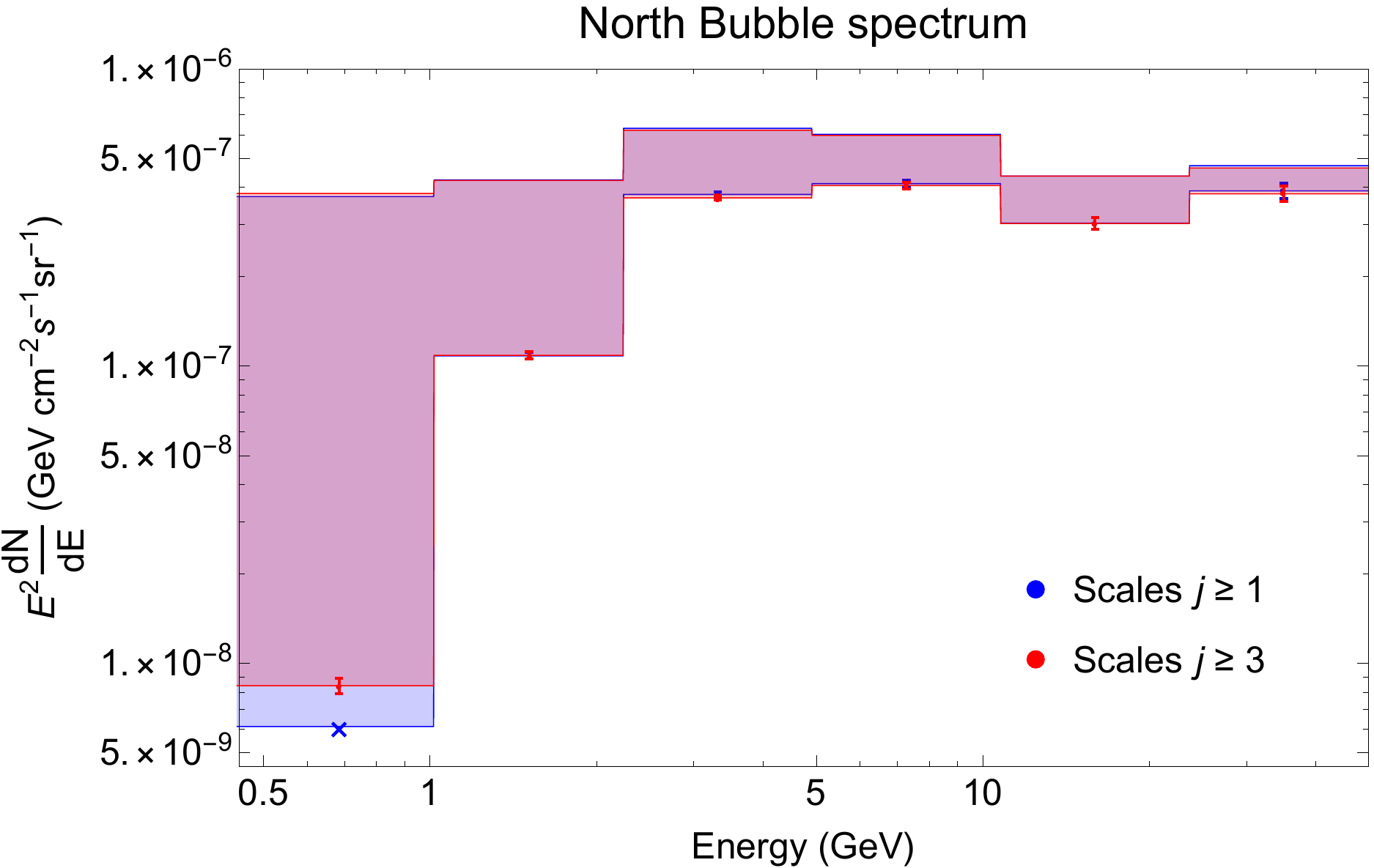}
\caption{Using PASS7 data (see~\ref{subsec:data}), the Southern (\textit{left} and Northern (\textit{right}) Bubbles' 
energy spectra. As with Figure~\ref{fig:sbubblespece2w3} we show  
results with all scales $j\geq1$ (blue) or $j\geq3$ (red). These two spectra agree with our PASS 8 Results and lead to the robust conclusion that the Bubbles are almost exclusively composed of diffuse emission.}
\label{fig:P7bubblespecrae2w3}
\end{figure*}

This work has focused on Pass 8 data. We briefly present here the basic results of section~\ref{sec:results} using instead
Pass 7 data from August 4th 2008 to August 14 2014 (see discussion in section~\ref{subsec:data}). We focus on the \textit{Fermi} 
Bubbles and the GCE. In Figure~\ref{fig:P7bubblespecrae2w3}, we give the Southern and Northern Bubbles spectra, derived 
using either all wavelet scales or just scales $j \geq 3$. The difference between the results in the two sets of scales is small. 
These spectra are in agreement with the Pass 8 spectra of Figure~\ref{fig:sbubblespece2w3}.

\begin{figure}[!ht]
\centering
\includegraphics[width=\linewidth]{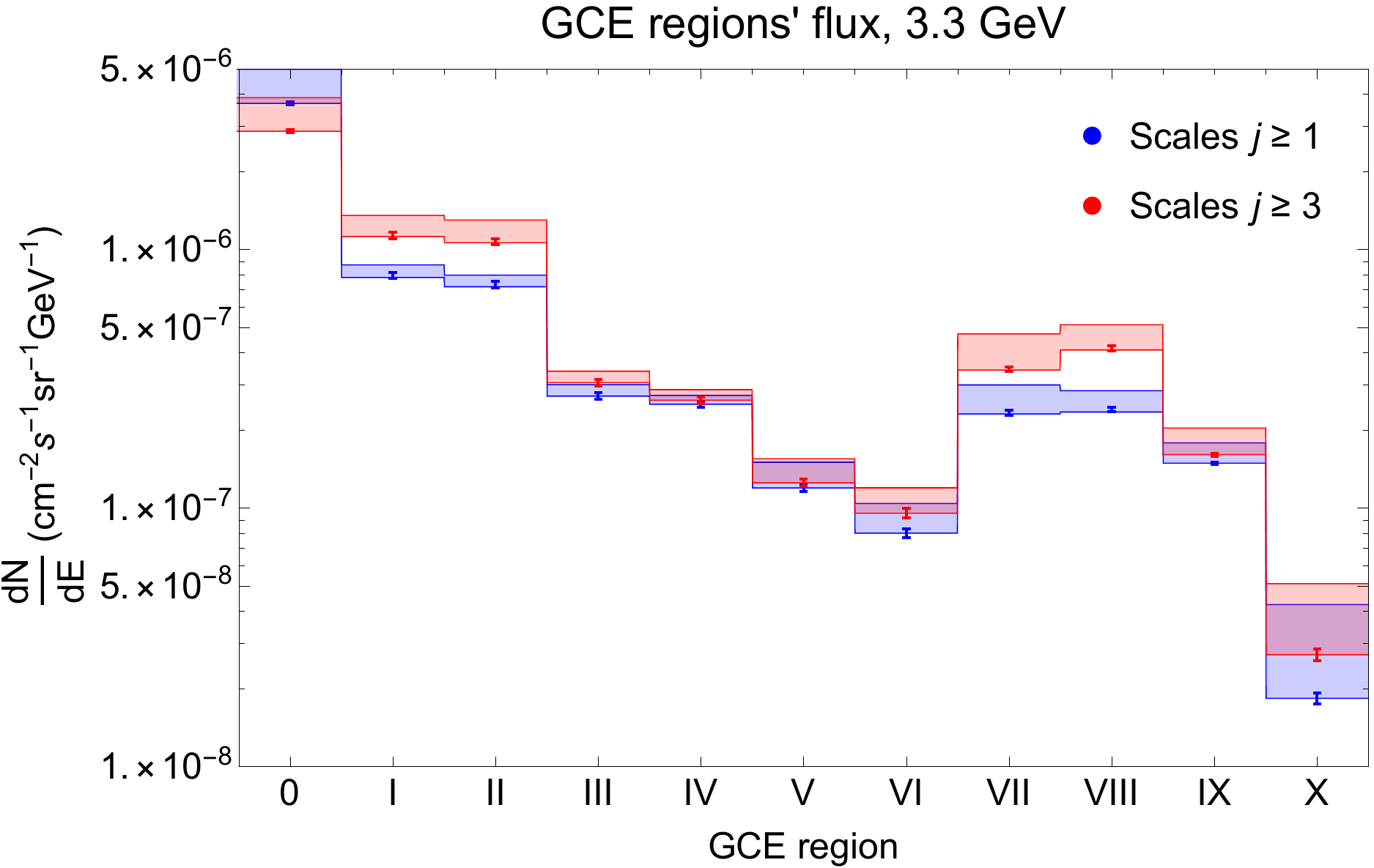}
\caption{Using PASS7 data, the flux around the Galactic center, at $E = 3.3$ GeV. 
As in Figure~\ref{fig:gcerege2} we show results with $j \geq 1$ (blue) or $j \geq 3$ (red). The difference between the fluxes in regions III, IV, V, VI, IX,
and X are up to $\sim$$10 \%$, while in regions 0, I, II, VII, and VIII they are between 30 and 100$\%$.}
\label{fig:P7gcerege2}
\end{figure}

In Figure~\ref{fig:P7gcerege2}, we show the equivalent result to Figure~\ref{fig:gcerege2}, for the GCE regions with Pass 7 data.
Like with Pass 8, regions III, IV, V, VI, IX, and X have little power in the smaller angular scales, with regions 0, I, II, VII, VIII having 
$\sim$30--50\% of their emission in the first two scales.	
	
\begin{figure*}[!h]
\centering
\includegraphics[width=.46\linewidth]{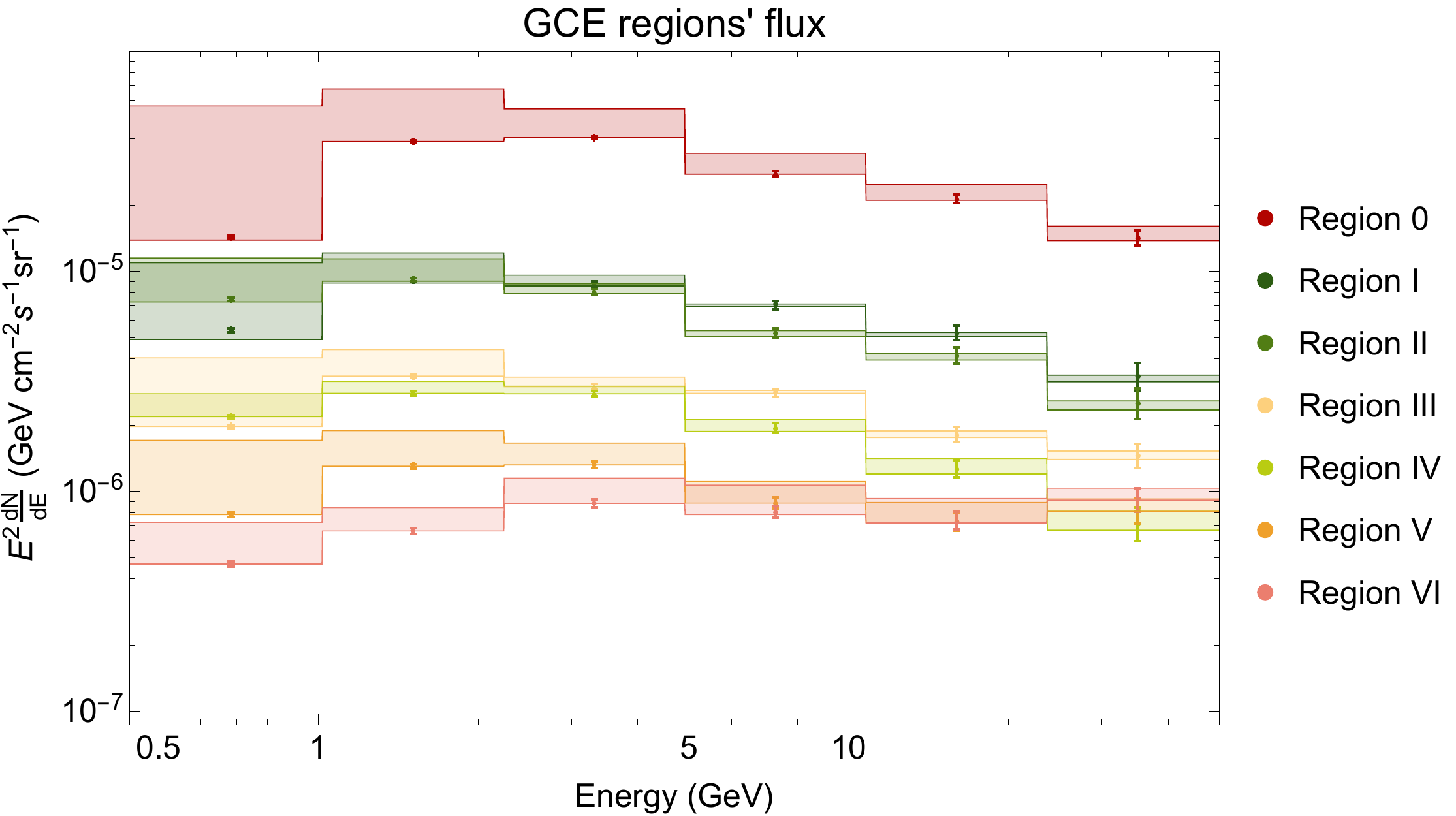}~~~~~
\includegraphics[width=.46\linewidth]{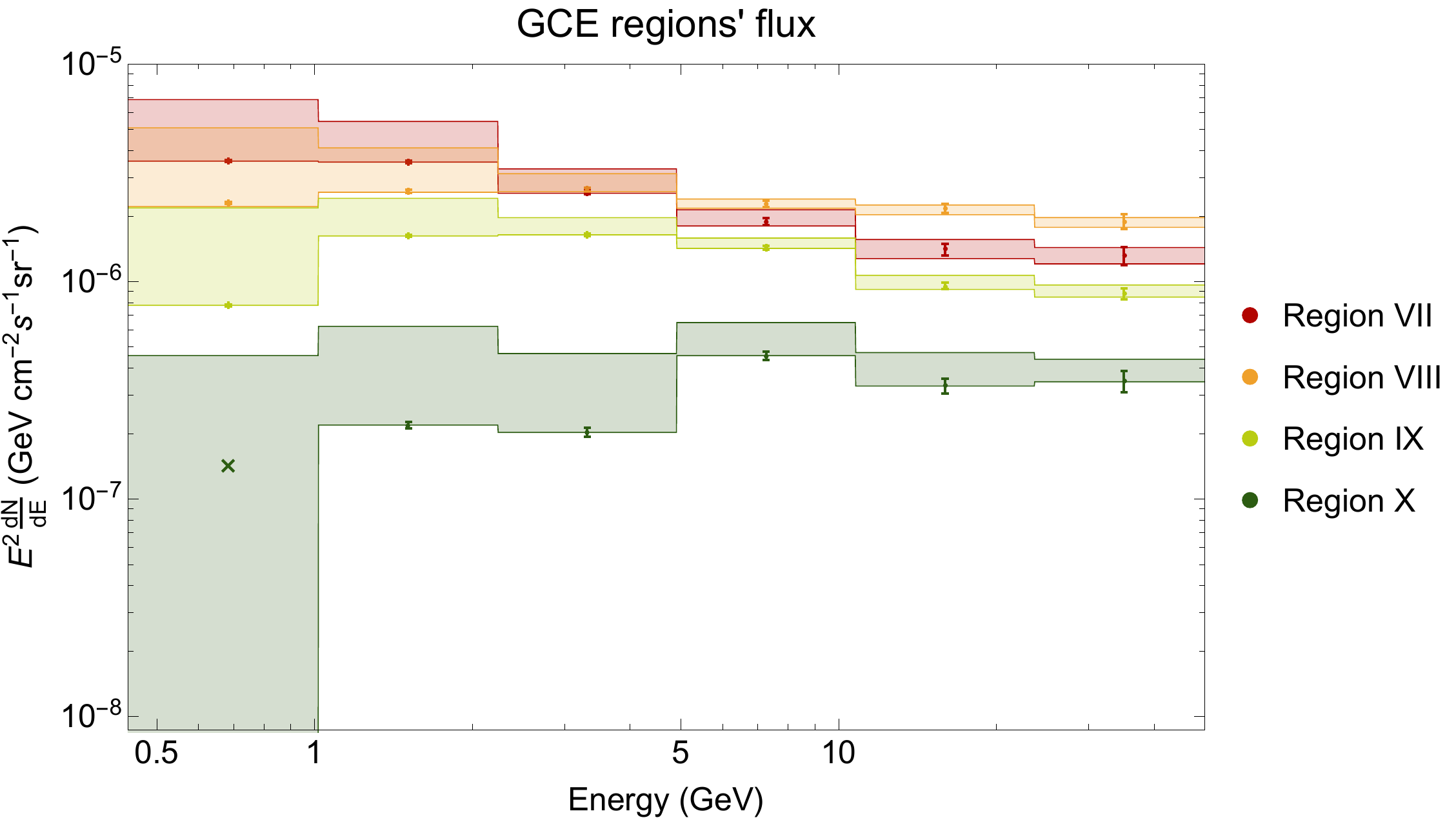}
\caption{Using PASS7 data, the flux spectra around the Galactic center. 
\textit{Top}, regions 0 and I--VI. \textit{Bottom}, regions VII--X. 
Our results agree with those in Figure~\ref{fig:gceregSpectra}.}
\label{fig:P7gceregSpectra}
\end{figure*}

Finally, in Figure~\ref{fig:P7gceregSpectra}, we give the spectra for the GCE in its subregions. Our spectra with Pass 7 are similar 
to those of Pass 8 given in Figure~\ref{fig:gceregSpectra}.	

\end{appendix}

\end{document}